\documentclass[aps,12pt,showpacs,prc,nofootinbib,floatfix,]{revtex4}

\usepackage{graphicx}

\usepackage{latexsym}
\usepackage{hhline}
\usepackage{amsmath}
\usepackage{amssymb}
\usepackage{wasysym}
\usepackage{epsfig}
\usepackage{ulem}

\sloppypar

\newcommand{\be}{\begin{equation}}
\newcommand{\ee}{\end{equation}}
\newcommand{\bea}{\begin{eqnarray}}
\newcommand{\eea}{\end{eqnarray}}
\newcommand{\ba}{\begin{array}}
\newcommand{\ea}{\end{array}}
\newcommand{\bi}{\begin{itemize}}
\newcommand{\ei}{\end{itemize}}

\newcommand{\refe}[1]{(\ref{#1})}
\newcommand{\nbox}[1]{\bf {#1}}

\newcommand{\g}{{\rm g}}

\newcommand{\mcl}{{\mathcal L}}

\renewcommand{\slash}{/ \!\!\!\!\,}

\newcommand{\foh}{\frac{1}{2}}

\newcommand{\DDelta}{\Delta(1232)}
\newcommand{\NN}{N^*}
\newcommand{\SSigma}{{\rm Im}\Sigma}

\newcommand{\fth}{\frac{3}{2}}

\newcommand{\done}{{\Delta_1}}
\newcommand{\dtwo}{{\Delta_2}}
\newcommand{\ch} {{\rm coher /incohr}}
\begin{document}

\title{$2\pi$  production in the Giessen coupled-channels model. 
\footnote{Supported by Transregio SFB/TR16, project B.7}}

\author{V. Shklyar}
\email{shklyar@theo.physik.uni-giessen.de}
\author{H. Lenske} 
\author{U. Mosel} 
\affiliation{Institut f\"ur Theoretische Physik, Universit\"at Giessen, D-35392
Giessen, Germany}

\begin{abstract}
	The coupled channels Lagrangian approach underlying the Giessen model (GiM) is extended 
%We present a coupled-channel Lagrangian approach (GiM) 
to  describe the $\pi N \to \pi N$, $2\pi N$  scattering in the resonance energy region.
As a feasibility study we investigate single and double pion production up to the second resonance region.
The $2\pi N$  production has been significantly improved  by using  the isobar approximation with   $\sigma N$ and $\pi \DDelta$ in the intermediate state.
The three-body unitarity is maintained up to interference pattern between the isobar subchannels.
The scattering amplitudes are obtained as a solution of the Bethe-Salpeter equation in the $K$ matrix approximation.
As a first application we perform a partial wave analysis of the  $\pi N \to \pi N$, $\pi^0\pi^0 N$ reactions in
the Roper resonance region.  We obtain  $R_{\sigma N}(1440)=27^{+4}_{-9}$\,\% and $R_{\pi \Delta}(1440)=12^{+5}_{-3}$\,\% for the $\sigma N$ and $\pi \DDelta$ 
decay branching ratios of $\NN(1440)$ respectively.
The extracted   $\pi N$ inelasticities and reaction amplitudes are consistent with the results from other groups.

\end{abstract}


\maketitle

 \section{Introduction}
The investigation of  properties of nucleon resonances remains  one of the primary goals of  modern hadron physics.
 The main information about the hadron spectra comes from the analysis of scattering data.
Coupled-channel approaches have   proven to be an  efficient tool to extract  baryon properties from experiment.
The Giessen coupled-channel model (GiM) \cite{Feuster:1998a,Feuster:1998b,Penner:2002a,Penner:2002b,shklyar:2004a,Shklyar:2006xw,Shklyar:2012js,Cao:2013psa,Shklyar:2005xg} 
has been  developed for a combined analysis of $(\pi/\gamma) N \to \pi N$, $2\pi N$, $\eta N$, $\omega N$ $K\Lambda$, $K\Sigma$  
 reactions to extract properties of nucleon resonances from pion- and photon-induced reactions.

Since the $\pi N \to 2\pi N$ reaction could account for up to 50\% of the $\pi N$ inelasticity this production channel had been included into the GiM calculations 
\cite{Feuster:1998a,Feuster:1998b,Penner:2002a,Penner:2002b,shklyar:2004a,Shklyar:2006xw,Shklyar:2012js,Cao:2013psa,Shklyar:2005xg}. 
However due to the complexity of the problem  the $2\pi N$ final state has been treated in a simplified way where only resonance  decays 
into a 'generic' $2\pi N$ final state were allowed. This simplified treatment allowed to maintain two-body unitarity and reproduce partial wave 
cross sections  extracted  by Manley et al in \cite{Manley:1984}. In view of the large contribution to the $\pi N$  inelasticity  it is  important to 
extend the calculations by treating three-body final states explicitly preserving three-body unitarity.

First, this approach would allow for the direct analysis of the $2\pi N$ experimental data. Since the corresponding Dalitz plots are found 
to be strongly non-uniform it is natural to assume that the main effect to the reaction comes from the resonance decays into isobar subchannels
\cite{Manley:1984}. 
The most important contributions are expected  to be from the intermediate  $\sigma N$, $\pi \DDelta$,  and $\rho N$ states. 
Analysis of the $\pi N\to 2\pi N$ reaction would therefore provide  very important  information about  the resonance 
decay modes into different  isobar final states. Presently  lattice simulations \cite{Edwards:2011jj,Dürr21112008} and  functional approaches 
\cite{SanchisAlepuz:2011jn}
succeeded  in calculation of the  spectrum of QCD. Therefore  unambiguous identification of the excited spectrum of baryons would provide 
an important link between theory and experiment. Similar to the constituent quark models \cite{Isgur,Capstick:1998uh}  the lattice QCD calculations demonstrate 
a much richer spectrum   \cite{Edwards:2011jj} of the  non-strange  sector of QCD than observed in scattering experiments so far. On the  
experimental side  most of  the non-strange baryonic states have been identified from the analysis of the elastic $\pi N$ 
data \cite{Arndt:2006bf,Cutkosky:1979fy,Hoehler:1993}. 
As pointed out in \cite{Isgur} the signal of excited states with a small  $\pi N$ coupling could be suppressed in  the elastic  $\pi N$ scattering. 
As a solution to this problem a series of photoproduction experiments has been done to accumulate enough data for study of the nucleon excitation spectra.  
However,   the results  from  the photoproduction reactions are still controversial. While recent investigations of the 
photoproduction reactions presented by the BoGa group \cite{Anisovich:2011fc} reported indications for some new resonances  not all of these  states are found in other calculations \cite{pdg}.
This raises a question about independent confirmation for the  existence of  such states from the investigations of  other  reactions.

Because of the   smallness  of  the electromagnetic  couplings   the largest contribution  to the  resonance self-energy comes from the hadronic decays. 
If the $N^*\to \pi N$ transition  is small one can expect sizable  resonance  contribution into remaining hadronic decay channels. 
As  a result the effect from  the resonance with a small
$\pi N$ coupling could still be significant in the inelastic pion-nucleon scattering: here the smallness of resonance coupling to the 
initial $\pi N$ states could be compensated by the potentially large decay branching ratio to other different  inelastic final states.  
 Such a scenario is realized e.g. in the case of the well known $\NN(1535)$ 
state. While the effect from this resonance to the elastic $\pi N$ scattering is only moderate  at the level of total cross section its  contribution 
to the $\pi N \to\eta N$ channel turns out to be 
dominant \cite{Shklyar:2012js}. Since the $\pi N \to 2\pi N$ reaction could account for up to 50 \% of the total  $\pi N$ inelasticity this channel becomes
very important not only for the investigation  of the properties of already  known resonances  but also for the  search for  the  signals of possibly  
unresolved  states. 

Another important issue in studies of the $2\pi N$  channel  is related to the possibility to investigate  
cascade transitions like ${N^*}'\to\pi N^*\to \pi\pi N$, where a massive state ${N^*}'$ decays via intermediate  excited  
$N^*$ or $\Delta^*$.
It is interesting to check whether such decay modes are responsible for  the large decay width of higher lying  mass states.
%So far  such decay  have only been considered in $\cite{Manley:1984}$  where authors considered resonance decay into $\pi N(1440)$ isobar channel. 
So far only the $\pi N^*(1440)$ isobar channel  has  been considered in \cite{Manley:1984} in a partial wave analysis (PWA)
 of the $\pi N\to 2\pi N$  experimental data \cite{Manley:1984}.

There are several complications in the coupled-channel analysis of   $2\to 3$ transitions.  
The  first one is the difficulty to perform  the partial-wave decomposition
of the three-particle state.  
 The second complication is related to the issue of  three-body unitarity. 
  For  a full dynamical treatment of the $2\to 3$  reaction the  Faddeev  equations have to be  solved.
This makes the whole problem quite difficult for  practical implementations.
Here we address both issues and present a coupled-channel approach for solving the $\pi N \to 2\pi N$ 
scattering problem in the isobar approximation.   
  In this formulation the $(\pi/\pi\pi) N\to  (\pi /\pi\pi) N$  coupled-channel equations are  reduced to the
two-body scattering equations  for  isobar production.  Such a description accounts by construction
for the full spectroscopic strength of intermediate channels and, in addition provides a considerable 
numerical simplification.
Three-body unitarity leads to a relation between the  imaginary part of the  
elastic scattering amplitude  and the sum of the total elastic and inelastic cross sections by the well known   optical theorem. Since in  the isobar approximation the pions in the $\pi\pi N$  channel are produced from the 
isobar subchannels all contributions to the total $\pi N \to \pi\pi N$ cross section  are driven by the isobar 
production. The optical theorem can be fulfilled if all discontinuities in isobar subchannels are taken into
account.
  In the present work the   three-body unitarity is maintained  up to 
interference term between the isobar subchannels.

 As a first application of our model we apply the developed approach for  the study 
of the $\pi^- p \to \pi^0\pi^0 n$ data in the first resonance energy region assuming the dominant $S_{11}$ and $P_{11}$ partial wave contributions in the $\sigma N$ and
$\pi \Delta$ reaction subchannels. 
The main purpose of this paper is to introduce the model and demonstrate the feasibility of treating two-pion dynamics in the framework 
of a large-scale coupled channels approach. For this aim, we restrict the calculations to the $\pi^0\pi^0 n$ channel, taking advantage of the fact that only isoscalar two-pion and  $\pi\Delta$ isobar channels are contributing to the process. We emphasize that this restriction is not a matter of principle but is only for the sake of a feasibility study. In particular, this means that at this stage we do not consider the $\rho N$ state  but postpone its inclusion into the numerical scheme to a later stage. Naturally, the results presented in the following are most meaningful for the energy region of the $N^*(1440)$ Roper resonance.  
%In the present work we do not include the  $\pi^+\pi^-N $ 
%reaction data due to possible effect from the $\rho$ meson dynamics which is missing in the present approach. The future extensions of the model including the isobar $\rho N$ channel

 The first resonance energy region  is of particular interest because of the sizable  effect from  $\NN(1440)$.  The dynamics of the Roper resonance turns out to be rich because of the two-pole structure reported in earlier studies \cite{Arndt:1985vj,Cutkosky:1990zh}, (see
 \cite{Arndt:2006bf,Doring:2009yv,Suzuki:2009nj} for the recent status of the problem. )
 The origin of the Roper resonance is also controversial. For example the calculations in the J\"ulich model  explain this state as a dynamically generated pole due to the strong attraction in the $\sigma N$ subchannel.
At the same time the Crystal Ball collaboration finds no evidence of   strong $t$-channel  
 sigma-meson production in their $\pi^0\pi^0$ data \cite{Craig:2003yd}. 
From the further analysis of the $\pi^0\pi^0$ production the effect of the sigma meson was found to be small \cite{Prakhov:2004zv}. 
On other hand the $pp\to pp\pi^0 \pi^0$ scattering experiment by CELSIUS-WASA collaboration  \cite{Skorodko:2008zzb}   finds the
 $\sigma N$ decay mode of the Roper resonance to be dominant.

The $2\pi N$  decay properties  of the Roper resonance  are  also under discussion. 
The recent multichannel analysis from the Kent group \cite{Shrestha:2012ep} gives the branching ratios for the $\sigma N$ and $\pi \DDelta$ decay modes 
${\rm R}_{\sigma N}^{N(1440)}=27\pm 1\%$  and  ${\rm R}_{\pi \DDelta}^{N(1440)}=6.8\pm 0.8\%$, respectively. At the same time  Anisovich et al \cite{Anisovich:2011fc} 
obtain a significantly larger decay fraction   for the $\pi\DDelta$ channel ${\rm R}_{\pi\DDelta}^{N(1440)}=21\pm 8\%$ and somewhat smaller for the $\sigma N $:
${\rm R}_{\sigma N}^{N(1440)}=17\pm 7\%$. 

In view of these problems we perform an  analysis of the Crystal Ball data \cite{Prakhov} 
assuming  dominant contributions from the $S_{11}$   and $P_{11}$  amplitudes in the isospin $I=\foh$
channel.  Since the effect from $\NN(1520)$ is expected to be important above 1.46 GeV 
we have limited the  present  calculations up to $\sqrt{s}=$1.46 GeV energy region. Using unitarity the  contribution  from $\NN(1520)$
to the total  $2\pi N$ cross section could be estimated. The effect from the latter state is thus can be taken into account  for  the 
error estimation of the extracted parameters  of the $\NN(1440)$ resonance. The  interaction kernel used in the scattering equation  is calculated from 
the corresponding  interaction   Lagrangians. Though the effect from the background terms are found to be small, the 
t-channel  pion exchange of the $\sigma$ meson production  turns out to be important  close to threshold.

The $\sigma N$ decay fraction of $\NN(1440)$ is found to be dominant. The extracted value $R_{\sigma N}\approx 27\%$ is about two times larger than
that of  $\pi\Delta$: $R_{\pi \Delta}\approx 12\%$. The extracted partial waves of the isobar production  are close to the single energy solutions(SES) from  the analysis of Manley et al 
\cite{Manley:1984} except for the sign at the real part of the $\sigma N$ reaction amplitudes. The calculated $S_{11}$ and $P_{11}$ inelasticities demonstrate 
a good agreement with the results from the GWU group \cite{Arndt:2006bf}.
 In the present study the Roper resonance is treated as a genuine pole.  An alternative scenario would be  to describe  $\NN(1440)$ in terms of a dynamical pole.
 However, if such a pole is generated by the, e.g., $t$-channel  exchange  in the isobar production channel the angular 
dependence of the reaction amplitude could be different from the case with genuine pole. One may hope that the detailed analysis of the $2\pi N$
production channels could help to disentangle  the different scenarios. The forthcoming  measurements of the $\pi N \to 2\pi N$ reaction
at HADES and JPARC facilities  provide a new possibility to solve these long standing problems in the non-strange baryon spectroscopy.

The paper is organized as follows: in Section  \ref{over_view} we present a  short overview of the partial wave analysis of the $\pi N \to 2\pi N$ reactions. 
The details of the Giessen Model (GiM) are presented in Section \ref{GiM_model}.  The impact of the isobar dynamics on the data analysis is presented in 
Section  \ref{section_isobar_contributions}.
The results of the calculations and the partial wave analysis are 
discussed in Section \ref{results}.
%Resonances with a small $\pi N$ coupling  have in general quite large total decay width.
%In case of inelastic  scattering the smallnest of resonance coupling to the initial $\pi N$
%states could be compensated by the potentially large decay branching ratio to the inelastc channel.  

 \section{Overview of the $\pi N \to 2\pi N$ reaction \label{over_view}}
Here we present a short overview of the analyses of the $\pi N \to 2\pi N $ reaction made so far. Further details can be found
in the  papers cited in the present section.  
One of the most   extensive studies of  $2\pi N$ production in the resonance energy region has been made by Manley et al in \cite{Manley:1984}.
There a partial wave analysis of the $\pi N\to2\pi N$ experimental data was performed within the isobar approximation.
The database consisted of old 241214 bubble chamber events in the energy region  1.320-1.93 GeV taken before 1984.
No $\pi^0 \pi^0 n$ data were available at the time. 
By binning the events into 22 energy bins and performing a sophisticated truncation scheme to reduce the number of independent 
amplitudes  partial wave contributions  were obtained for each isobar channel.  The dependence on the energy of the isobar
was neglected and  neither two- or three-body  unitary was  explicitly maintained.  The main result of the work  \cite{Manley:1984} are single energy solutions (SES) 
  extracted for each isobar channel in every energy bin. Since the   dependence on isobar subenergy  
was neglected in \cite{Manley:1984} the derived solutions are simple functions  of the c.m. energy.

In general the PWA of  experimental data  does not provide direct information about  N* spectra: it  only helps to disentangle contributions into
 the different partial wave amplitudes   using conservation laws for  total spin, parity and isospin. 
To investigate the reaction dynamics theoretical energy-dependent  amplitudes should be defined. 
A specific parameterization 
of the scattering amplitude could be used to construct the scattering amplitudes.
The non-resonant contributions can be 
parameterized in terms of smooth polynomial functions or by distant poles.
The dynamical approaches are based  on  solving   relativistic scattering equations to calculate the transition amplitudes.
These calculations pursue the description of the scattering process in terms of mesons and baryons as the  effective degrees of freedom of low-energy
QCD.  Since the interaction kernel  is obtained from the given Lagrangian densities  important  constraints, e.g. chiral symmetry, can be also respected.

 Since we describe the pseudoscalar vertices by derivative couplings \cite{Penner:2002a,Feuster:1998a} the Giessen model accounts for a central requirement of chiral symmetry,
at least at the minimal level of interaction vertices. 
In view of the large
energy range covered by our analyses we have been using in the past a mesonic picture by explicitly using scalar and
vector mesons as the relevant degrees of freedom rather than following the scheme of chiral perturbation theory and expressing those channels in terms of a perturbative multi-pion expansion.
In this paper, we are making  a first step towards a more detailed
description of meson production on the nucleon by treating explicitly the two-pion resonance nature of a selected subset of the heavy mesons. 
We emphasize again that we are well aware of the limitations of such a restricted approach, primarily intended as a feasibilty study of two-pion production in the frame work of a coupled channels approach. 
Our approach is based on a field-theoretical Lagrangian formulation, fully accounting for Lorentz-invariance and the relevant
internal symmetries.

The next step is to constrain the theoretical amplitudes to SES  and fix the resonance parameters. 
%Alternatively the corresponding  observables calculated from the
%theoretical amplitudes  can be directly compared to the experimental data to obtain the resonance parameters without the intermediate step in of obtaining SES.    
Alternatively the calculated amplitudes could also be  fitted directly to the data without an PWA analysis of experimental observables as an  intermediate step. 
 
Several studies \cite{Manley:1992yb,Shrestha:2012ep,Vrana:1999nt} have been made to extract the nucleon excitation spectra 
from the single-energy solutions (SES) derived in \cite{Manley:1984}. The KSU approach is based on multichannel parameterization 
of the scattering matrix in the form $S=(1+iK)/(1-iK)$ within K-matrix formalism \cite{Manley:1992yb} whereas the calculations \cite{Vrana:1999nt} utilize 
CMB ansatz  of Cutkosky et al \cite{Cutkosky:1979zv}. 
While both approaches are able to  maintain at least two-body unitarity their PWA amplitudes are fitted to the 
single energy solutions  from \cite{Manley:1984} which are obtained by neglecting this constraint.

A combined analysis of the $ \pi^0 \pi^0 N$ production channel  from  the $\gamma p$ and $\pi^- p$  scattering  has  been presented in \cite{Sarantsev:2007aa}.
The authors do not use the SES from  \cite{Manley:1984} but fit the calculated observables directly to the experimental data. One of the interesting 
conclusions made in \cite{Sarantsev:2007aa} is that in  photoproduction  the background contributions to the $\pi^0 \pi^0$  production are as 
large as the resonant one.  Note that the analysis of the two-pion photoproduction data is more involved due to 
 the complications related with the gauge invariance \cite{Haberzettl:2012ew}.
On the other hand the non-resonant terms play only a minor role  in the $\pi^- p\to \pi^0\pi^0 n$ scattering \cite{Sarantsev:2007aa}. As a result the latter 
reaction is better suited for an  investigation of the properties of the Roper resonance \cite{Sarantsev:2007aa}.
The decay width of $\NN(1440)$ is found to be $\Gamma_{\sigma N}^{N(1440)}=71\pm 17$~MeV  and $\Gamma_{\pi\DDelta}^{N(1440)}=59\pm 15$~MeV which 
leads to  the slightly larger $\sigma N$ decay fraction.   The updated analysis   \cite{Anisovich:2011fc} gives
 ${\rm R}_{\pi \DDelta}^{N(1440)}=21\pm 8\,\%$ and  ${\rm R}_{\sigma N}^{N(1440)}=17\pm 7\,\%$ which are different from \cite{Shrestha:2012ep}.

Several dynamical approaches have been developed to investigate the $\pi N\to 2\pi N$ scattering. 
The J\"ulich model \cite{Krehl:1999km,Ronchen:2012eg} obtains the scattering amplitudes  by solving the Lippmann-Schwinger equation where the two-pion production
is treated in the isobar approximation. One of the interesting results obtained in  J\"ulich model is that the Roper resonance 
could be represented by a dynamically generated pole due to the correlations in   the   $\sigma N$ subchannel. However  no direct 
comparison of their  calculations  to the 
$\pi N \to \pi \pi N$  experimental data has been made so far \cite{Ronchen:2012eg}.

An investigation of the properties of   $\NN(1440)$  has been presented in  \cite{Kamano:2006vm} where the authors also applied the isobar approximation.
A set of chiral constraints has been used to derive amplitudes for the  $\pi N\to\pi\DDelta$, $\sigma N$ transitions.
The results of calculations are compared with the total experimental $\pi N \to 2\pi N$  cross sections and the Crystal Ball measurements of the  $\pi^0\pi^0n$ 
production. Both the $\pi \DDelta$ and the $\sigma N$ decay modes of $\NN(1440)$ are found to be important \cite{Kamano:2006vm}.

There are several studies of  the $\pi N \to 2\pi N$ reaction within the chiral perturbation theory \cite{Fettes:1999wp,Bernard:1995gx,Bernard:1997tq,Mobed:2005av}.
In general the chiral calculations in the heavy baryon limit demonstrate a nice agreement with experiment in the low energy region.
One of the important results from the chiral calculations  is that the effect from one loop diagrams is negligible \cite{Fettes:1999wp}.
By fixing low energy constants from the comparison with the $\pi^- p \to \pi^-\pi^+ n$ experimental data  the predictions for the other charge transitions are given.
Though the contributions from excited states are encoded into low energy constant (LEC's) the analytical structure of the scattering amplitude could be quite 
different from the case when  resonance  are  explicitly included into calculations.  
Recent calculations including $\DDelta$-isobar are presented in \cite{Siemens:2014pma}.

In addition several meson exchange models have also been used to attack the problem \cite{Schneider:2006bd,Oset:1985wt}. 
The authors of  \cite{Schneider:2006bd} apply a tree-level parameterization to evaluate two-pion production. It is interesting that   the authors 
do not use the Breit-Wigner parameterization but obtain the vertices  and the propagators at the tree-level diagrams   by solving
dynamical equations. In addition to $\NN(1440)$ several
additional states have also been included into the calculations. Close to threshold the findings  of \cite{Schneider:2006bd} demonstrate a nice agreement with 
experiment. These results also support  a large contribution from the 
Roper resonance to the $\pi^- p \to \pi^+ \pi^- n$  and  $\pi^- p \to \pi^0 \pi^0 n$ 
reactions which is in  line with the conclusions of \cite{Kamano:2006vm,Sarantsev:2007aa}. Though the authors of \cite{Schneider:2006bd} do not give their
decay branching ratios for resonance decays,  one might  expect a sizable  ${\rm N^{\foh+} (1440)}\to\sigma n$ decay fraction  from the large  
$\g_{\sigma NN(1440)} \gg \g_{\pi \DDelta N(1440)}$  coupling constant in \cite{Schneider:2006bd}.

Another dynamical approach  to solve the coupled-channel problem for the two pion production is presented in \cite{Matsuyama:2006rp,Kamano:2008gr}. 
This approach is close in spirit to \cite{Ronchen:2012eg}.
 The model aims to go beyond the isobar approximation having  both dispersive contributions and unitarity cuts under control. 

Since the full calculations require  large  computation efforts  only a limited  amount  of $\gamma/\pi N\to \pi\pi n$ experimental data has been 
analyzed, see   \cite{Kamano:2013ona,Kamano:2013iva} and references therein.  
One of the complications reported in   \cite{Matsuyama:2006rp} 
is  the appearance  of the moving singularity in the scattering amplitude. Presently it is not clear whether the phenomena has a physical origin or is related to the treatment of  
the scattering problem  in the Euclidean space.

 \section{Coupled-channel unitary model 
 	for  $\pi N\to 2\pi N$ scattering\label{GiM_model}}

 \subsection{The issue of unitarity}
Unitarity is a one of the important key issues   in the partial wave analysis. This constraint is maintained in a coupled-channel 
treatment of the scattering problem. The requirement   that the sum of all transition probabilities should be 1 leads to the condition $SS^+=1$ for the
 scattering $S$-matrix. This gives 
\bea
-i(T-T^+) = TT^+ 
\label{unitar1}
\eea
for the  $T$-matrix in the  operator form. On the amplitude level it leads to a relation between the imaginary part of the elastic scattering amplitude and
the transition  probability summed over all elastic and inelastic asymptotic channels
\bea
{\rm Im} T_{ii} = \sum_j T_{ij}T^+_{ij} = \sum_j |T_{ij}|^2,
\label{unitar2}
\eea
where indices $i,j$ denote incoming  and outgoing asymptotic  final states,  e.g. $\pi N$, $2\pi N$ etc,  and  their   quantum numbers.
The summation in Eq.~\refe{unitar2} stands for   summation over spin, isospin  and integration  over intermediate particle momenta. From  the form 
of  relation Eq.~\refe{unitar2} follows that the scattering amplitude  $T$ is a matrix of dimension N$\times$N where N is  the number of all open channels
and independent spin-isospin combinations.

\begin{figure}
  \begin{center}
    \includegraphics[width=8cm]{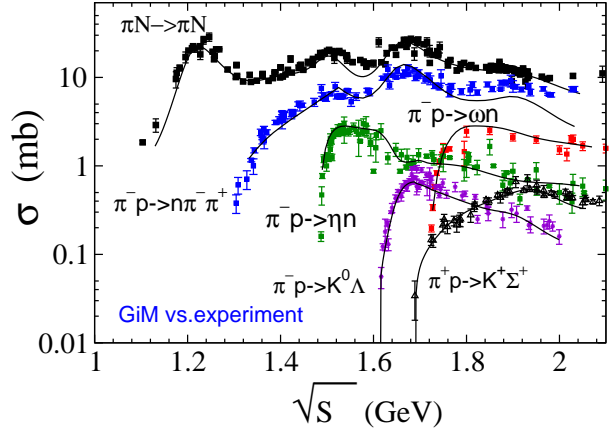}
    \caption{(Color online) total cross sections for pion-induced reactions calculated  in  Giessen model 
    \cite{shklyar:2004a,Shklyar:2006xw,Shklyar:2012js,Cao:2013psa,Shklyar:2005xg} vs. experimental data.   
      \label{unitar5}}
  \end{center}
\end{figure}

For the sake of  simplicity we consider here  $\pi N$ scattering below the $3\pi N$ threshold.  Then only $\pi  N$ and $2\pi N$ final states are important on the
right side of  Eq.~\refe{unitar2}; the electromagnetic processes can be neglected. 
 The  quantity $T_{ii}$  denotes a scattering amplitude  for elastic  transitions where all quantum 
numbers (including momenta)  of the particles in the 
in-state are identical to those in the out-state. This can only take place for the elastic scattering at 
forward directions. Then, 
for the elastic $\pi N$ scattering one can write   $T_{ii}= T_{\pi N}^{\rm els.}(0)$,  where  
$T_{\pi N}^{\rm els.}(0)$ is the $\pi N$ elastic scattering amplitude  
for forward angles. Eq.~\refe{unitar2} can be rewritten in the form of the optical theorem 
\bea
%{\rm Im} T_{\pi N}^{\rm els.}(0) = \frac{k^2}{4\pi } ( \sigma_{\pi, \pi} + \sigma_{\pi, 2\pi} ),
{\rm Im} T_{\pi N}^{\rm els.}(0) = \frac{k^2}{4\pi } ( \sigma_{\pi N \to \pi N} + \sigma_{\pi N \to 2\pi N} ),
\label{unitar3}
\eea
where $k$ is a c.m. momentum of the initial $\pi N$ state  and $\sigma_{\pi N \to \pi N }$ and   $\sigma_{\pi N \to 2\pi N }$  denote 
the total $\pi N \to \pi N$ and $\pi N \to 2\pi N$ cross sections respectively. 
For higher energies the right hand side of  equation Eq.~\refe{unitar3} can include contributions from other open channels. 
The importance of the optical theorem for the data analysis can be seen after the partial wave decomposition of 
Eq.~\refe{unitar3}:
\bea
{\rm Im} T_{\pi N}^{JP}= \frac{k^2}{4\pi } ( \sigma_{\pi N \to \pi N}^{JP} + \sigma_{\pi N \to 2\pi N}^{JP}),
\label{unitar4}
\eea
where the  subscripts $J$ and $P$ stand for the total spin and the parity. As a consequence  the imaginary part of the given
elastic $\pi N$  partial wave is a sum over all elastic and inelastic total partial wave cross sections. The results of GiM-calculations 
\cite{shklyar:2004a,Shklyar:2006xw,Shklyar:2012js,Cao:2013psa,Shklyar:2005xg} are shown in Fig.~\ref{unitar5}.   
All reactions turn out to be linked via  Eq.~\ref{unitar3} and its PWA-version,  Eq.~\refe{unitar4}. The largest inelastic
contribution to the left-hand part of Eqs.~(\ref{unitar3},\ref{unitar4}) comes from the $\pi N \to 2\pi N$ reactions. Hence it
is  very important 
that the $2\pi N$ channel is included in the PWA for the baryon resonance analysis. In our earlier  work the $2\pi N$ final
state was treated as a 'generic' channel to control inelasticities   associated with the $\pi N \to 2\pi N$ reaction, 
see \cite{Penner:2002a,Penner:2002b}. Here we discuss the extension of the model to incorporate the $2\pi N$ channel within the isobar 
approximation.

 \subsection{Giessen Model (GiM)  for the $\pi N \to \pi N$, $ 2\pi N$ transitions\label{scatt.equation} }

\begin{figure}
  \begin{center}
    \includegraphics[width=14cm]{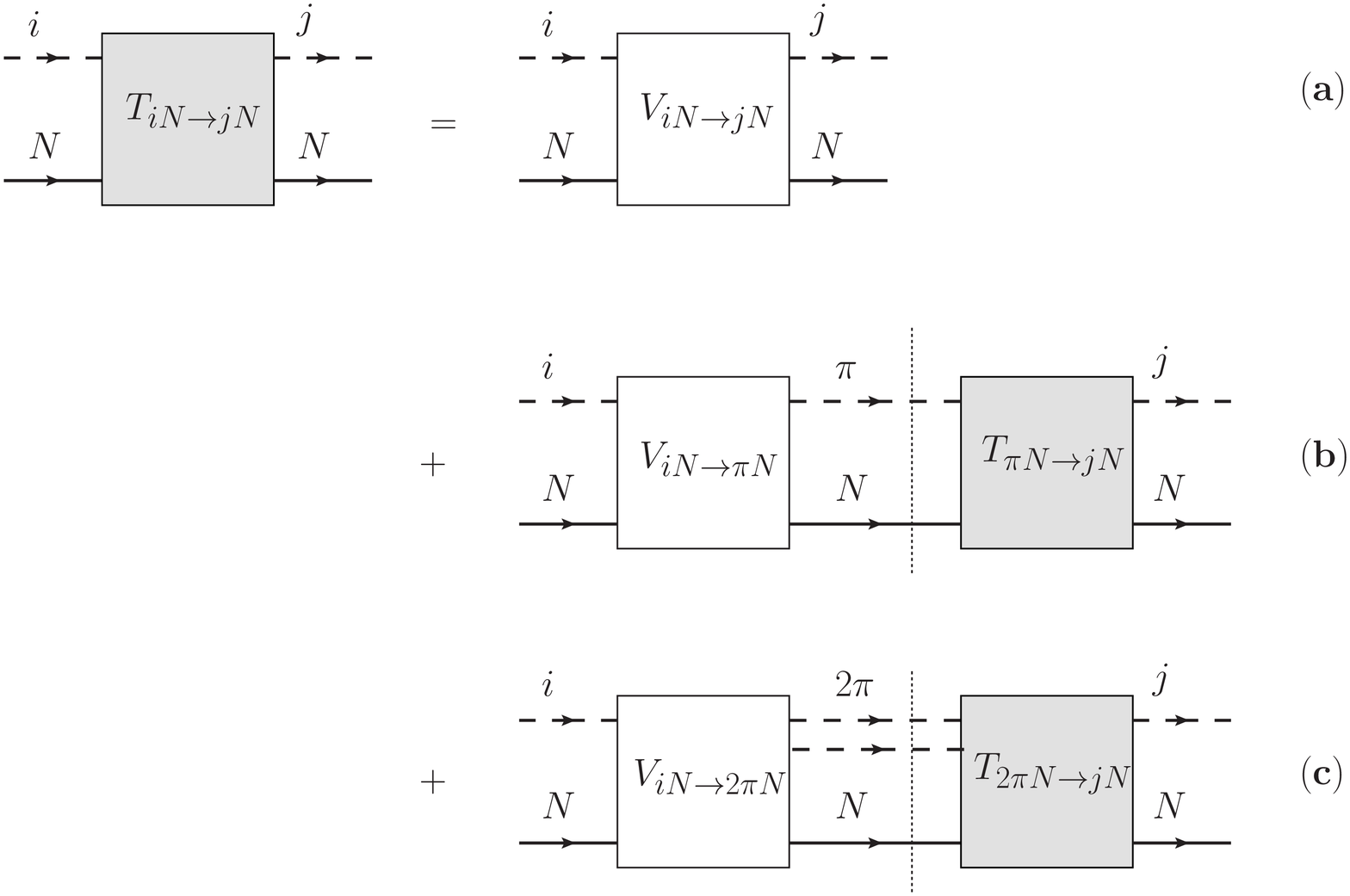}
    \caption{Scattering equation for the  process $i N \to j N$ where  $i, j$=$\pi N$,$2\pi N$;  $V_{i N \to j N}$ stands for the interaction kernel. 
      \label{unitar6}}
  \end{center}
\end{figure}

Though unitarity is a general property which is  independent of the specific form of the scattering equation to be solved 
it is easier to  consider the problem for the example of the Bethe-Salpeter equation in the ladder approximation. 
There are three amplitudes $T_{\pi N \to \pi N}$, $T_{\pi N \to 2\pi N}$
and $T_{2\pi N \to 2\pi N}$ which are obtained by solving the system of scattering equations. The structure of these equations  
is  represented diagrammatically in Fig.~\ref{unitar6}, where $V_{i N\to jN}$ denotes the interaction kernel.
In order to  fulfill the optical theorem,  Eqs.~(\ref{unitar3},\ref{unitar4}), the $\pi N$ and $2\pi N$ unitarity cuts have to be taken into account.
The optical theorem for the $2\to 2$ transitions has  thoroughly been  discussed in \cite{shklyar:2004a,Penner:2002a,Penner:2002b}. Here 
we concentrate on the last term  in the scattering equation which contains  the $2\pi N$ loop, see Fig.~\ref{unitar6}(c). 

\begin{figure}
  \begin{center}
    \includegraphics[width=14cm]{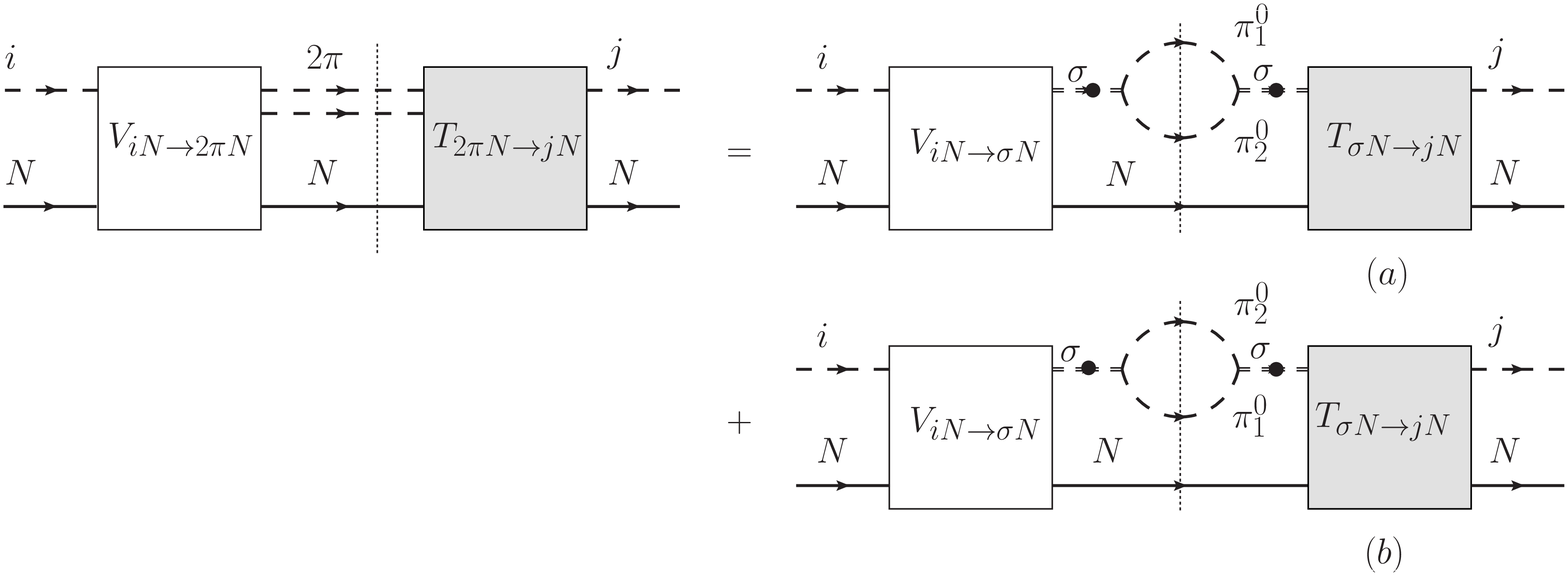}\\
\vspace*{5mm}
    \includegraphics[width=14cm]{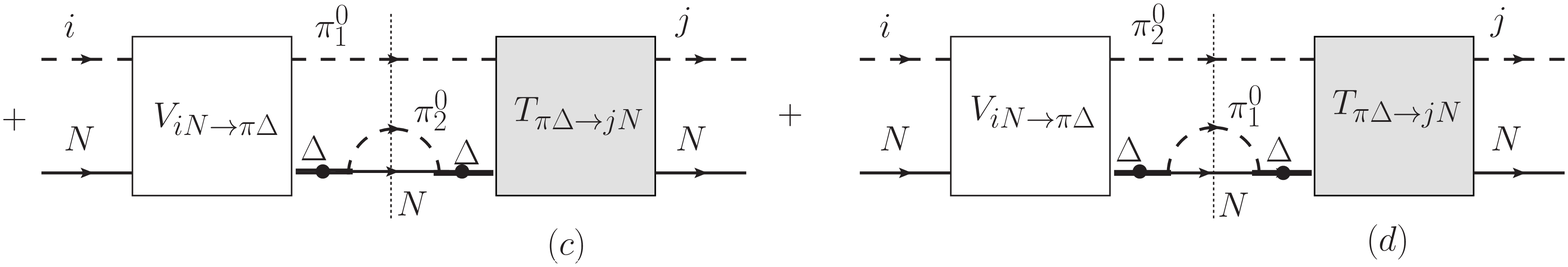}\\
\vspace*{5mm}
    \includegraphics[width=14cm]{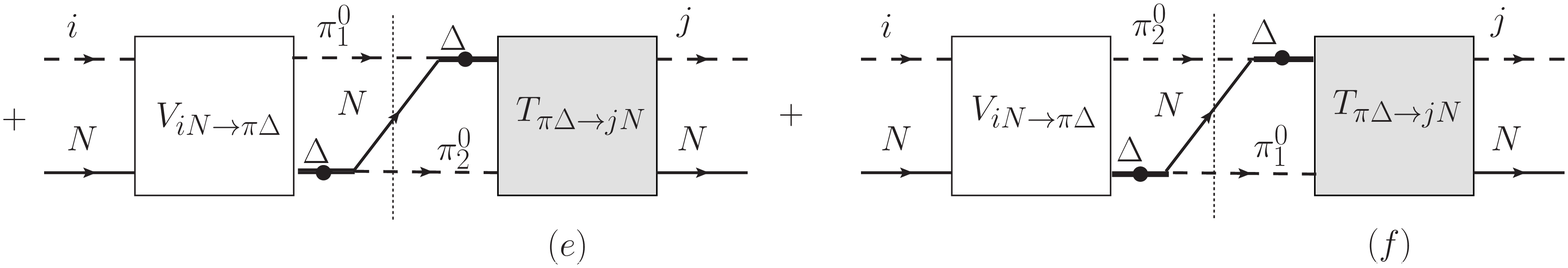}\\
\vspace*{5mm}
    \includegraphics[width=14cm]{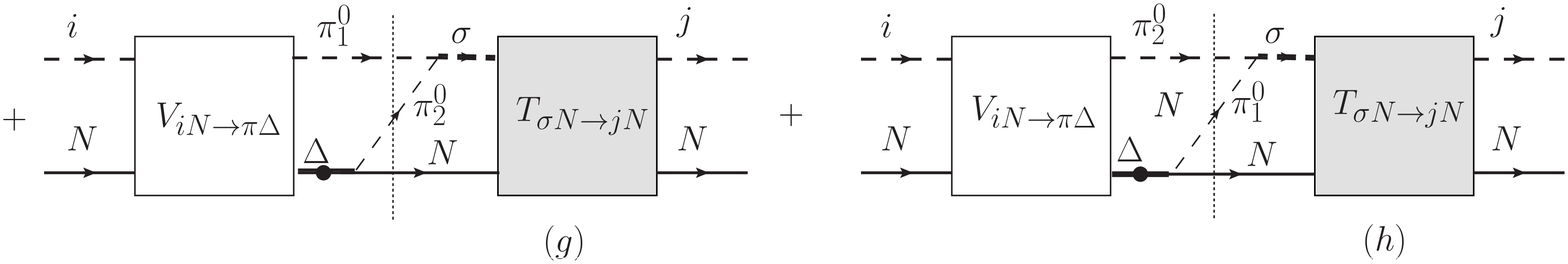}\\
\vspace*{5mm}
    \includegraphics[width=14cm]{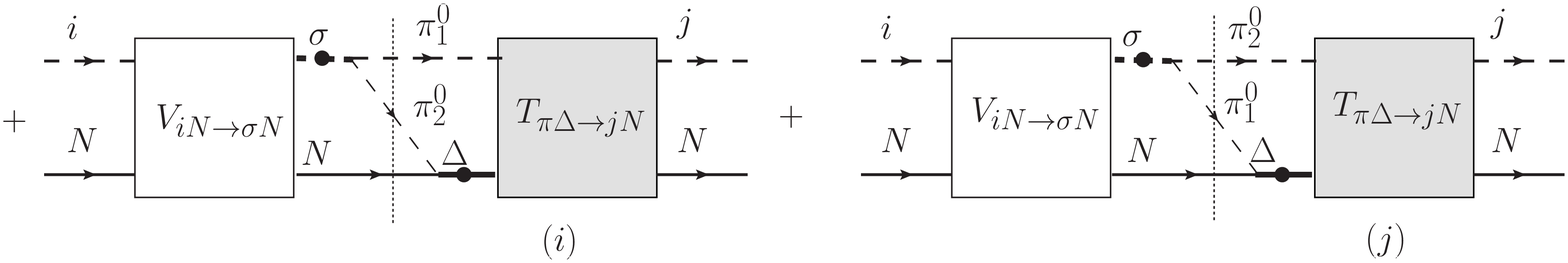}
    \caption{Representation of the graphs depicted in Fig.~\ref{unitar6} (c) in term of the isobar contributions.  
      \label{unitar7}}
  \end{center}
\end{figure}

\begin{figure}
  \begin{center}
    \includegraphics[width=14cm]{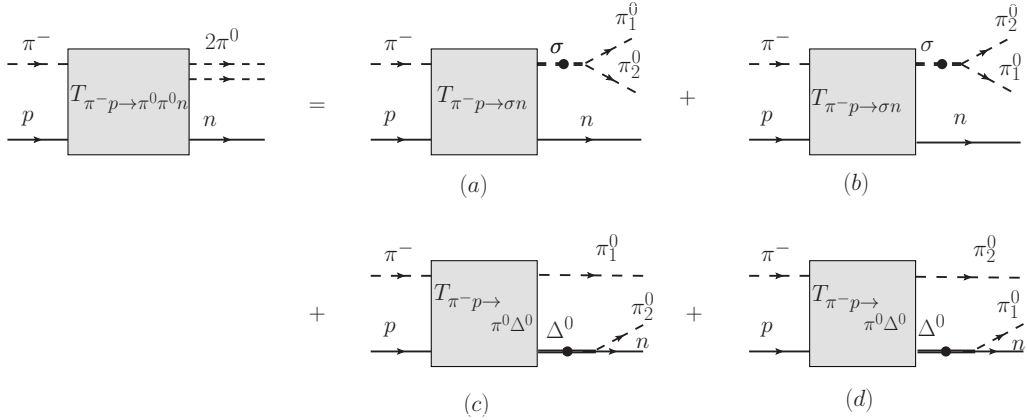}
    \caption{ The $\pi^- p\to\pi^0\pi^0$  reaction  expressed via  the isobar production amplitudes.  
      \label{unitar8}}
  \end{center}
\end{figure}

In the isobar approximation the main contribution to the $2\pi N$ final state comes from the decays of isobars. For,  e.g., the $\pi^- p \to \pi^0\pi^0 n$ 
reaction at low energies the main effect is expected to be from the $\sigma N$ and $\pi \DDelta$ subchannels. By taking care of the symmetrization of the $\pi^0 \pi^0$ state
the term (c) in Fig.~\ref{unitar6} can be rewritten via isobar production amplitudes as demonstrated in Fig.~\ref{unitar7}. From the representation
in Fig.~\ref{unitar7} it follows that the problem of solving the equations depicted in Fig.~\ref{unitar6} is reduced to the that of calculating the 
isobar production amplitudes 
%$T_{\pi N\to \pi\DDelta}$,  $T_{\pi N\to \sigma  N}$,  $T_{\sigma  N\to \sigma  N}$
\bea 
T_{\pi N\to \pi\DDelta}, ~T_{\pi N\to \sigma  N}  ~T_{\sigma  N\to \sigma  N} ~{\rm etc.}
\label{isobar_amplits_}
\eea
% etc.
 The $\sigma$-meson and $\DDelta$-isobar 
are treated as unstable particles with masses of $m_{\sigma}^2=(q_{\pi_1^0}+q_{\pi_2^0})^2$ and   $m_{\Delta}^2=(q_{\pi_1^0}+p_N')^2$ respectively.
Here $q_{\pi_1^0}$,  $q_{\pi_2^0}$, and ${p_N'}^2$ are final four-momenta of the final pions and the nucleon.
The  $\pi^- p \to \pi^0\pi^0 n$ transition rate  can be obtained from the isobar production amplitudes as shown in Fig.~\ref{unitar8}. Assuming that the properties 
of isobars  are known the problem of calculation of the $\pi N \to \pi \pi N$ transitions rates can be reduced to the problem of evaluation of the isobar production 
amplitudes Eqs.~\refe{isobar_amplits_}.

  %%%%%%%%%%%%%%%%%%%%%%
%%%%%%%%%%%%%%%%%%%%%%%%% checked!

To simplify the  discussion  we  first consider  the  situation where   the two-pion production  exclusively proceeds via the $\sigma N$ subchannel.  
In this case only the subprocesses  depicted in Fig.~\ref{unitar7}(a) and  Fig.~\ref{unitar7}(b)   contribute to the scattering equation presented  in Fig.~\ref{unitar6}.
The two-pion loop in the Fig.~\ref{unitar7}(a,b) stands for the rescattering process in the isoscalar subchannel accounting for correlations due to
the $\pi\pi$-interaction. In  the ladder approximation  the equation for the $\sigma$-meson  propagator reads 
\bea
G_\sigma(q_\sigma)= G_\sigma^0 (q_\sigma) + \int \frac{d^4k}{(2\pi)^4} G_\sigma^0 (q_{\sigma}) V_{\sigma\to 2\pi}(q_\sigma,k)G_{2\pi}(q_\sigma,k) 
V_{\sigma\to 2\pi}(q_\sigma,k) G_\sigma(q_{\sigma}),
\label{eq1.0}
\eea 
where $V_{\sigma\to 2\pi}$ is  the $\sigma \pi\pi$ decay   vertex,  $G_\sigma^0 (q_{\sigma})=(q_\sigma^2- m_{\sigma_0}^2+ i0)^{-1}$, 
and  $G_{2\pi}$ denotes the two-pion propagators. To keep the problem as simple  as possible the solution of  the $\sigma$-meson
propagator, Eq.~\refe{eq1.0}, is obtained in  the $K-$ matrix approximation \cite{Pearce:1990uj} which allows the propagator to be expressed in the 
form 
\bea
G_\sigma(p_\sigma)=  \frac{1}{q_\sigma^2 -m^2_{\sigma_0} -i\SSigma_{\sigma}(q_\sigma) },
\label{eq1.1}
\eea 
where only  the imaginary part of the $\sigma$-meson self-energy $\SSigma_{\sigma}$ is taken into account.  In the present  approximation the $\sigma$-meson can be understood 
as an unstable particle with the quantum numbers $\rm J^{IP}=0^{0+}$ and  the mass of $m_\sigma^2=(q_{\pi_1}+q_{\pi_2} )^2$. 
The $\sigma\pi\pi$-coupling is given in Appendix~\ref{Appendix_Lagrangians}.
%The $\sigma\pi\pi$-coupling and solution of the Eq.~\refe{eq1.0}
%are presented in  Section \ref{results}.

The scattering equation  for the isobar amplitudes  can be written in the form 

\bea
\left <f| T(k',k)| i  \right > 
      = \left<f|V(k',k)|i\right >  + \int\frac{d^4 q' }{(2\pi)^4} \left <f |T(k',q') G_{\pi N}(q') V(q',k)| i\right > \nonumber\\
       + \int\frac{d^4 q} {(2\pi)^4} \left <f |T(k',q) G_{\sigma N}(q) V(q,k)| i\right >, 
\label{eq1}
\eea
where $i,f$ denotes initial and final states $i,f=\pi N, \sigma N$.  $ G_{\pi N}(q')$ and   $G_{\sigma  N}(q)$ stands for the $\pi N$  and $\sigma N $
propagators respectively. $k$ and $k'$ stand for the c.m. momenta of the incoming and outgoing meson. Solving   Eq.~\refe{eq1} turns out to be technically complicated.  A considerable numerical simplification is obtained by 
the  $K-$matrix approximation 
which consists in neglecting the dispersive part in  Eq.~\refe{eq1}. 
The technique and the relevance of this approximation  was  thoroughly discussed
e.g. in \cite{Pearce:1990uj,Sauermann:1997}. This approximation also allows to solve the scattering equation  in  Minkowsky space.
The transformation of the first term in the right part of  Eq.~\refe{eq1} is presented \cite{Pearce:1990uj,Sauermann:1997,Feuster:1998a,Penner:2002a}, 
here we  only consider the second term in Eq.~\refe{eq1}. Its contribution corresponds to graphs displayed  in Fig.~\ref{unitar7}(a,b) 
and can be written in the form
\bea
 \int\frac{d q_0\, d^3q} {(2\pi)^4} <f|T(k',q) 
\,\frac{\slash p -\slash q +m_N}{ (p-q)^2 -m_N^2 +i0}\,\frac{1}{(q+p)^2-m_\sigma^2 + i\SSigma_\sigma \left((p+q)^2\right)}
\cdot V(q,k)| i >,
\label{eq2}
\eea
with $p=(\sqrt{s}/2,0,0,0)$ and $s$ being the Mandelstam  variable.  
The  discontinuity of the fermion propagator Feucan  easily be taken  into account \cite{Pearce:1990uj,Penner:2002a} which reduces Eq.~\refe{eq2}
to 
\bea
 \int\frac{d q_0 \,q \,d\Omega_q} {2(2\pi)^3}\sum_{\xi} T_{f,\sigma N}(k',q) 
\frac{1}{(q+p)^2-m_\sigma^2 + i\SSigma_\sigma \left((p+q)^2\right)}
 V_{\sigma N, i}(q,k).
\label{eq3}
\eea
where  $\xi$ denotes all quantum numbers of the intermediate particles states, $q=\sqrt{(\sqrt{s}/2-q_0)^2-m_N^2}$ is a c.m. 
momentum of  $\sigma N$ subsystem. The sigma meson propagator can be rewritten in the form

\bea
\frac{1}{\mu^2 -m_\sigma^2 +i\SSigma_\sigma(\mu^2)}=\frac{ (\mu^2-m_\sigma^2)}{(\mu^2 -m_\sigma^2)^2 + (\SSigma_\sigma (\mu^2))^2    }
 -{\rm i}\frac{\SSigma_\sigma(\mu^2)}{(\mu^2 -m_\sigma^2)^2 +(\SSigma_\sigma(\mu^2) )^2},\nonumber
\eea
where $\mu^2=(p+q)^2=2q_0\sqrt{s}$. The first term gives rise to the dispersive corrections and is neglected here. Substituting the second term into 
Eq.~\refe{eq3} and replacing the integration variable $q_0\to\mu^2_\sigma\sqrt{s}$ one gets 
\bea
-{\rm i} \int\frac{d \mu^2_\sigma } {4\sqrt{s}} \frac{q\,d\Omega_q}  {(2\pi)^3}\sum_{\xi} T_{f,\sigma N}(k',q) 
\frac{\SSigma_\sigma(\mu^2_\sigma)}{(\mu^2_\sigma -m_\sigma^2)^2 +(\SSigma_\sigma (\mu^2_\sigma))^2}
 V_{\sigma N, i}(q,k).
\label{eq4}
\eea
By defining the spectral function of the $\sigma$-meson in the form 
\bea
A_\sigma(\mu^2_\sigma)= \frac{1}{\pi} \frac{\SSigma_\sigma(\mu^2_\sigma)}{(\mu^2_\sigma-m_\sigma^2)^2 + (\SSigma_\sigma(\mu^2_\sigma))^2}
\label{eq4.1}
\eea
one can rewrite Eq.~\refe{eq4}  as follows
\bea
-{\rm i} \int\frac{q\,d\Omega_q} {8\sqrt{s}(2\pi)^2}
\int_{4m_\pi^2}^{\sqrt{s}-m_N} d\mu^2_\sigma 
A_\sigma(\mu^2_\sigma)
\sum_{\xi} T_{f,\sigma N}(k',q)  V_{\sigma N, i}(q,k).
\label{eq5}
\eea
%%%%%
%%%% checked
Up to the additional integral over $d\mu^2_\sigma A_\sigma(\mu^2)$ the quantity in Eq.~\refe{eq5} looks very similar to the rescattering in the two-body 
channel within the K-matrix approximation to BSE. Using the partial wave decomposition Appendix~\ref{Appendix_PWA} the integral over 
$d\Omega_q$ can be calculated analytically. Then Eq.~\refe{eq1} reduces  to  the equations for the partial wave scattering amplitudes in the closed form:  
\bea
T^{JP}_{f\,i}(\sqrt{s})=  K^{JP}_{f\,i}(\sqrt{s}) + {\rm i}\sum_{j}\int_{4m_\pi^2}^{\sqrt{s}-m_N} d\mu^2_j A_j (\mu^2_j)T^{JP}_{f\,j} K_{j\, i}^{JP}
\label{eq6}
\eea 
where $K=V$,  $f,i,j=\pi$, $\sigma$ and  $A_\pi (\mu^2_\pi)=\delta(\mu^2_\pi -m_\pi^2)$ and $A_\sigma(\mu^2_\sigma)$ is defined in Eq.~\refe{eq4.1}. 
Since the  two-pion discontinuities  are taken into account the three-body unitarity in the form of the optical theorem of Eqs.~(\ref{unitar3},\ref{unitar4})
is strictly fulfilled, see Appendix~\ref{AppendixA}.

As a next step we consider the $\pi\DDelta$ contribution to  the  two-pion production. The full $\pi^-p \to \pi^0\pi^0 n$ transition  
amplitude  corresponds to the graphs (c) and (d) in Fig.~\ref{unitar8}. It can be written  as a sum 
$T_{\pi^- p \to\pi^0\pi^0 n} = T^{c}+T^{d}$. The second term appears because of the symmetrization of the $\pi^0\pi^0$ final state.
While in the case of $\sigma N$ the effect of the symmetrization is trivial in the case of $\pi^0\pi^0 N$ production via  $\pi\DDelta$
it leads to complications because the isobar momentum in the diagrams (c) and (d) of Fig.~\ref{unitar8} is different.  
  Then the two-pion production cross section is defined by the integral over three body phase space
 with  the production probability 
\bea
|T_{\pi^- p \to\pi^0\pi^0 n}|^2 =  |T^{c}|^2+|T^{d}|^2 + T^{cd}.
\eea
 $T^{cd}$ is a non-vanishing interference term due the symmetrization of the $\pi^0\pi^0$ final state. 
When only (c) and (d) terms in Fig.~\ref{unitar7}  are included into the scattering equation the three-body unitarity 
is fulfilled up to the interference  $T^{cd}$ term.  This  can be demonstrated in a similar way as discussed in Appendix~\ref{AppendixA}. 
To take into account the effect of this interference the contributions from the (e) and (f)  terms in Fig.~\ref{unitar7} should also 
be included into scattering equation.  If 
 both the  $\sigma N$ and $\pi \DDelta$ isobar channels contribute to the reaction
it is also necessary to include contributions from the  diagrams   shown in Fig.~\ref{unitar7}(g-j). These terms correspond 
to the interference between  amplitudes of  the two-pion production via  the $\sigma N$ and $\pi \DDelta$ isobars. 
The  contributions from the  diagrams (e)-(j)  cannot be reduced to the simple integral form   of Eq.~\refe{eq6} but contain
 additional integration over kinematical variables. Since the data analysis requires  a large number of iterations the  evaluations of 
these amplitudes becomes numericaly very expensive. In the present study we neglect these contributions keeping only contributions from the 
diagrams (a)-(d) in Fig.~\ref{unitar7}. Then the scattering equation of the  isobar production  becomes
\bea
T^{JP}_{f\,i}(\sqrt{s})=  K^{JP}_{f\,i}(\sqrt{s}) + {\rm i}\sum_{j}\int_{\mu^2_{\min}}^{\mu^2_{\max}} d\mu^2_j A_j (\mu^2)T^{JP}_{f\,j} K_{j\, i}^{JP},
\label{eq6.1}
\eea 
where the  spectral function of the $\DDelta$-isobar is given by 
\bea
A_\Delta^i (\mu^2_\Delta)= \frac{1}{\pi}\frac{ {\SSigma}^i_\Delta(\mu^2_\Delta) }{(\mu^2_\Delta - m_\Delta^2)^2 +  ({\SSigma}_\Delta{(\mu^2_\Delta)})^2 },
\eea
and  $i=\fth,\foh$ denotes the spin projections of propagating the $\DDelta$, and   
$\SSigma_\Delta{(\mu^2_\Delta)}= {\SSigma}^\foh_\Delta{(\mu^2_\Delta)}+{\SSigma}^\fth_\Delta{(\mu^2_\Delta)} $.
The three-body unitarity in form of the optical theorem of  Eq.~\refe{unitar3} is therefore fulfilled    
up to interference  between different isobar production channels:
\bea
{\rm Im} T_{\pi N}^{\rm els.}(0) = \frac{k^2}{4\pi } ( \sigma_{\pi N \to \pi N} + \sigma^{\rm incohr}_{\pi N \to 2\pi N} ).
\label{eq7}
\eea  
Here the total  two-pion production cross section $\sigma^{\rm incohr}_{\pi N \to 2\pi N}$
 is calculated neglecting interference between $\sigma N \leftrightarrow \pi \DDelta$
and  $\pi \DDelta \leftrightarrow \pi \DDelta$ channels, see Appendix~\ref{Kinematics}.
The  cross section $\sigma^{\rm cohr}_{\pi N \to 2\pi N}$  is evaluated taking the above terms into account.
The size of the interference between the isobar contributions  can then be 
estimated by comparing $\sigma^{\rm incohr}_{\pi N \to 2\pi N} $ and $\sigma^{\rm cohr}_{\pi N \to 2\pi N} $.
In the present calculations the contribution from the $\pi \DDelta \leftrightarrow\pi \DDelta$
interference is found to be very small and comparable to the overall 1\% accuracy of the calculations. A somewhat larger effect
is observed for the  $\sigma N \leftrightarrow \pi \DDelta$ interference. However its contribution does not exceed the few percent level.
In Section~\ref{results} we present results of calculation for the  right and the left parts of 
Eq.~\refe{eq7} and compare the  difference between  $\sigma^{\rm incohr}_{\pi N \to 2\pi N} $ 
and $\sigma^{\rm cohr}_{\pi N \to 2\pi N} $. Note that the difference 
 $(\sigma^{\rm incohr}_{\pi N \to 2\pi N}  - \sigma^{\rm cohr}_{\pi N \to 2\pi N} )$ is not directly 
equal to the sum of contributions from  diagrams (e)-(j) in Fig.~\ref{unitar7}; the scattering equation should be re-iterated
once the (e)-(j) diagrams are taken into account. However  $(\sigma^{\rm incohr}_{\pi N \to 2\pi N}  - \sigma^{\rm cohr}_{\pi N \to 2\pi N} )$ 
can be used to estimate the size of such contributions which are found to be small in the present study. 
 The magnitude of this difference also 
indicates the size of the violation of the constraint Eq.~\refe{unitar4}. We discuss this isssue in Section~\ref{section_unitarity}.

Note  that  the $\sigma N \leftrightarrow \pi\DDelta$  interference is still important in the calculation of, e.g., the angular
distributions. However this  effect is small at the level of the total cross section, 
see discussion in  Section~\ref{results}.
%%%%%%%%%%%%% checked
%%%%%%%%%%%%%

\subsection{ Interaction kernel\label{int_kernel}}
 \begin{figure}[h]
  \begin{center}
    \includegraphics[width=13cm]{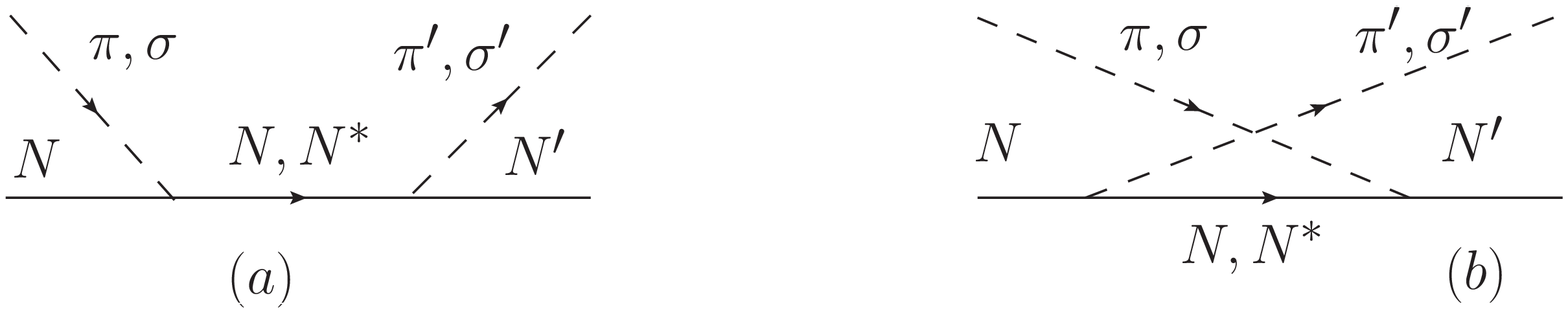}\\
\vspace*{1cm}
    \includegraphics[width=13cm]{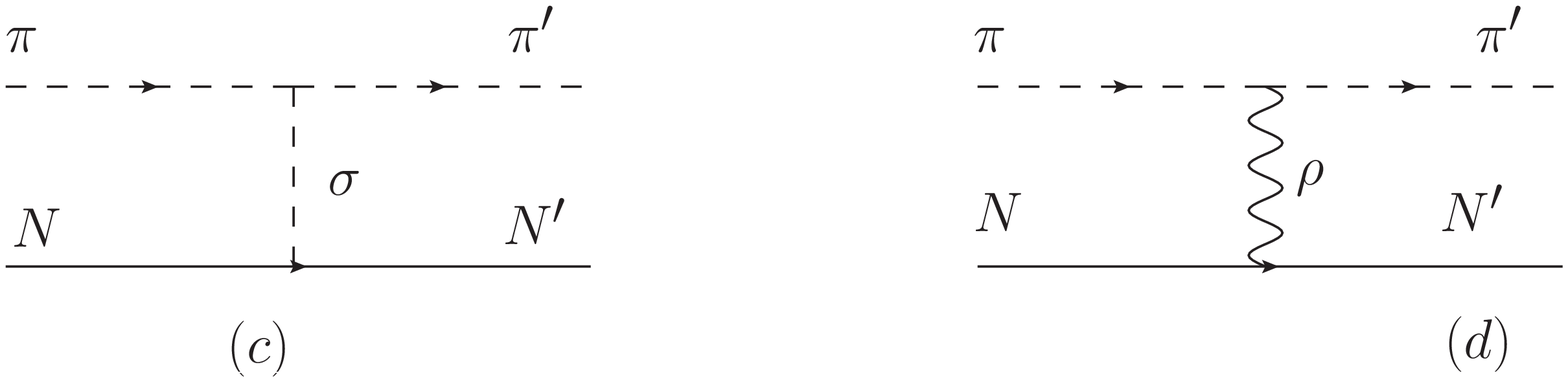}\\
\vspace*{1cm}
    \includegraphics[width=13cm]{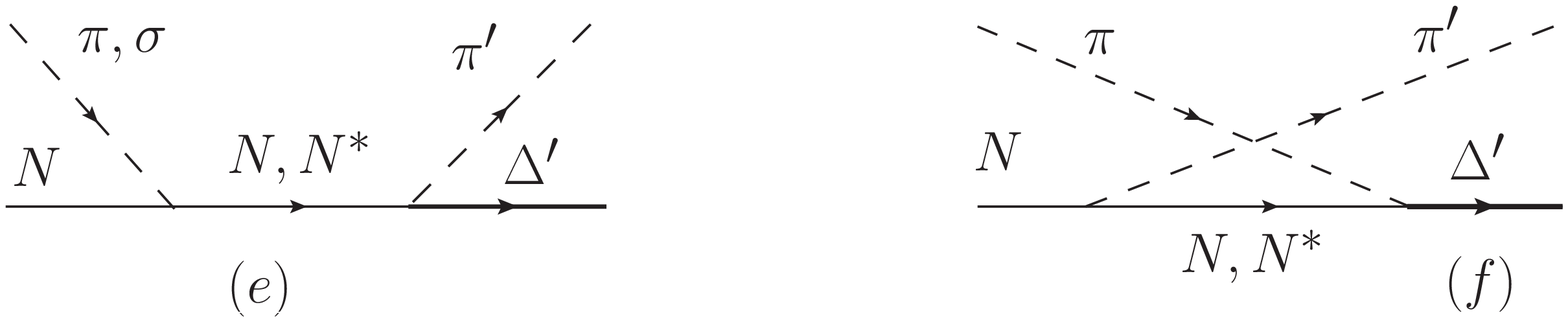}\\
\vspace*{1cm}
    \includegraphics[width=13cm]{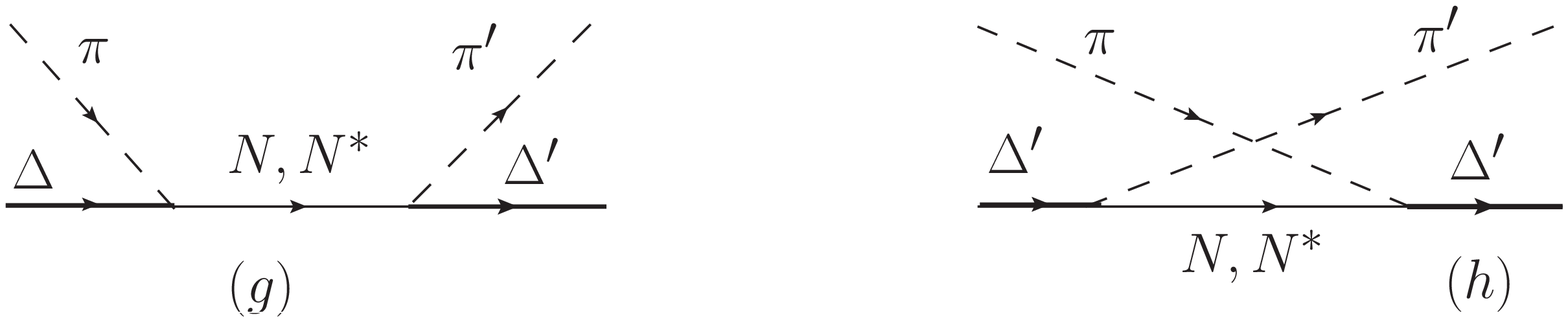}\\
\vspace*{1cm}
    \includegraphics[width=13cm]{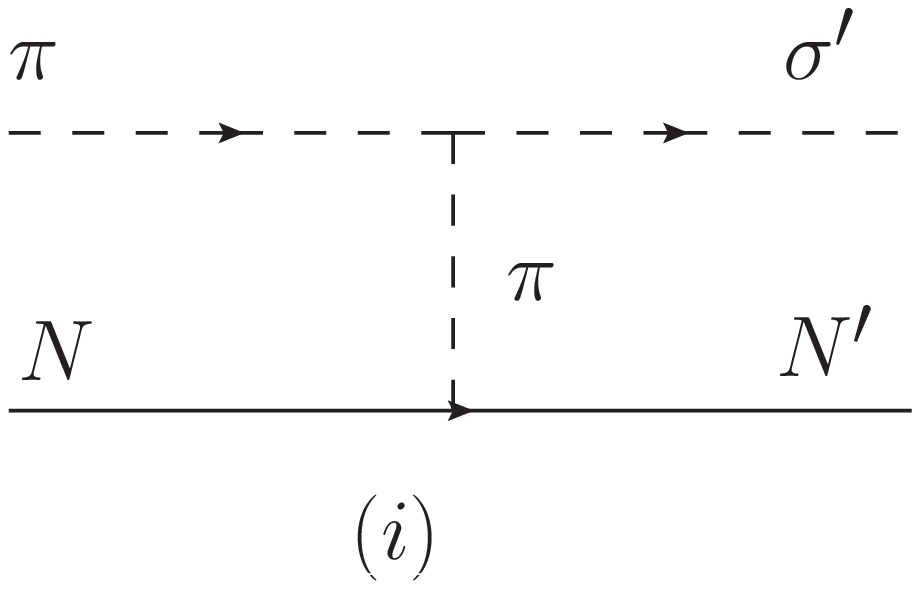}
    \caption{  Tree-level diagram contributions to the interaction kernel of the scattering equation. 
      \label{int_kernel_fig1}}
  \end{center}
\end{figure}
To solve the scattering equation,  Eq.~\refe{eq6.1}, the interaction kernel must be specified. It is constructed as a sum of 
contributions from the tree-level diagrams shown in Fig.~\ref{int_kernel_fig1}.  
 For the $(\pi/\sigma) N \to  (\pi/\sigma) N $ 
scattering the diagram in Fig.~\ref{int_kernel_fig1}(a) corresponds to the nucleon Born term and the 
resonances contributions. In the present study we concentrate on $I=\foh$ only and this  include  $\NN(1535)$  and $\NN(1440)$ states.
The additional graph  in Fig.~\ref{int_kernel_fig1}(i)  is responsible for the non-pole part  in the $\sigma N$
subchannel. The driving terms for the $(\pi/\sigma) N\to \pi\DDelta$ and $\pi \DDelta \to \pi\DDelta$ reactions 
 are constructed from the $s$- and $u$-exchanges processes
displayed  in Fig.~\ref{int_kernel_fig1}(e-h)  with the ground state or excited nucleon in the  intermediate state.

From the point of view of the  $\pi N \to\pi\pi N$ transitions  the processes  (a) and (e) in Fig.~\ref{int_kernel_fig1}
with the unstable baryon in the intermadiate state are   resonant ones.  The non-resonant term  is described by the (b)-(d),~(f), and (i) graphs in the same figure.  The corresponding 
Lagrangian densities are given in Appendix~\ref{Appendix_Lagrangians}. Each vertex in dressed by the formfactor  (e.g. for $s$-channel exchange )
\bea
F(q^2)=\frac{\Lambda^4}{\Lambda^4 + (q^2 - m^2)^2},
\eea
where $q^2$ is the square of the four-momentum of exchange particle and  $m$ is its mass.  
 To reduce the number of free parameters we use the same cutoff
	for all resonance decay channels $\Lambda_{N^*}$ = 1.95 GeV. The cutoff at the nucleon vertex 
	has been chosen to be  $\Lambda_N=$ 0.95 GeV.  
	For  the  $t-$channel  meson exchange we use  $\Lambda_t$=1.54 GeV.

The coupling constants used in the calculations  are listed  in Table~\ref{Int_kernel_tab1}.
The values of the coupling constants  $\g_{\pi NN}=12.8$ and   $\g_{\rho\pi\pi}=6.02$  are 
the same as in our previous calculations   
\cite{Penner:2002a,Penner:2002b,shklyar:2004a,Shklyar:2006xw,Shklyar:2012js,Cao:2013psa,Shklyar:2005xg}.  
The sign at  $\g_{\pi \Delta N}$  and its values are also taken in accordance with the results from 
\cite{Penner:2002a,Penner:2002b,shklyar:2004a,Shklyar:2006xw,Shklyar:2012js,Cao:2013psa,Shklyar:2005xg}.
The $\g_{\rho NN}$, $\kappa_{\rho NN}$, and $\g_{\sigma NN}$ coupling constants   were allowed to be varied during fit.
The  $\g_{\sigma\pi\pi}$ constant was  fixed from the requirement to reproduce the $\sigma$-meson decay properties listed in PDG.
With  $\g_{\sigma NN}=3.25$  and $m_\sigma=0.650$\,GeV the pole position $ z_0= (0.47 -0.19\,{\rm i})$ GeV   of the sigma meson is well reproduced \cite{pdg}. We also 
obtain $ z_0= (1.210 -0.5\,{\rm i})$ GeV for the pole position of $\Delta(1232)$.
Since the isobar parameters could be subjected to  small uncertainties  we also allowed for small
variation of $\g_{\pi N\Delta}$ and $\g_{\sigma \pi\pi}$ within a few percent during calculations.
This allows for a 
small variation of background 
contributions in the $\pi N \to \pi\pi N$ reaction. 

The resonance couplings to the  $N^* \to \pi N$,   $\sigma N$,   $\pi \DDelta$ final state are constrained by the direct comparison of the 
calculated amplitudes to the experimental data. These parameters   are discussed in Section~\ref{results} and Appedix~\ref{Appendix_Lagrangians} .

Within the  isobar approximation the contribution from 
the nucleon Born term % where pions are produced without forming an isobar  
is neglected.
 Here pions are produced from the nucleon intermediate state without forming an isobar.
These terms  can be regarded as a 'non-resonant background' to the $2\pi N$ production. Since the nucleon pole 
lies below the $2\pi N$ threshold no strong   effect is expected from this transition. At the same time the nucleon
Born term  also gives rise to the process
 \bea
\pi N \stackrel{N(938)}{\to}  \pi \DDelta\, \stackrel{\Delta\to\pi N}\longrightarrow \,\pi \pi N.
\label{yes_isobr}
\eea
 Because of the $\DDelta$-isobar dynamics the effect
from  the  'non-resonant' term  \refe{yes_isobr} is expected to be much larger  than in the process
without forming an  isobar. 
%\refe{no_isobr}.  

Our ansatz is also supported by the measurements in \cite{Prakhov:2004zv} where the mass distributions close to threshold 
are shifted to the higher invariant pion masses which is identified with the effect of $\sigma$-meson spectral function, see
discussion in Section \ref{results}.
%%%%%%%%%%%%%
%%%%%%%%%%%%% checked

\begin{table}[t]
  \begin{center}
    \begin{tabular}
      {l|r|l|r|l|r|l|r}
      \hhline{========}
      $g$ & value & $g$ & value & $g$ & value & $g$ & value \\
      \hhline{========}
$\g_{\pi NN}$ & $12.8 $ & $g_{\sigma NN}$ & 4.25  & $\g_{\rho NN}$ & $ -0.69 $ & $\kappa_{\rho NN}$ &  $5.99 $ \\
$\g_{\sigma  \pi\pi}$ & $3.25 $ & $\g_{\rho \pi\pi}$ & 6.02  & $\g_{\pi\Delta N}$ & $ -2.2 $ & &  \\
      \hline
%$g_{N\Sigma K}$  &   2.48 & $g_{N\Sigma K_0^*}$ & $-52.30$ & $g_{N\Sigma K^*}$  &   4.33 & $\kappa_{N\Sigma K^*}$  &  $-0.86$ \\

      \hhline{========}
    \end{tabular}
  \end{center}
  \caption{Nucleon and $t$-channel couplings. First line: $C$-calculations
Second  line: $S$-calculations.
    \label{Int_kernel_tab1}}
\end{table}

 \section{Isobar contributions to the $\pi N \to \pi^0\pi^0 N$ reaction \label{section_isobar_contributions}}
In this section we discuss the impact of various observables on the partial-wave  analysis. 
For unpolarized measurements  the full information about the reaction dynamics is encoded into 4-fold differential 
cross sections, see e.g. Eq.~\refe{kinematics_eq4}. In practice however the  experiment often provides only a limited set of observables such as 
angular or mass distributions. 
This raises the question how different reaction 
channels can be extracted  from  experimental data. 
%This rises a question of the unambigous identification of the isobar  
%contributions from the experimental data.
Hence it is  an important issue to disentangle different decay modes of the  same resonance.

\begin{figure}
  \begin{center}
    \includegraphics[width=8cm]{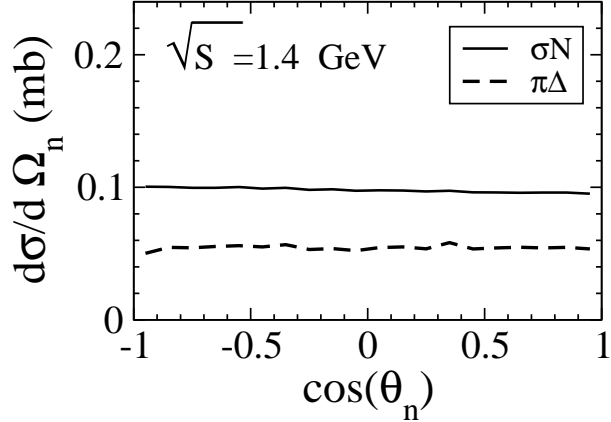}
    \caption{Reaction $\pi^- p \to \pi^0 \pi^0 n$: differential cross section as a function of the nucleon scattering angle.
     Solid line: effect of the  $\NN(1440)$ resonance  decay into the $\sigma N$ subchannel, dashed line:  $\NN(1440)$ decay into the $\pi\DDelta$ final state.
      \label{results_fig3}}
  \end{center}
\end{figure}

In \cite{Prakhov:2004zv}  measurements of the differential cross sections as a function  of the nucleon scattering angle have been reported. 
The contributions from the $\sigma N$ and $\pi \DDelta$ isobar channels  to this observable are shown in Fig.~\ref{results_fig3} at the fixed energy $\sqrt{s}=1.4$ GeV.
The calculation is done   assuming only a $J^P=\foh^+$-wave  contribution  to the production mechanism.
Though the angular distribution is known to be very important for the partial wave analysis the separation 
between the $\pi \DDelta$ and $\sigma N$ subchannels turns out to be difficult.
Both distributions are only weakly dependent  on the nucleon scattering angle which indicates that 
each isobar 
subchannel is produced in the $J=\foh$ partial wave. However, any further separation between  $\sigma N$ and $\pi\DDelta$
subchannels is hardly possible.

On  the other hand a great part of the information on the reaction dynamics in encoded in the experimental  mass distributions. 
We  first discuss the influence of the $\sigma$-meson spectral function on the results of the data analysis. 
This quantity appears explicitly  in the scattering equation  Eq.~\refe{eq6.1} and  implicitly in the squared   modulus  of the $\pi\pi$ production amplitude   in form of 
the product of the  propagator  and the decay vertex of the isobar, see Fig.~(\ref{unitar8}) (a,b) and discussion in Appendix~\ref{Kinematics}. 

\begin{figure}
  \begin{center}
    \includegraphics[width=8cm]{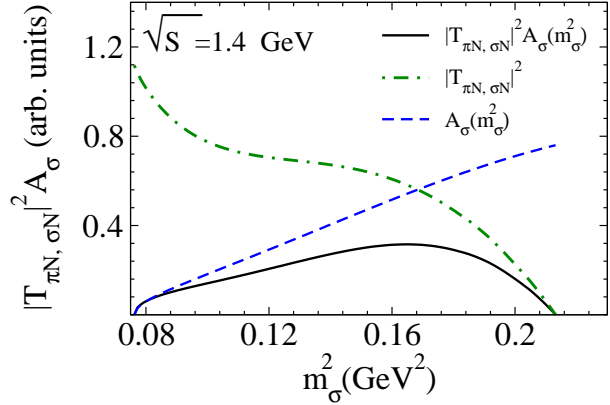}
    \caption{(Color online) dash-dotted line: modulus of the average production amplitude squared $|T_{\pi N, \sigma N}|^2$ as a function of the invariant isobar mass $m_{\sigma}^2=(q_1 + q_2)^2$,
      where $q_1$ and $q_2$ are pion momenta; dashed line: spectral function of the sigma meson $A_{\sigma}(m_{\sigma}^2)$ as a function $m_{\sigma}^2$. Solid line: 
      $|T_{\pi N, \sigma N}|^2  A_{\sigma}(m_{\sigma}^2)$. All quantities are normalized by the arbitrary factors to put 
      them together on the same figure.
      \label{results_fig1}}
  \end{center}
\end{figure}

First, it  leads to an additional dependence of the production amplitude on the isobar mass. 
This is different to, e.g., the parameterization used in  work of Manley et al \cite{Manley:1984} where single energy solutions are assumed to be
functions only of the c.m. energy. The dependence on the isobar mass for the  $\pi N \to \sigma N$ production amplitude is demonstrated in 
Fig.~\ref{results_fig1}. The isobar production amplitude has a maximum at $m^2_{\sigma,{ \min}}=  4m_\pi^2$ 
and vanishes for maximal values of   $m^2_{\sigma,{\max}} =(\sqrt{s} -m_N)^2$.  The latter effect would correspond to the $\sigma N$ reaction 
threshold  if the  $\sigma$-meson were a  stable particle
with  mass of $m_{\sigma,{\max}}$.
The  spectral function  demonstrates the opposite behavior: for the  energy  at hand $\sqrt{s}=1.4$ GeV  
it is maximal   for the maximal allowed invariant  $\sigma$-meson mass and vanishes
at  $m^2_{\sigma,{ \rm min}}$. The contribution from the $\sigma N$ isobar channel to the two-pion production  cross section
is  defined by the product of 
the modulus  squared of the reaction amplitude and the $\sigma$-meson spectral function. This quantity is shown  in Fig.~\ref{results_fig1} by the solid line. 
It demonstrates 
 a rapid variation as a function of the $\sigma$-meson mass with the maximum lying in the interval   
$[m^2_{\sigma,{\rm min }}$, $m^2_{\sigma,{max}}]$. The position of the maximum
is  defined by the spectral function of the $\sigma$-meson, the  c.m. energy $\sqrt{s}$, and the dynamics in the $\sigma N$ channel.
For the energies at hand the resulting distribution  is  shifted  to  higher masses. A similar behavior is also seen in the experimental mass  distributions close to the 
$2\pi N$ threshold. This allows to draw a  conclusion on the important contributions
from the $\sigma N$ subchannel to the $\pi^0\pi^0 n$ final state, as discussed below.

\begin{figure}
  \begin{center}
    \includegraphics[width=8cm]{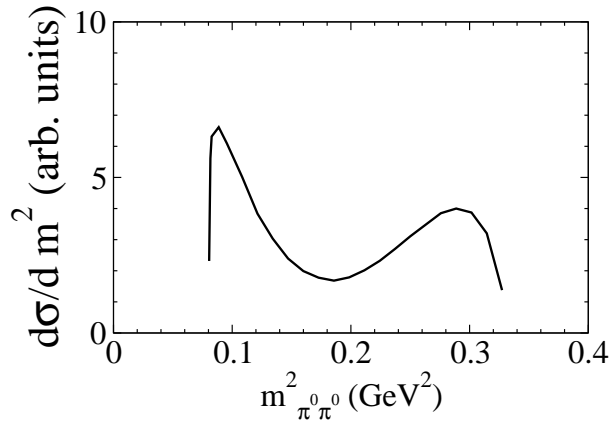}
    \caption{Invariant mass distribution of the  $\pi^0\pi^0 n$  system produced via $\pi\DDelta$ subchannel in the $P_{11}$ partial wave.
      \label{results_fig2}}
  \end{center}
\end{figure}

The analysis of the mass distribution  ${{\rm d} \sigma }/{\rm d} m_{\pi^0\pi^0}^2$ 
for the pions produced from the $\pi\DDelta$ isobar subchannel turns out 
to be more complicated. For the sake of simplicity we neglect for the moment
 the effect  of  the symmetrization for the two-pion final state.
Let the first pion be produced in the $\pi N \to \pi \DDelta$ transition
and the second one  in the $\DDelta$ isobar decay. The invariant mass $m_{\pi^0\pi^0}^2$ 
is a function of the angle  between the two pions. In the c.m. of initial particles the momentum of the first
pion is opposite to the momentum of the $\DDelta$-isobar. Hence the angular dependence  between the two pions can be translated into the  dependence
on the angle  between the second pion and the direction of the  isobar momentum. The latter  is defined by the spin structure of the decay vertex.  
In Fig.~\ref{results_fig2} the  ${{\rm d} \sigma }/{\rm d} m_{\pi^0\pi^0}^2$ mass distribution is shown 
for pions coming  from the $\pi\DDelta$ subchannel which is produced in the $J^P=\foh^+$-wave.  
Therefore, only  
 $\lambda_\Delta=\pm \foh$ helicity combinations contribute 
at the $\Delta\to \pi N$ decay  vertex. For the decay at rest 
the transition probability behaves as 
\bea
(1+3\cos^2{\theta_\pi})
\label{angl_dep}
\eea
where  $\theta_\pi$ is the angle between the momentum of the
final pion  and the direction of the isobar momentum.  Eq.~\refe{angl_dep}
exhibits a  symmetric distribution with two maxima at $\theta_\pi=0,\pi$.
In the center of mass sytem
of the initial $\pi N$ particles the produced $\DDelta$ isobar has a non-vanishing momentum. 
 Therefore the  dependence Eq.~\refe{angl_dep}
 becomes more complicated when the decay of the isobar is considered
in the full three-body kinematics taking into account the effect  of the symmetrization 
of the $\pi^0\pi^0$ states. 
However, even in this case the two maxima structure is clearly visible
 in the mass spectra  shown in Fig.~\ref{results_fig2}.
A similar behavior is also found in the calculations of \cite{Kamano:2006vm,PhysRevC.66.065201}.

Comparison of the  ${{\rm d} \sigma }/{\rm d} m_{\pi^0\pi^0}^2$ mass distributions for the $\sigma N$ and $\pi\DDelta$  subchannels in 
Fig.~\ref{results_fig1},  and Fig.~\ref{results_fig2}
 demonstrates very different mass  dependencies for the  isobar subchannels produced in the same partial wave. Hence the analysis of this observable 
becomes crucial for the identification of the resonance decay in  the various isobar   subchannels.
%%%% checked
%%%%

 \subsection{Non-resonant contributions into the $\pi N \to \pi^0\pi^0 $ reaction} 
\begin{figure}
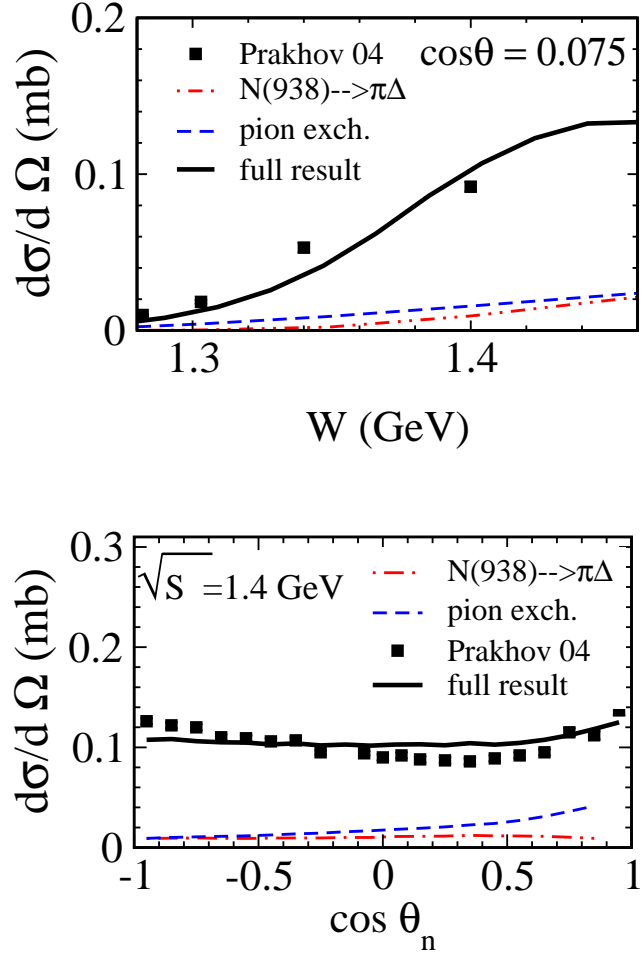

  \begin{center}
    \includegraphics[width=8.4cm]{fig17.eps} \\
\vspace*{1cm}
   \includegraphics[width=8.4cm]{fig18.eps}
    \caption{(Color online) non resonant contribution to the $\pi^0\pi^0 n$ production as a function of the c.m. energy (upper panel) and the nucleon
    scattering angle (lower panel). Dashed line: t-channel pion exchange. Dash-dotted line: Nucleon $s$-channel contribution to the 
    $\pi\DDelta$ isobar channel. Solid line: full model.
     The  experimental data denoted by  Prakhov\,04, are taken from \cite{Prakhov:2004zv}. 
      \label{Non_resonant_fig1}}
   \end{center}
\end{figure}

The non-resonant  part of the interaction kernel consists of the $s$- and $u$-channel nucleon Born terms 
and  the $t$- channel pion exchange for  the $\pi N \to \sigma N$ transition, see diagram (i) in  Fig.~\ref{int_kernel_fig1}. 
Since the $\g_{\pi NN}$, $\g_{\sigma \pi\pi}$, and $\g_{\pi N\Delta}$ couplings  are fixed (see Section \ref{int_kernel}) 
the size of these contributions
can be easily estimated. The result of the calculations without formfactors at the interaction
vertices  is shown  in Fig.~\ref{Non_resonant_fig1} vs. the data from \cite{Prakhov:2004zv}. 
At low energies  the $t$-channel pion exchange
gives rise to the s-wave scattering  and the final $\sigma N$ system is produced in the $J=\foh$ state. Therefore
the  differential cross section  demonstrates  only a  very weak angular dependence.   
We conclude that the $t$-channel pion exchange is  responsible for the description of the $\pi^0\pi^0 n$ data close to threshold.
However this mechanism starts to  underestimate the data at energies  above 1.3 GeV where 
the excitation of the Roper resonance is expected.  With increasing  c.m. energy  the $t$-channel exchange
 starts to give rise to  higher partial waves.  
This enhances the   calculated cross section at forward angles as
seen in the right panel of Fig.~\ref{Non_resonant_fig1}.
Note that the overall effect from  pion exchange is found to be smaller than  would be expected from the large 
$\g_{\sigma\pi\pi}$ and $\g_{\pi N\Delta}$ coupling constants. 
This is because  the $\sigma N$ contribution to the differential
 cross section at hand can be represented as an integral of the modulus squared  of the isobar production  amplitude 
multiplied by the  $\sigma$-meson spectral function, 
$|T_{\pi N\to \sigma N}|^2 A_\sigma$,   over the invariant two-pion mass  $m_{\pi^0\pi^0}^2=m^2_\sigma$ (see discussion in Appendix~\ref{AppendixA}). 
The dependence  of $|T_{\pi N\to \sigma N}|^2$ and $ A_\sigma$ on  $m^2_\sigma$  were shown in Fig.~\ref{results_fig1}. 
These quantities  demonstrate  
an opposite  behavior at  higher invariant masses: while the spectral function rises the isobar production amplitude declines.
This reduces the total effect  which is shown by the solid line in Fig.~\ref{results_fig1}. We conclude that realistic calculations should 
account for the   dependence on the dynamical isobar mass
both in the production amplitude and for the propagation and decay of the $\sigma$-meson.

The nucleon Born term also gives an important contribution to the $\pi^0\pi^0 n$ production through the coupling to the
$\pi \DDelta$ isobar channel. However, close to the $2\pi N$  threshold its effect  turns out 
to be smaller than that of the pion-exchange; the same conclusion has been drawn   in   \cite{Kamano:2006vm}.
Note that the non-resonant contributions  discussed above are fixed up to a cutoff at the interaction vertex which has to be constrained during the fit. 

Another source of the non-pole components in the interaction kernel for  isobar production comes from  the nucleon  
coupling to  $\sigma N$, see diagrams (a) and (b)  in Fig.~\ref{int_kernel_fig1}. The $\g_{\sigma NN}$ coupling constant 
and the cutoff are  fixed during  the fit. However, the same  vertex together with the $\rho$-meson exchange is also responsible for  the   description 
of the $S_{11}$- wave amplitude of the $\pi N$ elastic scattering at low energies. The corresponding diagrams are shown in Fig.~\ref{int_kernel_fig1}(c,d).
This provides an additional  constraint on the size of $\g_{\sigma NN}$. The obtained value is given in Table~\ref{Int_kernel_tab1}.

 The  calculated nucleon Born  term
 contribution to the $\sigma N$ production,  
see diagrams Fig.~\ref{int_kernel_fig1}(a,b), turns out to be significantly smaller than other 
'non-pole' terms and we do not show it  in Fig.~\ref{Non_resonant_fig1}.

 \section{Results and discussion \label{results}}
 \subsection{Database}

Here we briefly discuss the reaction data base used in the calculations. 
To simplify the analysis the  $S_{11}$ and $P_{11}$  $\pi N$ partial waves  are directly constrained by the single energy solutions (SES)
derived by GWU(SAID)    \cite{Arndt:2006bf}. 

The experimental data on the $\pi^- p \to\pi^0\pi^0 n$ reaction are taken from \cite{Prakhov:2004zv}. These measurements provide  high statistics data 
on  the angular distributions ${\rm d} \sigma /{d\Omega_{\pi\pi}}$ 
where $\Omega_{\pi\pi}$ is the scattering angle of the $\pi\pi$ pair (or the  final nucleon  in c.m.). 
This data are accompanied by the corresponding statistical and systematical errors. No such information is available for  the  mass distributions in
\cite{Prakhov:2004zv}. These observables are provided in a form of weighted events without systematic and statistical uncertainties. To use them 
in the data analysis we rescale them with the requirement that the integrated distributions  should reproduce the total cross section of the $\pi^- p \to 2\pi^0 n$
reaction.  We also assign  about  10\%  error bars to  each mass bin to perform  the $\chi^2$ minimization. 
Starting from 1.46 GeV the excitation of  $\NN(1520)$ starts to be important. Already at this energy a small contribution from the 
spin $J=\fth$ partial wave could modify the angular and mass distributions. Because of this reason  we  do not try to fit the data 
above 1.46 GeV.

\subsection{Elastic $\pi N$ scattering}
\begin{figure}[h]
  \begin{center}
    \includegraphics[width=17cm]{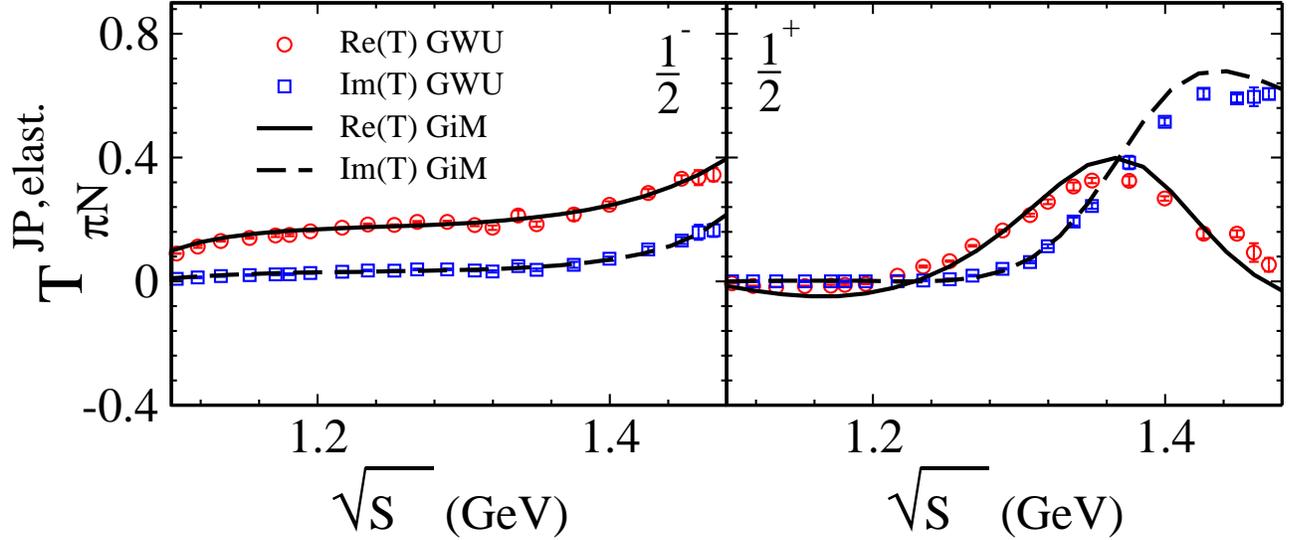}
    \caption{(Color online) the elastic $\pi N$ partial wave amplitudes vs. the energy independent solutions
 from the GWU analysis \cite{Arndt:2006bf}. 
      \label{piNelastic_fig1}}
   \end{center}
\end{figure}

The resonance couplings are constrained  by simultaneous descriptions of the  $S_{11}$ and $P_{11}$ $\pi N$ single energy solutions from GWU(SAID)
and the data from the Crystal Ball measurements \cite{Prakhov:2004zv}.  The results of our calculations are
 shown in Fig.~\ref{piNelastic_fig1} in comparison with the $\pi N$ elastic scattering amplitudes from GWU group
 \cite{Arndt:2006bf}. 

\begin{figure}
  \begin{center}
    \includegraphics[width=17cm]{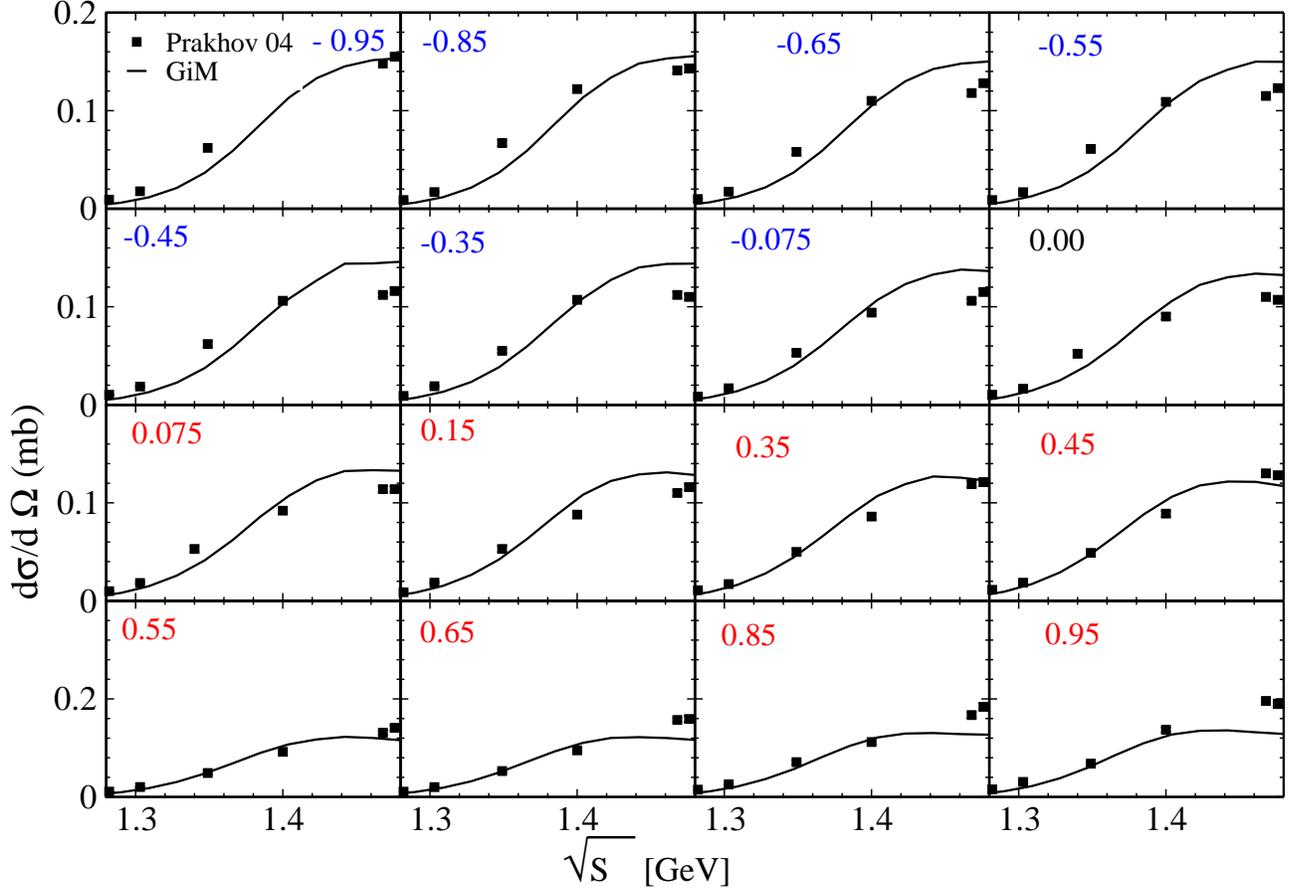}
    \caption{(Color online) reaction $\pi^-p \to\pi^0\pi^0n$: differential cross section as a function of the c.m. energy at fixed scattering angles vs
  the experimental data from \cite{Prakhov:2004zv}. The numbers in the upper left corner give $\cos{\theta_N}$. 
      \label{2piN_dif_fig1}}
   \end{center}
\end{figure}
\begin{figure}
  \begin{center}
    \includegraphics[width=17cm]{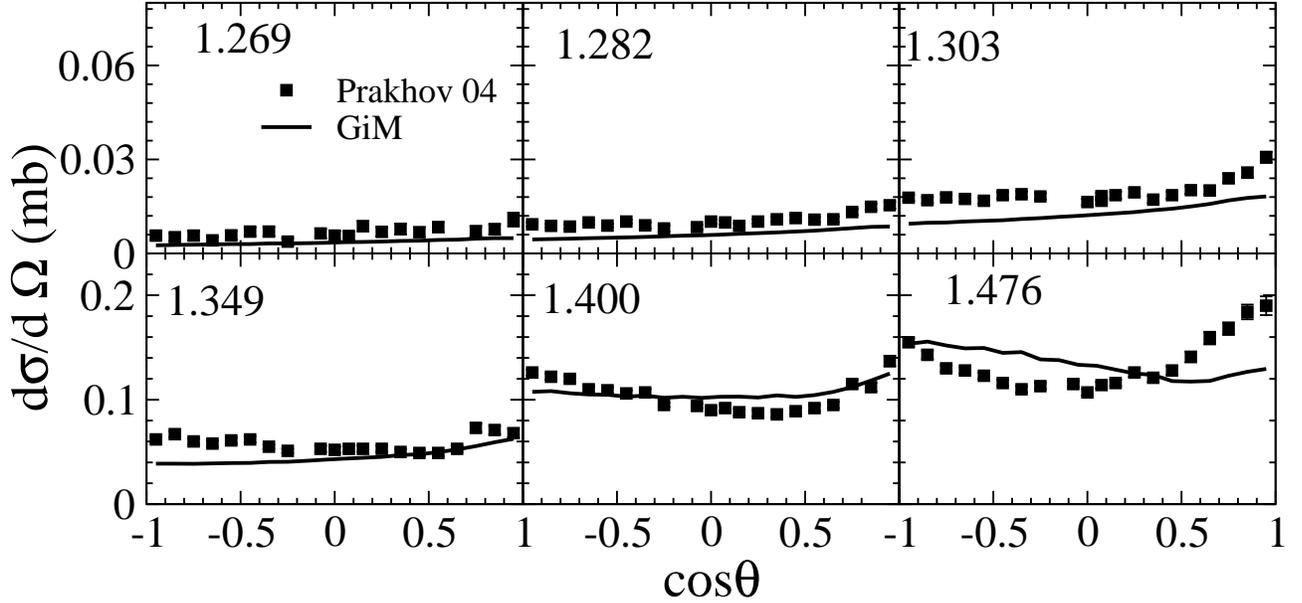}
    \caption{Reaction $\pi^-p \to\pi^0\pi^0n$: differential cross section as a function of the c.m. energy at fixed scattering angles vs
  the experimental data from \cite{Prakhov:2004zv}. The numbers in the upper left corner give the c.m. energy $\sqrt{s}$.
      \label{2piN_d3d_fig1}}
   \end{center}
\end{figure}

The present calculations  demonstrate the good description of  SES in the whole energy region. The small rise in the $S_{11}$ partial wave amplitude is 
due to the tail of the  $\NN(1535)$ resonance. The $t$-channel $\rho$- and $\sigma$-meson exchanges are found to be important for the description of the
real part of the $S_{11}$ amplitude at low energies.

\subsection{Reaction $\pi^- p \to \pi^0\pi^0 n$}
The calculated   differential cross sections  are shown in Fig.~\ref{2piN_dif_fig1} in comparison with the Crystal Ball data as a function of the c.m. energy.
The measurements demonstrate a rapid rise of the cross sections  at  the energies 1.3-1.46 GeV.
Similar to \cite{Kamano:2006vm,Sarantsev:2007aa} we identify this behavior as an indication for the strong contribution coming  from the Roper resonance. 
%Indeed the resulting $\pi N$ inelasticities from GWU(SAID) \cite{Arndt:2006bf} 
%analysis indicate that the main contributions to the inelastic transitions at energies
%at hand are expected to be from the $P_{11}$ partial wave. 
Indeed, the resulting $\pi N$ inelasticities from  the GWU(SAID) \cite{Arndt:2006bf} 
analysis indicate that the  $P_{11}$ partial wave dominates 
the inelastic transitions at these energies. 
The inelasticity  from the $S_{31}$ channel is about  three times less than that from $P_{11}$. 
At the same time the $\Delta({1620})$ is strongly coupled to the $2\pi N$ final state through the $\pi\DDelta$ decay \cite{pdg}. Since 
the contribution from the $\sigma N$ 
subchannel is found in the present work  to be about twice as large than that of $\pi\DDelta$ is it safe to neglect the possible effect from the $\Delta({1620})$
resonance in the  first approximation. We also allow the $\NN(1535)$ resonance decays to the $\pi \DDelta$ and $\sigma N$ isobar final states 
which are  however  found to be negligible. At energies close to 1.5 GeV the obtained cross section slightly overestimates the 
experimental data at backward and underestimates them at forward scattering angles.  This is a region where the $\NN(1520)$ starts to play a dominant role. We
conclude that the contribution from the  $D_{13}$ partial wave should be included for the successful  description of the data at 1.5 GeV.
The  effect from the missing spin $J=\fth$ amplitude is also seen in the angular distributions at the energy 1.476 GeV presented  
in   Fig.~\ref{2piN_d3d_fig1}.   The experimental data demonstrate the increase at forward and backward angles which is not 
fully reproduced by the present calculations. We conclude that the missing contributions from the $\NN(1520)$ 
resonance could be responsible for the effect. The impact of this resonance on the data analysis is estimated in 
Section \ref{Nstar_param}.
\begin{figure}
  \begin{center}
    \includegraphics[width=16cm]{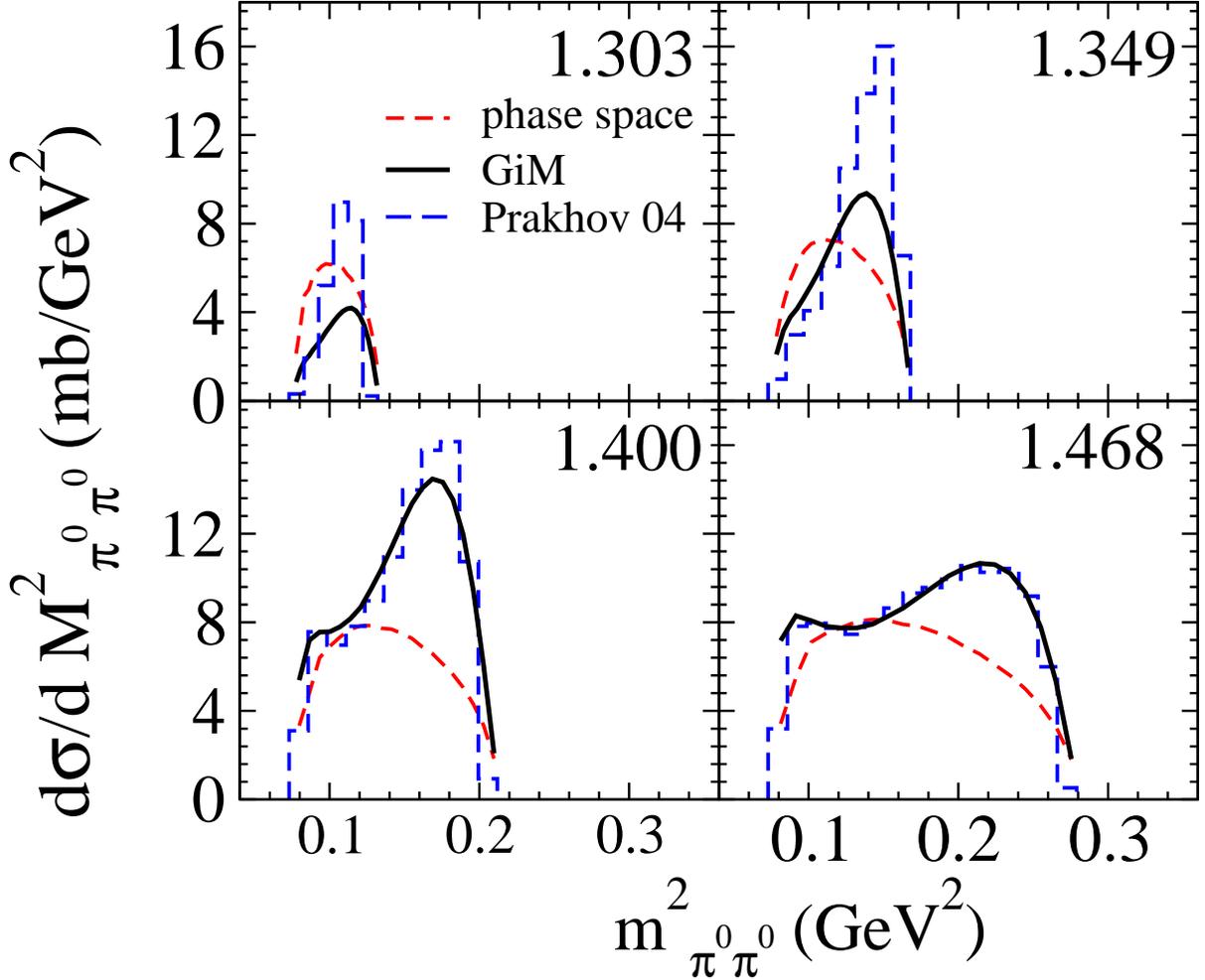}
    \caption{(Color online) reaction $\pi^-p \to\pi^0\pi^0n$: differential cross section as a function of  $m^2_{\pi\pi}$ at fixed c.m. energies vs
  the experimental data from \cite{Prakhov:2004zv} (dashed). 
      \label{2piN_mass_fig1}}
   \end{center}
\end{figure}
\begin{figure}
  \begin{center}
	    \includegraphics[width=16cm]{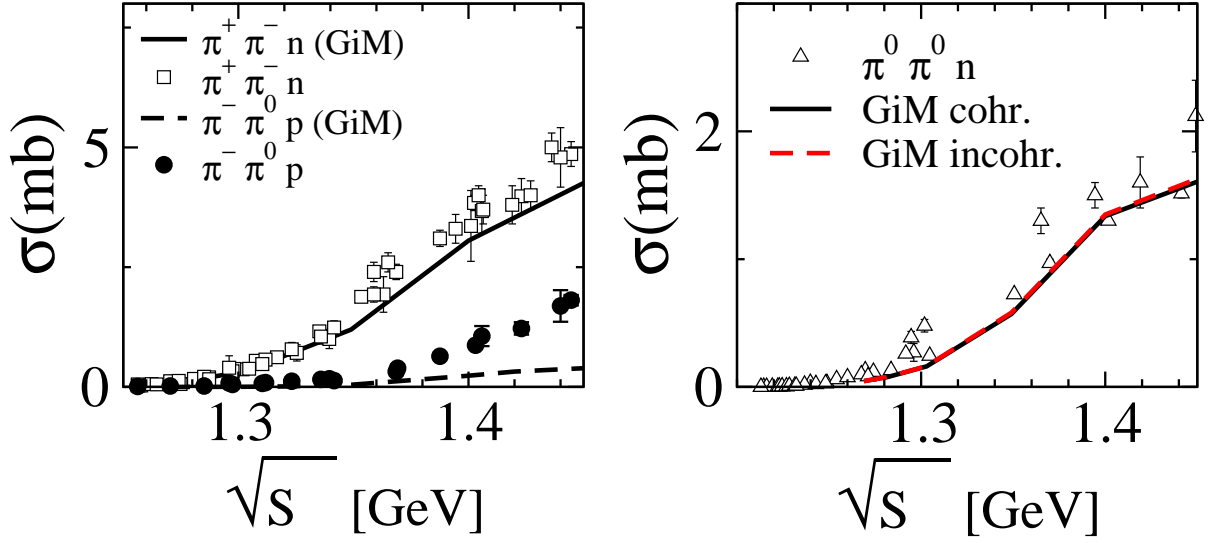}
    \caption{(Color online) total cross section  $\pi^-p \to\pi^+\pi^-n, \,\pi^-\pi^0 p$  (left) and $\pi^-p \to\pi^0\pi^0n$(right) vs experimental  data \cite{pipi_Data}. 
      \label{2piN_tot_fig1}}
   \end{center}
\end{figure}

\begin{figure}
  \begin{center}
    \includegraphics[width=8cm]{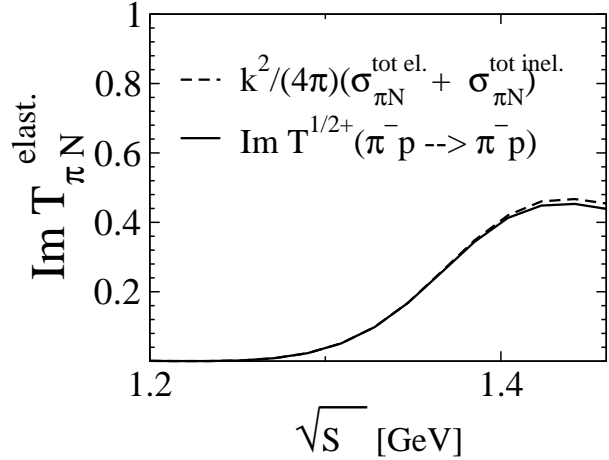}
    \caption{The left (solid line)   and right part (dashed line) of the optical theorem  Eq.\refe{unitar4}.
      \label{2piN_unitar_fig}}
   \end{center}
\end{figure}

At lower energies the angular distributions show a moderate dependence on the nucleon scattering angle. As discussed in Section 
\ref{section_isobar_contributions} the separation of the  $\sigma N$ and $\pi \DDelta$ isobar channels from this observable 
turns out to be difficult.
The difference between the production mechanism is expected to be more pronounced in the invariant mass distributions.
They are shown in Fig.~\ref{2piN_mass_fig1}. Close to threshold the Crystal Ball data 
demonstrate a shift to the higher invariant masses  for all energies up to 1.5 GeV whereas the  three-body  phase space 
tends to have a maximum at lower $m_{\pi^0\pi^0}^2$. 
Since the imaginary part of the $\pi N$ $P_{11}$   elastic amplitude at the energy $\sqrt{s}=$1.303 GeV, see the right panel
of Fig.~\ref{piNelastic_fig1} ,  
is small the   effect from the Roper resonance  is also expected to be  small.  
 As discussed in  Section \ref{section_isobar_contributions} the effect from
the nucleon Born term in the $\pi\DDelta$ channel  is less significant  close to threshold.
In the present calculations the main contributions to the $\pi^- p \to \pi^0\pi^0 n$
reaction close to threshold are driven by  $t$-channel pion exchange. This mechanism produces the invariant
distributions which are shifted to the higher  $\pi^0\pi^0$ invariant masses. However, the present calculations do not 
completely follow the experimental data at 1.303 and 1.349 GeV.
It is interesting that the calculations of \cite{Kamano:2006vm}  also underestimate the mass distributions at the same energy though 
using a different ansatz for the non-resonant part of the production amplitude.
The missing  contributions  are also seen in the angular cross sections at the same energies shown in Fig.~\ref{2piN_d3d_fig1}.
However in the latter case the effect is less pronounced since it is smeared out over a large kinematic region. 
 It is not clear whether  the missing strength is associated  with the neglected $S_{31}$-partial wave  contributions
or whether it could be  related to the  underestimation of the $\pi \pi$ correlations in the  isoscalar  channels.
We  postpone this problem to a future  study.

In the region of the Roper resonance our calculations are able to describe the mass distributions rather satisfactorily.
Also in this region the production strength is shifted to higher invariant masses $m_{\pi^0\pi^0}^2$. 
At the same time a peak at small  $m_{\pi^0\pi^0}^2$ 
becomes also visible. In \cite{Prakhov:2004zv} the authors identify this behavior with a strong decay of the $\NN(1440)$ state 
into the final $\pi \DDelta$ subsystem. At the same time a large decay branching ratio into the latter final state would lead to the more
pronounced two-peak structure as demonstrated  in Fig.~\ref{results_fig2}. In the present calculations the fit tends  to decrease 
the magnitude of  the $ \pi\DDelta$ production and compensate it by  enhancing the strength  into $\sigma N$.
The obtained decay branching ratio of $\NN(1440)$   for the $\sigma N$ channel is about twice  as large as for the $\pi\DDelta$. 

Both  the small peak at small and the broad  structure at large invariant masses are well reproduced 
indicating an important interplay between the $\sigma N$ and $\pi \DDelta$ production mechanism.  
 It is interesting that the isoscalar correlations in the $\pi\pi$ rescattering are also found
to be necessary  \cite{Kamano:2006vm} in order to reproduce the
asymmetric shape of the mass distributions. Hence the result of the present study and those from 
\cite{Kamano:2006vm} are opposite to the conclusion drawn in  \cite{Prakhov:2004zv} where no effect from the $\sigma N$ production
is found. 
 Though the $\pi\DDelta$ production
produces a two-peak structure only the first one  at small  $m_{\pi^0\pi^0}^2$ is visible  at energies 1.4-1.468 GeV. Within 
the present calculation the second peak at high  $m_{\pi^0\pi^0}^2$ is not seen because of the 
large $\sigma N$ contributions. In the present study 
 $\pi^0 \pi^0 n$ production is calculated as a coherent sum of  isobar contributions. Though the interference effect are important 
they are found to be very small  at the level of the total cross sections.

The results for the total cross sections are shown in Fig.~\ref{2piN_tot_fig1}. The present calculations  demonstrate
a very good description of the experimental data in the region of the Roper resonance.

\subsection{Unitarity\label{section_unitarity}}
 Unitarity is a one of the key issues in the baryon resonance analysis.
It relates the imaginary part of the elastic $\pi N$ scattering to the elastic 
and inelastic  total partial wave cross sections  in form of  the optical theorem Eqs.~(\ref{unitar3},\ref{unitar4}).
In this study the three-body unitarity is strictly maintained only up to interference terms between the isobar channels. 
 This raises the question to which extent the 
interference between the $\pi \DDelta$ and $\sigma N$ channels violates  the constraint of Eq.~(\ref{unitar4}).
The difference between the coherent and incoherent production is shown on right panel of Fig.~\ref{2piN_tot_fig1}. 
The solid(dashed) line corresponds to the case where the total cross section is calculated  taking into account (neglecting) 
the interference between isobar channels. Both curves almost  coincide indicating a very small 
difference between coherent and incoherent production in the present calculations. Note that the
 interference could still have a   visible impact on 
 the e.g. angular distributions. However being integrated  over the three-body phase space its  effect is found to be small 
in the total cross sections. This indicates that the contributions from the graphs (e)-(j) in  Fig.~\ref{unitar7},  which are 
neglected in the present study, are expected to be  small.

The left and the right parts of the optical theorem of Eq.~\refe{unitar4} for the $P_{11}$ partial 
wave are shown in Fig.~\ref{2piN_unitar_fig}.
The solid curve corresponds to the  imaginary part of the elastic $\pi^- p\to \pi^-p$ scattering . It can be  evaluated from 
the imaginary part of the $\pi N$ partial wave amplitude shown in Fig.~\ref{piNelastic_fig1} as follows:
\bea
{\rm Im}~ T^{\foh^+}_{\pi^-p\to\pi^-p}(\sqrt{s})= \frac{2}{3}\,{\rm Im}~ T^{\foh^ +}_{\pi N}(\sqrt{s}),
\eea
where  $\frac{2}{3}$ stands for an isospin factor.

To check the optical theorem the $\pi^- p \to \pi^-\pi^+ n$ and  $\pi^- p \to \pi^-\pi^0 p$ total cross sections have been calculated, see left panel of  Fig.~\ref{2piN_tot_fig1}.
Since the isospin $T_1^\fth$ contributions which are neglected in the present calculations
could be significant  in these reactions \cite{Manley:1984} the description of 
these channels in terms of only  $\pi \Delta$ and $\sigma N$ cannot be  fully complete. 
Therefore,  the results in the left panel of Fig.~\ref{2piN_tot_fig1} are  inelastic flux into these 
channels exclusively produced by the $\pi \Delta$ and $\sigma N$ ( for $\pi^-\pi^+ n$) 
in the $P_{11}$ wave channel. Note also that only the sum of these two quantities is fixed  here  due to the optical theorem. The effect from missing contributions is more pronounced in
the  $\pi^- p \to \pi^-\pi^0 p$ scattering. In fact, in the present calculations this reaction
is completely dominated by  $\pi \Delta$  which is clearly not enough to account for the total inelasticity in this channel.

The dashed curve in Fig.~\ref{2piN_unitar_fig} represents the sum of the total  $P_{11}$-wave cross sections for the
elastics $\pi^- p \to \pi^-p$ and all inelastic $\pi^- p \to \pi^0n$, $\pi^0\pi^0n$, $ \pi^+\pi^- n$, and   $  \pi^0\pi^- p$
transitions multiplied by the   normalization factor $k^2 (4\pi)^{-1}$ according to Eq.~\refe{unitar4}. 
The contributions to the $2\pi N$ final states are calculated coherently. 
% Since the incoherent production  fulfills the 
% condition of the optical theorem by construction 
 For the total $\pi N \to2\pi N$ cross section evaluated incoherently the condition of the optical theorem is
fulfilled by  construction. For this quantity the right part of   Eq.~\refe{unitar4} practically coincides with the left part of  Eq.~\refe{unitar4} and therefore is not 
shown here.
The effect of the interference  between the isobar channels in  the total cross sections is found to be small. The comparison of the left and right hand side  of the  Eq.~\refe{unitar4} 
in Fig.~\ref{2piN_unitar_fig} demonstrates that the condition of the  optical theorem for the three-body unitarity is fulfilled  with good accuracy.
 
Since  unitarity relates both $\pi N$ elastic and $2\pi N$ reactions one could expect that all $2\pi N$ total cross sections should also be explained once unitarity is fulfilled. We stress here that in the present work  we limited our calculations to  the 
$IJ^P$=$\foh\foh^\pm$ partial waves. Thus,  in these spin-isospin channels the constraints of Eq.~\refe{unitar4} 
are fulfilled  for each term up to a very small interference effect as discussed above. Therefore the remaining  isospin  $T^\fth_1$ component contributed by the  $\rho N$ channel not included here,  could be still important for the description of the $\pi^-\pi^0p$ and $ \pi^+\pi^- n$ final states.

\subsection{Resonance parameters \label{Nstar_param}}
The properties of the nucleons resonances are defined by the corresponding  pole positions and residues.  
However, for unstable particles the residues become  dependent on  the invariant isobar masses. 
One of the  possible ways to overcome this problem 
is to define these  quantities  as an integral over the corresponding  isobar spectral function. Due to the complexity of the structure of the
isobar amplitudes the  poles and residues 
of the $2\pi N$ amplitude will be discussed  elsewhere. Here we provide the    Breit-Wigner  parameters of the  resonances. 
Also in this case the width of the resonance decay into the isobar final state depends on the invariant mass of the unstable particle.   Therefore the 
quantities of interest  are calculated as  an integral over the corresponding spectral function, see Appendix~\ref{Appendix_Lagrangians}.

The extracted resonance properties are listed in~Table \ref{tab2} in comparison with the results from other studies. 
	The errors of the extracted resonance parameters have been obtained by combining results from several fits with $\chi^2$ values within
	$5\%$ deviation from the minimal value.	
	 The obtained validity ranges derived with this method are in general larger than those extracted  from the correlation matrix.

Since the 
analysis is done for energies below the  $\eta N$  production threshold 
the parameters of  $\NN(1535)$ cannot be  fully constrained. The mass is found to be in a wide range with the central value larger than in our
previous calculations \cite{Shklyar:2012js}. The decay branching ratio $R_{\pi N}$ into the  $\pi N$ final state is, however, very close to our  previous  result \cite{Shklyar:2012js}.
We obtain almost zero values for the $R_{\sigma N}^{N(1535)}$ and $R_{\pi \DDelta}^{N(1535)}$. This conclusion is in line with  other findings, see Table~\ref{tab2}. 

The mass of the Roper resonance is  lower than that found in our previous calculations. We obtain a quite large 
total decay width of $\NN(1440)$ which is, however, smaller than  that 
extracted in  previous work \cite{Shklyar:2012js}.
The decay strength for   the $\pi N$ channel is very close to the values given by PDG and other groups.  The obtained  branching ratio $R_{\sigma N}=27\%$  agrees very well 
with the recent  result of the KSU group \cite{Shrestha:2012ep}. However they find an almost twice lower  branching ratio for the 
 decay of  $\NN(1440)$ into the $\pi\DDelta$
subchannel. At the same time they get a slightly larger value for the $R_{\pi N}=64.8\%$ which should be compared with   $R_{\pi N}=61\%$ derived here.
The remaining decay flux of  about 1.5\%  is associated with the $\rho N$ isobar final state  \cite{Shrestha:2012ep}.  
 An opposite conclusion is drawn by the BoGa group \cite{Anisovich:2011fc}. They find a larger decay strength  for the $\pi\DDelta$ subchannel.
{  The $\sigma N$ decay flux of the Roper
resonance is also found to be large: 17\%.  This values  is somewhat smaller than the  $\pi\DDelta$ decay fraction.
 \begin{table}[t]
  \begin{center}
    \begin{tabular}
      {l|l|l|r|c|c|l}
      \hhline{=======}
      $N^*$ & mass & $\Gamma_{tot}$ & $R_{\pi N}$ & $R_{\sigma N}$ & $R_{\pi\DDelta}$& Reference \\
      \hhline{=======}
$\NN(1535)$\,$\foh^-$ &\, $1.544^{+6}_{-23}$ \, &\, $127^{+30}_{-9}$    \,&~ $36^{+4}_{-3}$       \, &~ $0^{+1}$           \, &~ $ 0^{+1}$             \,&\, this work\\
                  &\, $1.526^{+2}_{-2}$  \, &\, $131^{+12}_{-12}$   \,&~ $35^{+3}_{-3}$       \, &~  ng              \, &~   ng                  \,&\,  GiM12 \cite{Shklyar:2012js}\\
                  &\, $1.535^{+10}_{-10}$\, &\, $150^{+25}_{-25}$   \,&~ $45^{+10}_{-10}$     \, &~ $2^{+1}_{-1}$      \, &~ $ 0^{+1}$             \,&\,  PDG12\cite{pdg}\\
                  &\, $1.519^{+5}_{-5}$  \, &\, $128^{+14}_{-14}$   \,&~ $54^{+5}_{-5}$       \, &~ ng                \, &~ $ 2.5^{+1.5}_{-1.5}$  \,&\,  BoGa12\cite{Anisovich:2011fc}\\
                  &\, $1.538^{+1}_{-1}$  \, &\, $141^{+4}_{-4}$     \,&~ $37^{+1}_{-1}$       \, &~ $1.5^{+0.5}_{-0.5}$\, &~ $ 2.5^{+1.5}_{-1.5}$  \,&\,  KSU\cite{Shrestha:2012ep}\\
     \hline
$\NN(1440)$\,$\foh^+$ &\, $1.478^{+17}_{-27}$\, &\,  $569^{+30}_{-240}$ \, &\, $61^{+2}_{-7}$       \, & $27^{+4}_{-9}$      \,& $12^{+5}_{-3}$         \, &\, this work     \\
                  &\, $1.515^{+15}_{-15}$\, &\,  $605^{+90}_{-90}$  \, &\, $56^{+2}_{-2}$       \, &  ng                \,&   ng                  \, &\,  GiM12 \cite{Shklyar:2012js}\\
                  &\, $1.440^{+30}_{-20}$\, &\,  $300^{+150}_{-100}$\, &\, $65^{+10}_{-10}$     \, & $15^{+5}_{-5}$      \,& $ 25^{+5}_{-5}$        \, &\,  PDG12\cite{pdg}\\
                  &\, $1.430^{+8}_{-8}$  \, &\,  $365^{+35}_{-35}$  \, &\, $62^{+3}_{-3}$       \, & $17^{+7}_{-7}$      \,& $ 21^{+8}_{-8}$        \, &\,  BoGa12\cite{Anisovich:2011fc}\\
                  &\, $1.412^{+2}_{-2}$  \, &\,  $248^{+5}_{-5}$    \, &\, $64.8^{+0.9}_{-0.9}$ \, & $27^{+1}_{-1}$      \,& $ 6.5^{+0.8}_{-0.8}$   \, &\,  KSU\cite{Shrestha:2012ep}\\
                  &\, $1.458^{+12}_{-12}$  \, &\,  $363^{+39}_{-39}$    \, &\,ng \, & ng     \,& $ 40.5^{+17.5}_{-17.5}$   \, &\,  JM \cite{Mokeev:2012vsa} \\
      \hline
%$g_{N\Sigma K}$  &   2.48 & $g_{N\Sigma K_0^*}$ & $-52.30$ & $g_{N\Sigma K^*}$  &   4.33 & $\kappa_{N\Sigma K^*}$  &  $-0.86$ \\
      \hhline{=======}
    \end{tabular}
  \end{center}
  \caption{Breit-Wigner resonance parameters obtained in the present study. The decay branching ratios are given in percents.  The relevance intervals are shown by  the upper (lower) 
   subscripts. 'ng' - not given.
    \label{tab2}}
\end{table}

Both in the BoGa and  the present analysis the extracted parameters have large error bars.  Within the 
validity limits the results of this study  are overlapping  with the findings 
of \cite{Anisovich:2011fc}.  
The result of the JM \cite{Mokeev:2012vsa} 
analysis of the CLAS electroproduction data demonstrates a 
large $\pi \DDelta$ decay fraction of $\NN(1440)$. 
The central  value of 40\% is about 6 times larger than that obtained in   KSU calculations \cite{Shrestha:2012ep}. 
Using the quoted  values $R_{\rho N}^{N(1440)}<2\%$ with  the lower bound for   $R^{N(1440)}_{\pi \DDelta}=40.5 -17.5=23$\% from  \cite{Mokeev:2012vsa} and taking   
$R^{N(1440)}_{\pi N}=61\%$ as dictated by the analysis of the $\pi N$ inelasticities on gets  
 $R_{\sigma N}^{N(1440)}<16\%$ as an upper limit for the
$\sigma N$ branching ratio of   $\NN(1440)$.  Since  the decay properties of $\NN(1440)$ listed in Table~\ref{tab2} are obtained using different 
theoretical frameworks  and different reaction database it is not clear whether the difference between various analysis could be adresses to the model 
assumptions or related to a lack of experimental input. One may hope that the combined analysis of photon and pion incuduced reaction would help 
to pin down the parameters of $ \NN(1440)$.

One of the largest sources of uncertainties in the present calculations is related to the  possible influence  of the 
$\NN(1520)$ state on the $\pi^- p \to \pi^0\pi^0 n$ production. 
 Since the contribution from  the $\NN(1520)$ resonance is neglected in the present 
calculation we translate this effect into  errors  of
the extracted resonance parameters.
The contribution from $\NN(1520)$  can be estimated from the comparison of the $\pi N$ inelasticities in 
the $J=\foh^+$ and $J=\fth^-$ partial waves.  These quantities are evaluated in \cite{shklyar:2004b}  and shown in Fig.~\ref{2piN_inelas}. 
The $J^P=\fth^{-}$ inelastic cross section
rapidly rises starting from 1.42 GeV indicating the importance  of the  $\NN(1520)$ state at energies above 1.46 GeV.   To estimate the influence of the $J=\fth^-$ partial wave on the width of N(1440) we construct  an additional data set where the original   $\pi ^- N \to\pi^0\pi^0 N$ experimental data are scaled with the common scaling factor $f_s(\sqrt{s})=(\sigma^{\foh\foh^+}_{\rm inel.} -\sigma^{\foh\fth^-}_{\rm inel.})/\sigma^{\foh\foh^+}_{\rm inel.} $. Here 
$\sigma^{IJ}_{\rm inel.}$ is a total $\pi N$ inelastic partial wave cross section
for the given isospin $I$ and spin $J$ as shown in Fig.~\ref{2piN_inelas}. 
Then the parameters of  N(1440) are again
extracted by making an additional fit to the scaled data.
 The deviation from the original
parameters  indicates the influence of the  J$^{P}$I=$\foh^-\fth$ wave on the Roper resonance
parameters.
Taking this effect into account we obtain a large error
interval for the total width of $\NN(1440)$.   This  is a very conservative estimate of the  effect of  the  $\NN(1520)$ state.
We stress that in the case of the   large  $\rho N$ decay fraction of  $\NN(1520)$  \cite{pdg}
its  actual impact on the  $\pi^0 \pi^0 n$ production could be  smaller than 
concluded from the simple comparison of the $\pi N$ inelasticities. 
The contributions from isospin $I=\fth$ partial waves are found to be small, see Fig.~\ref{2piN_inelas}.

\begin{figure}
	\begin{center}
		\includegraphics[width=9.5cm]{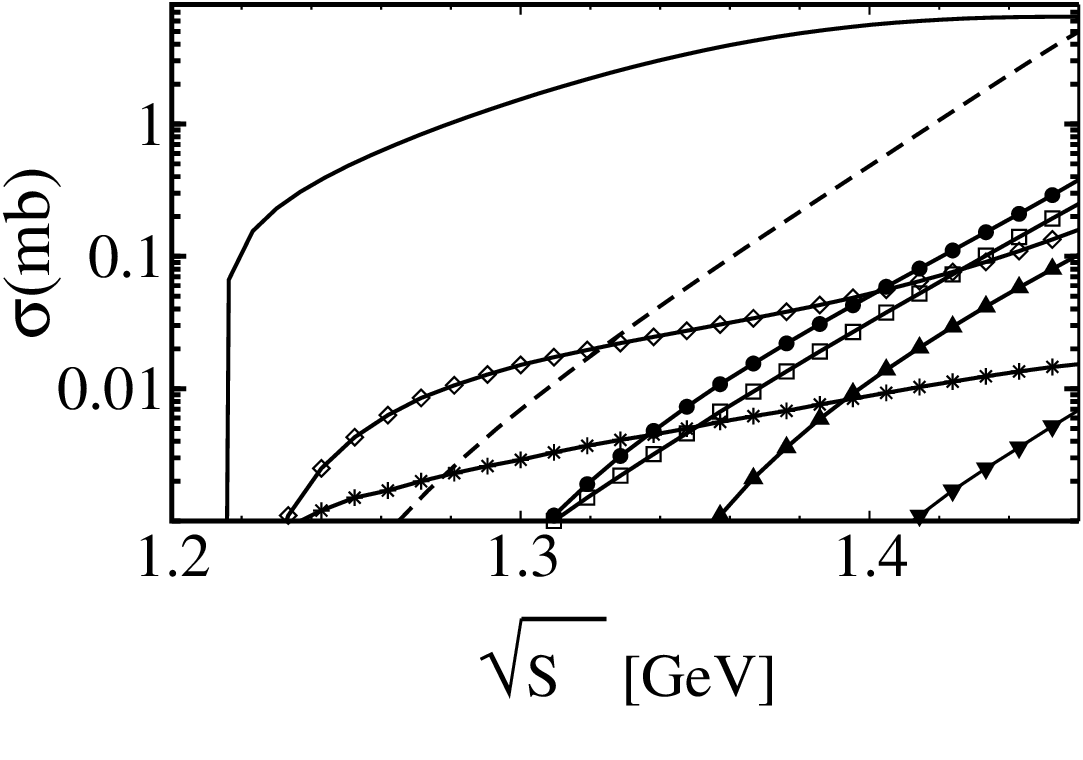}
		\caption{Comparison of the $\pi N$ partial wave inelasticities. Notation is as follows:
				$(\blacktriangle\hspace*{-0.5mm}\textendash\hspace*{-0.3mm}\blacktriangle)$: J$^{\rm P}$I=$\foh^-\fth$,\,
			$(\blacktriangledown\textendash\blacktriangledown$): $\fth^+ \fth$,\,
			$(\ast\textendash\ast)$: $\foh^+\fth$,\,
            $(\square\textendash\square$): $\fth \fth^-$,\,
  	  		$(\Diamond\textendash\Diamond)$: $\fth \fth^+$,\,
	  	  		$(\bullet\textendash\bullet)$: $\foh \foh^-$,\,
	  	  			$(\textendash\textendash\textendash)$: $\foh \foh^+$,\,
	  	  				$(\textendash\,\textendash)$: $\foh \fth^-$.
			\label{2piN_inelas}}
	\end{center}
\end{figure}

%%%%%%%%%%%%%%  checked
%%%%%%%%%%%%%%

 \subsection{Partial wave analysis of the $\pi N\to \pi N$, $2\pi N$  reactions }
The inelastic partial wave  cross sections calculated in this work are shown  in Fig.~\ref{PWA_isobar_fig1} in comparison with the results  
obtained from the SES extracted in \cite{Manley:1984}.  The energy-dependent solutions  from the latter work are also shown in the same figure.  
Our results demonstrate  larger inelastic 
contributions in  the $\sigma N$ channel   than those  extracted  by Manley et. al. On the other hand the agreement in  the $\pi\Delta$ subchannel is good.  
The difference between the GiM results and those from  \cite{Manley:1984} is also visible in  Fig.~\ref{PWA_isobar_fig2} where the 
total $P_{11}$ $\pi N$ inelasticities are presented in comparison with the results from the GWU analysis \cite{Arndt:2006bf}.
Above 1.4 GeV the  $2\pi N$ cross section from   \cite{Manley:1984} tends to be lower than the $P_{11}$ $\pi N$ inelasticity extracted in 
by the GWU group \cite{Arndt:2006bf}. This could be an indication for the inelastic 
contributions from  the e.g. $3\pi N$ channel. The difference between the $\pi N$ inelasticity  
 and the  $2\pi N$ reaction cross section  could amount up to 1.5mb 
at   $\sqrt{s}=1.46$GeV. 
In the  present study the possible effect from the $3\pi N$ production has been neglected and the whole inelastic flux 
moves into the $\sigma N$ channel. Thus we obtain   a larger $\sigma N$ contributions above 1.4 GeV as in the analysis  of \cite{Manley:1984}, 
see the left panel of  Fig.\ref{PWA_isobar_fig1}.  Obviously  conclusions on effects from the $3\pi N$ channel   can only be drawn when  
this final state  is  explicitly  included in calculations preserving  the unitarity constraint.

The $P_{11}$ $\pi N$ inelasticity calculated from GiM amplitudes is generally lower than that
obtained  from the GWU analysis. The reason is that the real 
and imaginary parts of the elastic  $\pi N$  amplitudes   tend to be  slightly larger than  the $P_{11}$ GWU solution, see Fig.\ref{PWA_isobar_fig2}. 
Due to unitarity this leads to somewhat lower inelastic reaction cross section than obtained in  \cite{Arndt:2006bf}. Note that in the present 
study the combined analysis of $\pi N\to \pi N$,  $2\pi N$    transitions is made assuming only $S_{11}$ and $P_{11}$ partial wave contributions. 
The inclusion of higher partial waves and additional decay channels (e.g. $\rho N$) could  lead to  the re-distribution of the inelastic flux
 between  the various  partial wave amplitudes of the $2\pi N$ production.  Thus further 
extensions of the model   are required  for a more  accurate extraction of the  partial wave contributions.

%%%%% HELP%%%  1\S+\N\n2\v{0.55}\h{-0.5}-

 \begin{figure}
   \begin{center}
     \includegraphics[width=16cm]{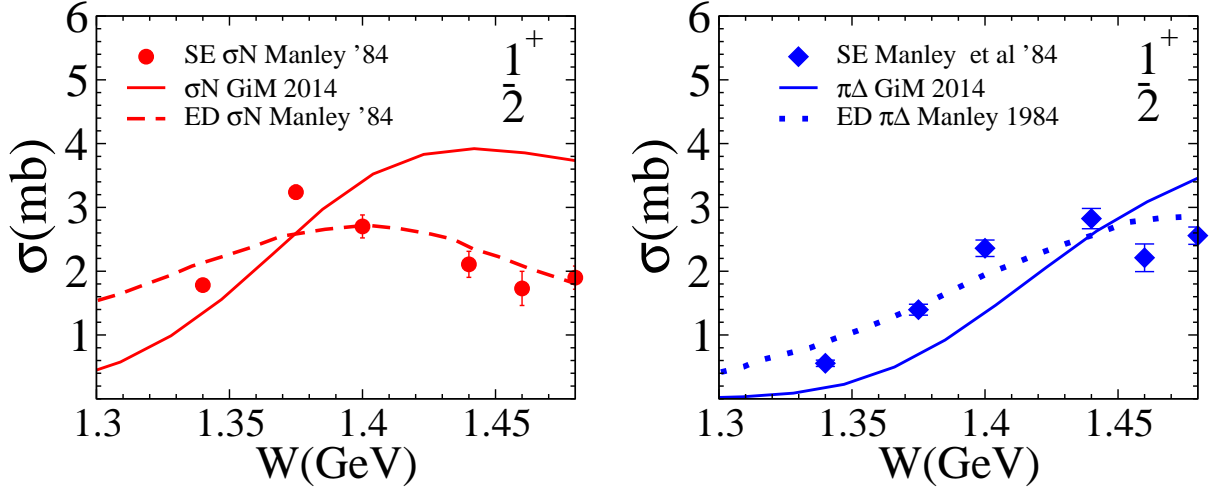}
     \caption{(Color online) $J^P={\frac{1}{2}}^+$ reaction cross sections $\sigma_{\pi N\to\sigma N}$ (left) and  $\sigma_{\pi N\to\pi\DDelta}$ (right) 
in comparison with the single energy(SE) and energy-dependent  (ED) results from 
 Manley'84\cite{Manley:1984}.
       \label{PWA_isobar_fig1}}
   \end{center}
 \end{figure}

 \begin{figure}
   \begin{center}
     \includegraphics[width=16cm]{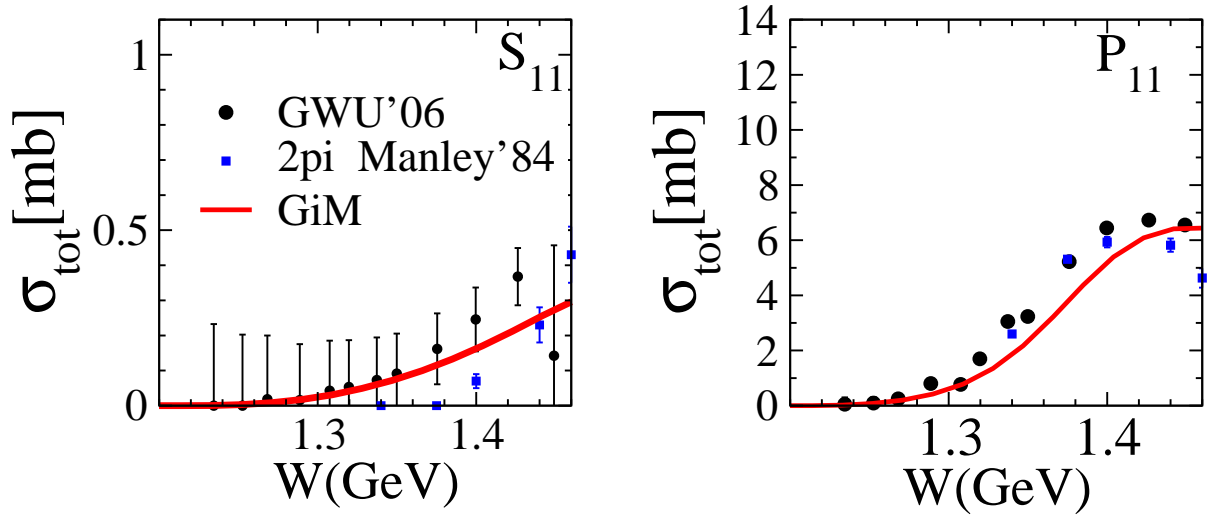}
     \caption{(Color online) the $S_{11}$ (left) and  $P_{11}$ (right) $\pi N$ inelasticities  vs the results from GWU  \cite{Arndt:2006bf} and the $2\pi$ cross section from  Manley et al \cite{Manley:1984}.
       \label{PWA_isobar_fig2}
 }
   \end{center}
 \end{figure}
Since  the $ \sigma N$ and $\pi \Delta$ partial wave amplitudes obtained in  this study have an additional dependence on the isobar mass the 
 direct  comparison of  our results  with the SES from   \cite{Manley:1984} is difficult. The reason is that the dependence on the isobar mass
is neglected in \cite{Manley:1984}. These amplitudes are normalized to give the reaction cross section in the form 
\bea
\sigma^{JP}_i(\sqrt{s}) = \frac{4\pi }{k^2}(J+\foh)  |T_i^{JP}(\sqrt{s})|^2,
\label{Manley_reaction_cr_sec}
\eea
with $i=\pi\DDelta$, $\sigma N$.
The same quantity in  the GiM calculations is given in terms of the   integral over the  isobar mass $\mu_i:$
\bea
\sigma^{JP}_i(\sqrt{s}) = \frac{4\pi }{k^2}(J+\foh) \int_{\mu_{\rm min}^2}^{\mu_{\rm max}^2} 
|T^{JP,\, {\rm GiM}}_i(\sqrt{s},\mu^2_{\gamma_i})|^2 A_{\gamma_i} (\mu^2_{\gamma_i})d\mu^2_{\gamma_i},
\label{GiM_reaction_cr_sec}
\eea
where $A_{\gamma_i} (\mu^2)$ is a spectral function
of the isobar $\gamma_i=\sigma, \DDelta$.   If  $T^{JP,\,{\rm GiM}}_i(\sqrt{s},\mu^2)$ had no $\mu^2$-dependence the Eq.~\refe{GiM_reaction_cr_sec} would reduce
to the form  which is similar to Eq.~\refe{Manley_reaction_cr_sec}
\bea
\sigma^{JP}_i(\sqrt{s}) = \frac{4\pi }{k^2}(J+\foh)  |T^{JP,\, {\rm GiM}}_i(\sqrt{s}) N_i(\sqrt{s})|^2,
\label{Manley_reaction_cr}
\eea
up to the  additional normalization factor 
\bea
N_i(\sqrt{s}) = \sqrt{  \int_{\mu_{\rm min}^2}^{\mu_{\rm max}^2}  A_{\gamma_i} (\mu^2_\gamma)d\mu^2_{\gamma_i} }.
\label{Manley_reaction_cr3}
\eea
This factor takes into account   the propagation and decay of an isobar. In  Eq.~\refe{Manley_reaction_cr_sec} it is absorbed into 
the normalization of the reaction amplitudes.
 \begin{figure}
   \begin{center}
     \includegraphics[width=17cm]{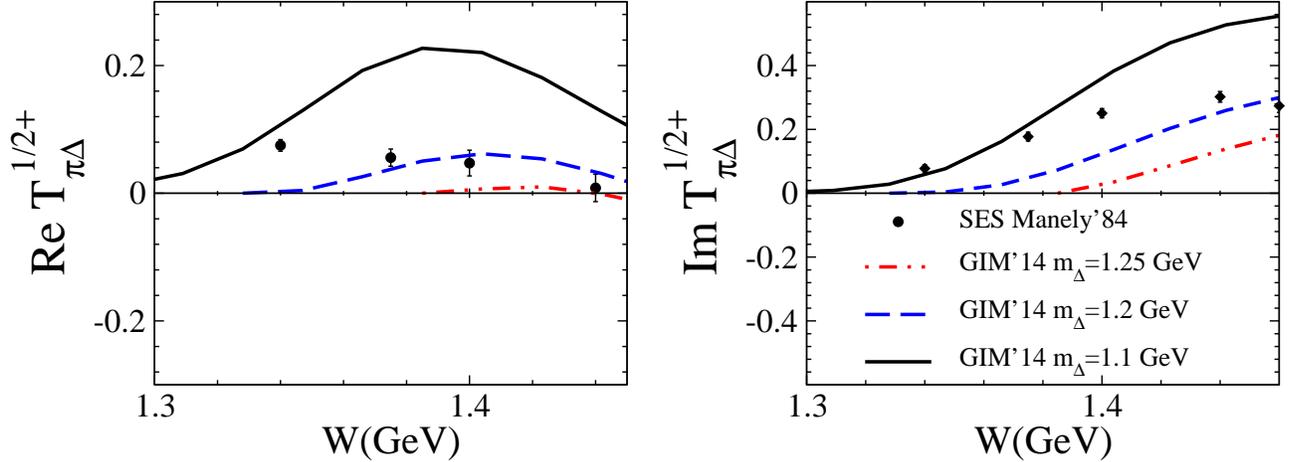}
\caption{(Color online) $J^P=\frac{1}{2}^+$ $\pi N \to\pi\DDelta$  amplitude:  the real(left) and imaginary (left) parts of the isobar production amplitudes 
  for different values of the isobar-mass.  The GiM amplitudes are normalized according Eq. \refe{Manley_reaction_cr4}.  
 The SES from  Manley et al \cite{Manley:1984} are presented by filled circles.
            \label{PWA_isobar_fig3}
}
   \end{center}
 \end{figure}
\begin{figure}
   \begin{center}
     \includegraphics[width=17cm]{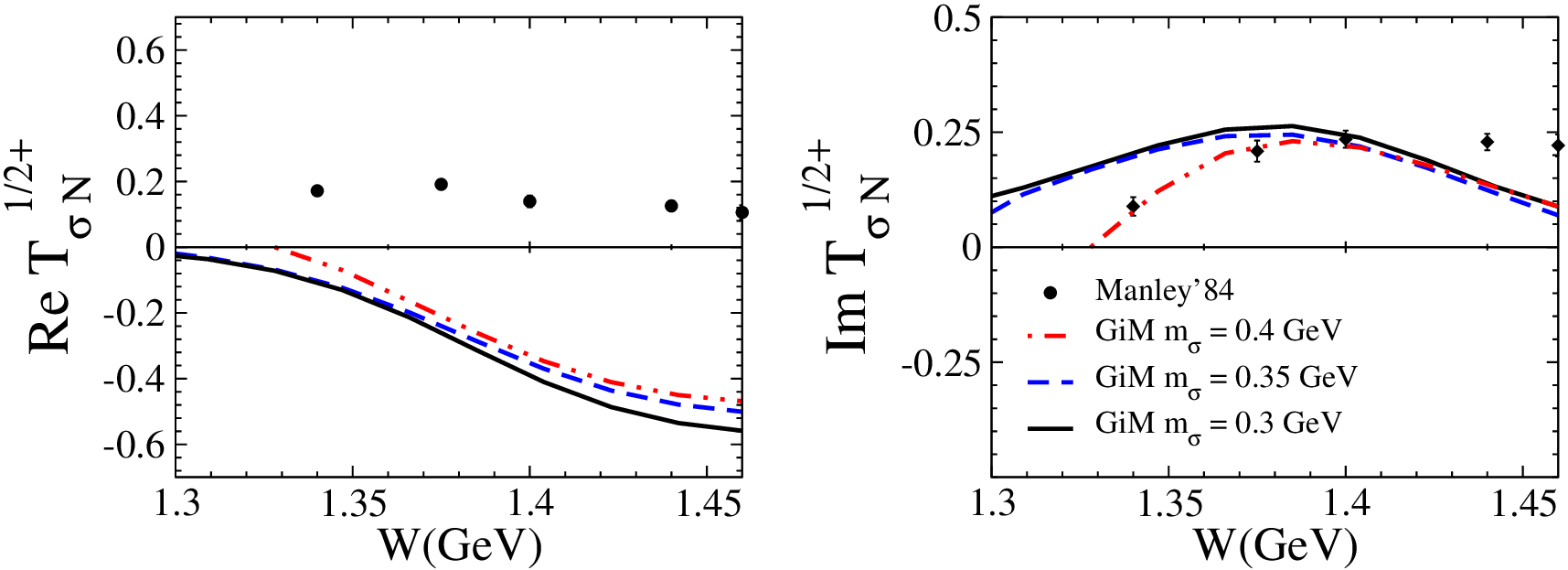}
     \caption{ (Color online) $J^P=\frac{1}{2}^+$ $\pi N \to\sigma N$ amplitude; notation is same as in  Fig.~\ref{PWA_isobar_fig3}.
       \label{PWA_isobar_fig4}}
   \end{center}
 \end{figure}
 \begin{figure}
   \begin{center}
     \includegraphics[width=17cm]{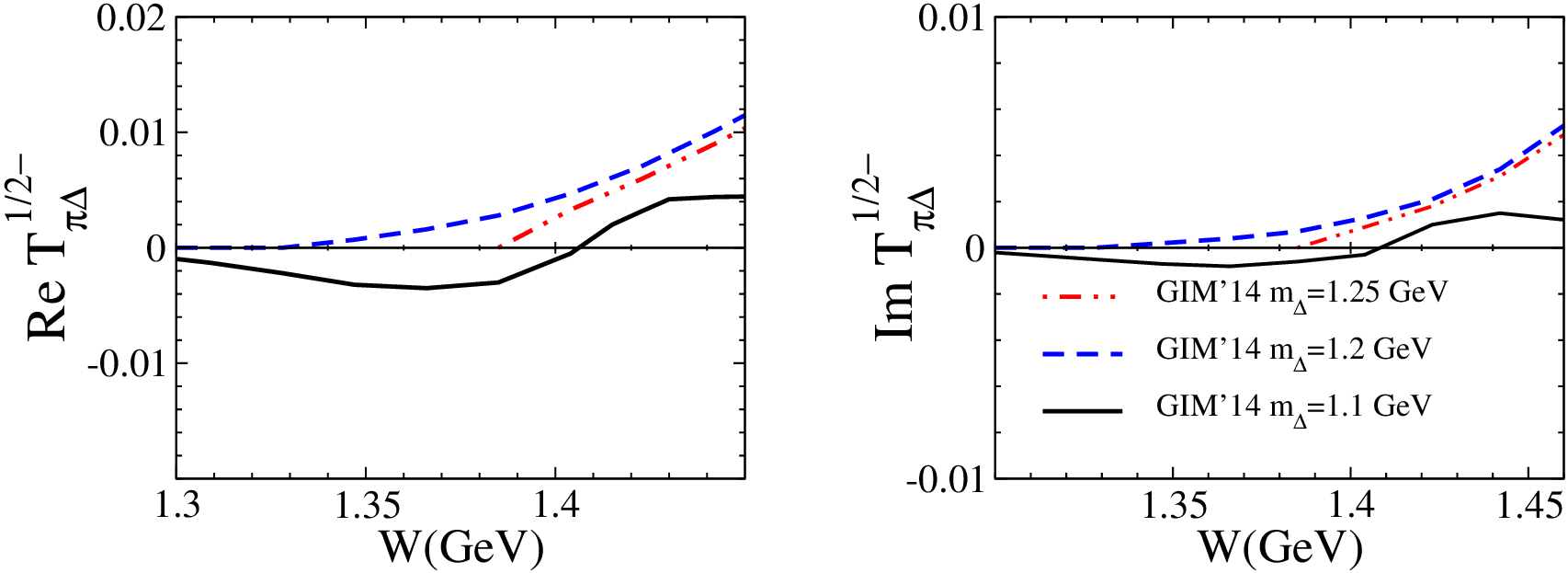}
     \caption{ (Color online) $J^P=\foh^-$ $\pi N \to\pi\DDelta N$ amplitude; notation is same as in  Fig.~\ref{PWA_isobar_fig3}.
       \label{PWA_isobar_fig5}}
   \end{center}
 \end{figure}
\begin{figure}
   \begin{center}
     \includegraphics[width=17cm]{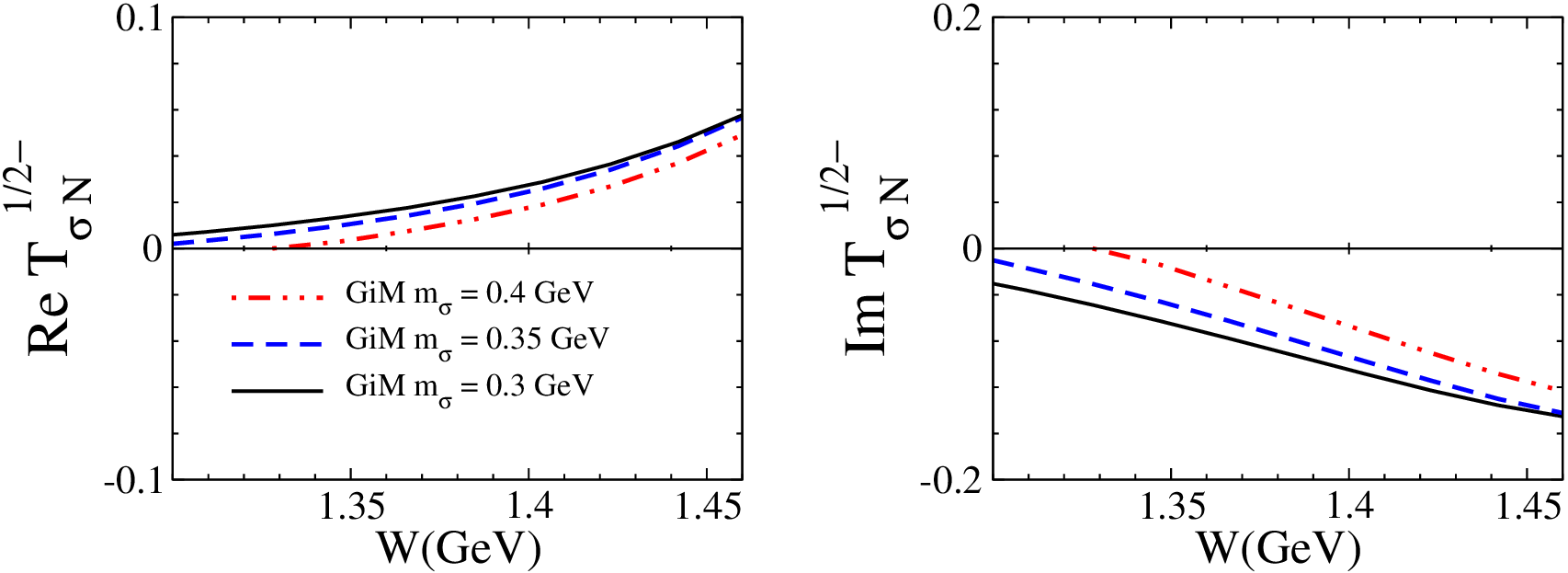}
     \caption{ (Color online) $J^P=\foh^-$ $\pi N \to\sigma N$ amplitude; notation is same as in  Fig.~\ref{PWA_isobar_fig3}.
       \label{PWA_isobar_fig6}}
   \end{center}
 \end{figure}

To compare  our  results  with those of Manley et al. \cite{Manley:1984} we therefore multiply our isobar amplitudes  by the factor $N_i(\sqrt{s})$:
\bea
T^{JP,\,{\rm GiM}}_i(\sqrt{s}, \mu^2_i) \to 
T^{JP,\,{\rm GiM}}_i(\sqrt{s}, \mu^2_i) N_i(\sqrt{s})
\label{Manley_reaction_cr4}
\eea
In Fig.~\ref{PWA_isobar_fig3} and Fig.~\ref{PWA_isobar_fig4} the  $J^P=\frac{1}{2}^+$ reaction  amplitudes  as defined in Eq.~\refe{Manley_reaction_cr4}
are presented in comparison with the
SES from  \cite{Manley:1984}. Except for  the real part of the  $\sigma N$ amplitude we find a good agreement with the results from \cite{Manley:1984}.
The major difference is the sign of the   Re$T^{\foh^+}_{\sigma N}$ amplitude. While  Re$T^{\foh^+}_{\sigma N}$ extracted in \cite{Manley:1984}
 is positive in the energy region at hand the real part of the GiM-amplitude  for the $\pi N \to \sigma N$ transition is negative.
The reason for this difference is unclear.  The absolute magnitude of  Re$T_{\sigma N}^{\foh^+}$  also tends to be larger than that of  \cite{Manley:1984}.
This  effect can be attributed to the additional $J^P=\foh^+$ inelastic flux found in \cite{Manley:1984} as discussed above. 
The inclusion of the $3\pi N$ channel would  bring an additional constraint to check the contribution from this channel.

Our calculations demonstrate that the dependence on the isobar masses cannot be neglected.
Though the $\sigma N$ amplitude factorized in the form Eq.~\refe{Manley_reaction_cr4} is a  smooth function of $\mu_\sigma^2$
above 1.38 GeV  the dependence on the isobar mass becomes more visible at lower energies. Thus, e.g., the  imaginary part  of $T_{\sigma N}^{\foh^+}$
vanishes for $\sqrt{s}< m_N + \mu_\sigma$. The mass dependence of $T_{\pi\Delta}^{\foh^+}$ shown in Fig.\ref{PWA_isobar_fig3} is more drastic: both
imaginary and real parts of the reactions amplitudes exhibit  fast variations as  functions of $\mu_{\DDelta}$.  
Except  for the sign at the Re$T_{\sigma N}^{\foh^+}$  amplitudes our calculation demonstrate in general good agreement with the SES from  \cite{Manley:1984}.
This agreement  is quite  remarkable taking into account the difference in  theoretical ansatz and in the reaction database used in the analysis.

The isobar amplitudes  in the $J^P=\foh^-$ partial wave are presented in Fig.~\ref{PWA_isobar_fig5} and Fig.~\ref{PWA_isobar_fig6}. The SES solutions from \cite{Manley:1984}
are not available at these energies.  In the present study we obtain almost vanishing  $\sigma N$ and $\pi\DDelta$
decay branching ratio of the  $\NN(1535)$ resonance. As a result the magnitude of the real and imaginary parts of the
$J^P=\foh^-$ $\pi\DDelta$ reaction amplitude are very small. The corresponding values for the $\foh^-$-wave for the  $\pi N\to \sigma N$ reaction 
are  found to be larger, see Fig.~\ref{PWA_isobar_fig6}. The dominant contribution to this amplitudes comes from the non-resonant t-channel
pion exchange. However the overall effect from the $J^P=\foh^-$  reaction amplitudes in the $2\pi N$ production is very small.  
This is also consistent with 
the $S_{11}$ $\pi N$   inelasticity shown in the left panel of Fig.~\ref{PWA_isobar_fig2} which is about an order of magnitude smaller than that of 
the $J^P=\foh^+$ scattering.  

  \section{Conclusion}
We develop a coupled-channel Lagrangian approach for the partial wave analysis of the $\pi N \to\pi N$, $\pi\pi N$ reactions.
The $\pi \pi N$ production is treated in the isobar approximation.  In this framework the optical theorem for the three-body unitarity 
is strictly fulfilled up to interference between the isobar channels.   The Bethe-Salpeter equation is solved to
obtain the reaction amplitudes. 
  Assuming  dominant contributions from the $S_{11}$
 	and $P_{11}$ partial waves in the $\sigma N$ and $\pi\Delta$ production channels
 	 we have performed a partial wave analysis of the $\pi N \to \pi N$ and $\pi^- p \to \pi^0\pi^0 n$ reactions 
 	 especially well suited to study the properties of the Roper resonance.
The  calculations demonstrate a good description of   both reactions.
We conclude that the invariant $\pi^0\pi^0$ mass distributions  play a crucial role in the separation   of the isobar contributions.
The  $\pi^- p \to \pi^0\pi^0 n$ reaction close to   threshold is dominated by the
$\sigma N$ production due to the $t-$channel pion exchange. The nucleon Born term contribution to the $\pi\DDelta$ channel is found to be
less significant. Similar effects are also found  in the independent study of \cite{Kamano:2013ona}.

For the decay branching ratios of $\NN(1440)$ we obtain  $R_{\sigma N}^{N(1440)}=27^{+4}_{-9}\%$ and  $R_{\pi \DDelta}^{N(1440)}=12^{+5}_{-3}\%$. 
Our value for  $R_{\sigma N}^{N(1440)}$ coincides with  the result of Shrestha and Manley  \cite{Shrestha:2012ep}. On other hand the 
central value of $R_{\pi \DDelta}^{N(1440)}=12^{+5}_{-3}\%$ is almost twice larger than  those  derived by these authors:
$R_{\pi \DDelta}^{N(1440)}=6.5^{+0.8}_{-0.8}$\%\cite{Shrestha:2012ep}.

The comparison of our results with the  parameters extracted by the  BoGa group 
$R_{\sigma N}^{N(1440)}=17^{+7}_{-7}\%$ and  $R_{\pi \DDelta}^{N(1440)}=21^{+8}_{-8}\%$ demonstrates that despite on the visible  difference 
in the central values these quantities could still coincide within their error bars. 
The extended analysis of the $\pi\pi N$ which includes higher partial waves would help to reduce the uncertainties of the 
extracted resonance properties.

The present calculations demonstrate a good agreement with the $S_{11}$ and $P_{11}$ $\pi N$ 
inelasticites from the GWU analysis.   We extract the  $\foh^-$- and $\foh^+$-partial wave   amplitudes
of the $\sigma N$ and $\pi\DDelta$  production. The obtained partial waves have an additional dependence on the isobar masses.
The extracted amplitudes are also in good agreement with the results of  Manley et al \cite{Manley:1984},
except for the sign of the real part of the $\sigma N$ amplitude.

In the present work the Roper resonance is described as a genuine  pole in contrast to the dynamical pole generated  by 
the  correlations in the $\sigma N$ subchannel as reported  in \cite{Ronchen:2012eg}. 
Both calculations  demonstrate a very good description of the  $P_{11}$ $\pi N$ elastic scattering amplitudes. 
This rises a question how   these scenarios could be identified  in experiment. 
The study of the $\pi\pi N$  reaction provides a chance to explore these possibilities  in more details.
If the pole associated with the Roper resonance is dynamically generated due to a strong $t-$ channel exchange in the $\sigma N$
channel one could also expect a substantial contribution from this mechanism to 
the higher partial waves of the $\pi N \to \pi\pi N$  production. This effect could  be more pronounced
with increasing   scattering energy.  At the same time the genuine pole  produces in general only minor 
'background contributions' due to  $u$-channel exchange; the major  effect is seen in the $J^P=\foh^+$ partial wave.
The angular distributions could be also different: the strong $t$-channel exchange ordinarly  gives rise at forward
angles  which can also be identified in the angular distributions. At the same time the $u-$channel mechanism
is more important  at backward angles. 
Therefore, a detailed analysis of the $2\pi N$  reaction could help to disentangle various scenarios. 
  
This program  cannot   be accomplished without a new generation of the high statistic $\pi N \to \pi \pi N $ scattering  data.
New measurements at  the HADES \cite{hades} and JPARC \cite{jparc} facilities would help to resolve to problem.

\section*{Acknowledgement}
Work supported in part by DFG, grant Le439/7 and SFB/TR 16.

\begin{appendix}

\section{\label{AppendixA} Three-body unitarity for  the  $\sigma N$  isobar channel}
Here we demonstrate the maintenance of the tree-body unitarity Eq.~\refe{unitar3} when the pions are produced via $\sigma N$ subchannel.
For the sake of clarity isospin indices are omitted. 
The scattering equation  Eq.~\refe{eq6} for the $(\pi/\sigma) N \to (\pi/\sigma) N$ transitions can be rewritten 
in the form:
% 
%Using  the coupled channel equation Eq.~\refe{eq6} for the isobar transition  amplitudes 
\bea
T^{JP}_{f\,i}(\sqrt{s})=  K^{JP}_{f\,i}(\sqrt{s}) + {\rm i} \,T^{JP}_{f\,\pi N} K_{\pi N\, i}^{JP}
+{\rm i}\int d\mu^2 A_\sigma (\mu^2)T^{JP}_{f\,\sigma N} K_{\sigma N\, i}^{JP}.
\label{appendixA_eq1}
\eea 
%and 
By replacing  the integral in Eq.~\refe{appendixA_eq1} by  summation one gets
\bea 
\int_{4 m_\pi^2}^{(\sqrt{s}-m_N)^2} d\mu^2 A_\sigma(\mu^2)\, T^{JP}_{f\,\sigma N}(\sqrt{s},\mu^2)\, K_{\sigma N\, i}^{JP}(\sqrt{s},\mu^2) =\nonumber\\
 \sum_l \Delta\mu^2_l\,\,A_{\sigma_l}(\mu_l^2)\,\, T^{JP}_{f\,\sigma_l N}(\sqrt{s},\mu^2_l)\, K_{\sigma_l N\, i}^{JP}(\sqrt{s},\mu^2_l).
\label{appendixA_eq1.1}
\eea
By introducing the  amplitudes and interaction kernel in the form
\bea
\tilde T_{\pi N, \,\sigma_l N} &=& T_{\pi N,\,\sigma_l N}(\sqrt{s},\mu^2_l) \sqrt{ \Delta\mu^2_lA_{\sigma_l}(\Delta\mu^2_l) },\nonumber \\
\tilde K_{\pi N, \,\sigma_l N} &=& K_{\pi N,\,\sigma_l N}(\sqrt{s},\mu^2_l) \sqrt{ \Delta\mu^2_lA_{\sigma_l}(\Delta\mu^2_l) },\nonumber \\
\tilde T_{\sigma_j N, \,\sigma_l N} &=&\sqrt{ \Delta\mu^2_j A_{\sigma_j}(\mu_j^2) }\,\, T_{\sigma_j N,\,\sigma_l N}(\sqrt{s},\mu_j^2,\mu^2_l)\,\, \sqrt{ \Delta\mu^2_lA_{\sigma_l}(\mu_l^2) },\nonumber\\
\tilde K_{\sigma_j N, \,\sigma_l N} &=&\sqrt{ \Delta\mu^2_j A_{\sigma_j}(\mu_j^2) }\,\, K_{\sigma_j N,\,\sigma_l N}(\sqrt{s},\mu_j^2,\mu^2_l)\,\, \sqrt{ \Delta\mu^2_lA_{\sigma_l}(\mu_l^2) },\nonumber\\
&...,&
\label{appendixA_eq2}
\eea
the integral Eq.~\refe{appendixA_eq1.1} reduces  to the  following sum:
\bea 
\int_{4 m_\pi^2}^{(\sqrt{s}-m_N)^2} d\mu^2 A_\sigma(\mu^2)\, T^{JP}_{f\,\sigma N}(\sqrt{s},\mu^2)\, K_{\sigma N\, i}^{JP}(\sqrt{s},\mu^2)
= \sum_l \tilde T^{JP}_{f\,\sigma_l N}\,  \tilde K_{\sigma_l N\, i}^{JP}.
\label{appendixA_eq1.2}
\eea

Defining the matrices $[\tilde T^{JP} ]$

 \be
    [\tilde T^{JP}] = \left( \begin{array}{llll}
      {     T^{JP}_{\pi N, \,\pi N}}        &{ \tilde T^{JP}_{\pi N,\, \sigma_1 N}}        & {\tilde T^{JP}_{\pi N,\, \sigma_2 N}} \,\,  \cdots \\
       { \tilde T^{JP}_{\sigma_1 N,\, \pi N} }   &{\tilde  T^{JP}_{\sigma_1 N,\, \sigma_1 N}  } & {\tilde T^{JP}_{\sigma_1 N,\, \sigma_2 N}  } \,\, \cdots \\
       { \tilde T^{JP }_{\sigma_2 N,\, \pi N} }    &{\tilde T^{JP}_{\sigma_2 N,\, \sigma_1 N}}  & {\tilde T^{JP}_{\sigma_2 N, \, \sigma_2 N} }\,\, \cdots \\
        \cdots & \cdots &\cdots\hspace*{0.8cm}\cdots
      \label{appendixA_eq3}
      \end{array} \right)
    \ee
 and $[\tilde K^{JP} ]$ 

 \be
    [\tilde K^{JP}] = \left( \begin{array}{llll}
      {     K^{JP}_{\pi N, \,\pi N}}        &{ \tilde K^{JP}_{\pi N,\, \sigma_1 N}}        & {\tilde K^{JP}_{\pi N,\, \sigma_2 N}} \,\,  \cdots \\
       { \tilde K^{JP}_{\sigma_1 N,\, \pi N} }   &{\tilde  K^{JP}_{\sigma_1 N,\, \sigma_1 N}  } & {\tilde K^{JP}_{\sigma_1 N,\, \sigma_2 N}  } \,\, \cdots \\
       { \tilde K^{JP }_{\sigma_2 N,\, \pi N} }    &{\tilde K^{JP}_{\sigma_2 N,\, \sigma_1 N}}  & {\tilde K^{JP}_{\sigma_2 N, \, \sigma_2 N} }\,\, \cdots \\
        \cdots & \cdots &\cdots\hspace*{0.8cm}\cdots
      \label{appendixA_eq4}
      \end{array} \right)
    \ee
the equations Eq.~\refe{appendixA_eq1}  get the matrix form
\bea
[\tilde T^{JP}]=[\tilde K^{JP}]+i [\tilde K^{JP}][\tilde T^{JP}].
\label{appendixA_eq5} 
\eea
The solution Eq. \refe{appendixA_eq5}  can be represented as 
\bea
[\tilde T^{JP}]=\frac{[\tilde K^{JP}]}{1-i [\tilde K^{JP}]}.
\label{appendixA_eq5.1} 
\eea

It is well known \cite{Penner:2002a,Sauermann:1997,Pearce:1990uj,Feuster:1998a} that the structure of  Eq.~\refe{appendixA_eq5.1} guarantees
the  maintenance of the two-body unitarity. In the present case it read as 
 \bea
{\rm Im }\,  T^{JP}_{\pi N \to \pi N} = \frac{k^2}{4\pi} \left( \sigma^{JP}_{\pi N\to  \pi N}+ \sum_{j}\sigma^{JP}_{\pi N \to \sigma_j N}  \right),
\label{appendixA_eq6}
 \eea
provided that the interaction kernel  $[\tilde K^{JP}]$ is hermitian. The first term in brackets of the right side of the Eq.~\refe{appendixA_eq6}
denotes the total  $\pi N$ elastic partial wave cross section and the second one is a sum of all inelastic partial wave cross sections.
It can be rewritten as
\bea
\sum_{j}\sigma^{JP}_{\pi N \to \sigma_j N} & =& \frac{4\pi}{k^2}\sum_j  |\tilde T^{JP}_{\pi N, \,\sigma_j N}|^2 \nonumber\\
 & =& \frac{4\pi}{k^2}\sum_j  | T^{JP}_{\pi N, \,\sigma_j N} (\sqrt{s},\mu^2_j)|^2 \,\Delta\mu^2_j\, A_\sigma(\mu^2_j)\nonumber\\
&=&\frac{4\pi}{k^2} \int_{4m_\pi^2}^{(\sqrt{s}-m_N)^2  } d\mu^2 A_\sigma(\mu^2) | T^{JP}_{\pi N, \,\sigma_j N} (\sqrt{s},\mu^2_j)|^2 \nonumber\\
&=&\sigma^{JP}_{\pi N\to\sigma N},
\label{appendixA_eq6.1}
\eea
where $\sigma^{JP}_{\pi N\to\sigma N}$ is a total $\sigma$-meson production cross section for the given total spin $J$ and parity $P$.
It remains to show that $\sigma^{JP}_{\pi N\to\sigma N}=\sigma^{JP}_{\pi N\to \pi\pi N}$ where pions are exclusively produced from the $\sigma$-meson decay.
The total cross section can be written in the form
\bea
\sigma^{JP}_{\pi N \to \pi\pi N} =  \frac{(2\pi)^4} {4 \sqrt{ (q_\pi\, p_N)^2- m_N^2 m_\pi^2  }} \int 
\frac{d^3 q_1'}{2E_1'(2\pi)^3 }
\frac{d^3 q_2'}{2E_2'(2\pi)^3 }
\frac{d^3 p_N'}{2E_N'(2\pi)^3 }\nonumber\\
\times \,| \overline{T^{JP}_{\pi N \to \pi \pi N}}|^2  \delta^4 (p_N +q'_\pi - q'_1 - q_2 - p_N'),
\label{appendixA_eq7}
\eea
where $p_N'$, $q_1'$, and  $q_2'$,  are four-momenta of the final nucleon  and the pions respectively, $p_N$ and  $q_\pi$ are momenta of the initial nucleon and 
the pion. The expression Eq.\refe{appendixA_eq7} can be rewritten in the form
\bea
\sigma^{JP}_{\pi N \to \pi\pi N}& =&  \frac{1} {8\pi \sqrt{ (q_\pi\, p_N)^2- m_N^2 m_\pi^2  }} \int 
d \mu^2 \,dF_2(s,m_N^2,s_{\pi\pi})\,dF_2(\mu^2,m_\pi^2,m_\pi^2)
| \overline{T^{JP}_{\pi N \to \pi \pi N}}|^2 ,\nonumber\\
\label{appendixA_eq8}
\eea
where the two-body phase spaces are given as 
\bea
dF_2(s,m_N^2,\mu^2) &=&  \frac{d^3 k }{2E_k(2\pi)^3 }\frac{d^3 p_N'}{2E_N'(2\pi)^3 } (2\pi)^4 \delta^4 (p_N +q_\pi - k - p_N'),\nonumber\\
dF_2(\mu^2,m_\pi^2,m_\pi^2) &=&  \frac{d^3 q_1' }{2E_1'(2\pi)^3 }\frac{d^3 q_2'}{2E_2'(2\pi)^3 }(2\pi)^4 \delta^4 (k-q_1' -q_2'),\nonumber\\
\label{appendixA_eq9}
\eea
with $\mu^2=E_k^2-{\bf k}^2=(q_1'+q_2')^2$. The transition amplitude $T_{\pi N \to \pi \pi N}$ is given by the expression
\bea
  T_{\pi N \to \pi \pi N} = T_{\pi N,\,  \sigma N }(p_N',q_\sigma')\, G_{\sigma}( q_\sigma'^2) \, V_{\sigma\pi\pi}(q'_\sigma, q_1',q_2'),
\eea
where  $T_{\pi N,\,  \sigma N }(p_N',q_\sigma')$  is a amplitude of the isobar production,
$G_{\sigma}( q_\sigma'^2)$  stands for the $\sigma$-meson propagator Eq.~\refe{eq1.1},  and $V_{\sigma\pi\pi}(q_\sigma', q_1',q_2')$
denotes $\sigma \pi\pi$ decay vertex. 

Since the  two-particle phase space is invariant under inhomogeneous Lorentz  transformations
the integrals over $d F_2(s_{\pi\pi},m_\pi^2,m_\pi^2)$ and $dF_2(s,m_N^2,s_{\pi\pi})$ can independently be evaluated  in separate reference frames.
The integration over $ dF_2(\mu^2,m_\pi^2,m_\pi^2)$ is evaluated in the $\sigma$-meson rest frame:
\bea
(2\pi)^4\int F_2(\mu^2,m_\pi^2,m_\pi^2)  |V_{\sigma\pi\pi}(q'_\sigma, q_1',q_2')|^2 = 2\sqrt{\mu^2}\, \Gamma_{\sigma\to\pi\pi}(\mu^2)=
2\Sigma_\sigma(\mu^2),
%(2\pi)^4\int F_2(\mu^2,m_\pi^2,m_\pi^2)  |V_{\sigma\pi\pi}(q'_\sigma, q_1',q_2')|^2 = 2\sqrt{\mu^2}\, \Gamma_{\sigma\to\pi\pi}(\mu^2)=
%2\sqrt{E_\sigma}\, \Gamma_{\sigma\to\pi\pi}'(\mu^2),
\label{appendixA_eq9.1}
\eea
%where $\Gamma_{\sigma\to\pi\pi}(\mu^2)$ is a decay width of the sigma meson in its rest frame, $\Gamma_{\sigma\to\pi\pi}'(\mu^2)$
%and $E'_\sigma$  are decay width  and the energy of the $\sigma$-meson in the  $\sigma N$  c.m. reference frame.
where  we  use relation between the decay width of the $\sigma$-meson   $\Gamma_{\sigma\to\pi\pi}(\mu^2)$
and the $\sigma$-meson self-energy $\Sigma_\sigma(\mu^2)$ calculated in the ladder approximation to  DSE, see Section \ref{scatt.equation}.
Using the result of Eq.~\refe{appendixA_eq9.1} and definitions  Eqs.~(\ref{eq1.1},\,\ref{eq4.1})  the intergral Eq.~\refe{appendixA_eq8} becomes
\bea
\sigma^{JP}_{\pi N \to \pi\pi N} =  \frac{1} { 4\sqrt{ (q_\pi\, p_N)^2- m_N^2 m_\pi^2  }} \int 
d \mu^2 \,dF_2(s,m_N^2,\mu^2)
| \overline{T_{\pi N \to \sigma N}}|^2 A_\sigma(\mu^2).
\label{appendixA_eq10}
\eea
Since $\Sigma_{\sigma}(\mu^2)$ is invariant under inhomogeneous Lorentz transformations one can evaluate Eq.~\refe{appendixA_eq10}
in the $\sigma N$ c.m. reference frame which gives 
\bea
\sigma^{JP}_{\pi N \to \pi\pi N} =  \frac{4\pi} { k^2} \int_{4\pi^2}^{\sqrt{s} -m_N} 
d \mu^2  | \overline{T_{\pi N \to \sigma N}^{JP}}|^2 A_\sigma(\mu^2)=\sigma^{JP}_{\pi N \to \sigma N}
\label{appendixA_eq11}
\eea
where only contributions which the total spin $J$ and parity $P$ have been taken into account.
Hence Eq.~\refe{appendixA_eq6} reads as 
 \bea
{\rm Im }\,  T^{JP}_{\pi N \to \pi N} = \frac{k^2}{4\pi} \left( \sigma^{JP}_{\pi N\to  \pi N}+ \sigma^{JP}_{\pi N \to \pi\pi N}  \right)
\label{appendixA_eq12}
 \eea
from which follows  that the  condition of the optical theorem Eq.~\refe{unitar3} is fulfilled.

\section{Kinematics of the  $\pi N\to 2\pi N$  reaction\label{Kinematics}}
The differential  cross section for the $\pi N\to \pi \pi N$ transition can be written as
\bea
\sigma_{\pi N \to \pi\pi N}^{\rm cohr/incohr} =  \frac{(2\pi)^4} {4 \sqrt{ (q_\pi\, p_N)^2- m_N^2 m_\pi^2  }} \int 
\frac{d^3 q_1'}{2E_1'(2\pi)^3 }
\frac{d^3 q_2'}{2E_2'(2\pi)^3 }
\frac{d^3 p_N'}{2E_N'(2\pi)^3 }\nonumber\\
\times \,| \overline{T_{\pi N \to \pi \pi N}^{\rm coher /incohr}}|^2  \delta^4 (p_N +q'_\pi - q'_1 - q_2 - p_N'),
\label{kinematics_eq1}
\eea
where $p_N'$, $q_1'$, and  $q_2'$  are four-momenta of the final nucleon  and the pions, $p_N$ and  $q_\pi$ are momenta of the initial nucleon and 
the pion.   The quantities $ \overline{ |T_{\pi N \to \pi \pi N}^{\rm coher /incohr}|^2}$  are defined as
\bea
\overline{|T_{\pi N \to \pi \pi N}^{\rm coher}|^2}&=&  \frac{1}{2} \sum_{s_i s_f}| T_{s_i,\,s_f}^{a}+ T_{s_i,\, s_f}^{b}
+ T_{s_i,\, s_f}^{c}+ T_{s_i,\,s_f}^{d}
 |^2,   \nonumber\\
\overline{|T_{\pi N \to \pi \pi N}^{\rm incoher}|^2}&=& \frac{1}{2} \sum_{s_i s_f}\left(| T_{s_i,\,s_f}^{a} |^2+ |T_{s_i,\, s_f}^{b} |^2
+ |T_{s_i,\, s_f}^{c} |^2+ |T_{s_i,\,s_f}^{d} |^2 \right),
\label{kinematics_eq2}
\eea
where $s_i$ and $s_f$ are spin projections (helicities) of the initial and final nucleon respectively and the amplitudes 
\bea
T_{s_i,\, s_f}^{a}&=&T_{s_i,\, s_f}(\sqrt{s}, q_\sigma',p_N')\, G_\sigma(q_\sigma')\, V^{\sigma\pi\pi}(q_\sigma, q_1',q_2'), \nonumber\\
T_{s_i,\, s_f}^{b}&=&T_{s_i,\, s_f}(\sqrt{s}, q_\sigma',p_N')\, G_\sigma(q_\sigma')\, V^{\sigma\pi\pi}(q_\sigma, q_2',q_1'), \nonumber \\
T_{s_i,\, s_f}^{c}&=& \sum_{s_\dtwo} T_{s_i,\, s_\dtwo}(\sqrt{s}, p'_\dtwo,q_1')\, G_\Delta(p'_\dtwo)\, V^{\Delta\pi N}_{s_\dtwo,\,s_f }(p'_\dtwo, p_N'), \nonumber \\
T_{s_i,\, s_f}^{d}&=& \sum_{s_\done} T_{s_i,\, s_\done}(\sqrt{s}, p'_\done,q_2')\, G_\Delta(p'_\done)\, V^{\Delta\pi N}_{s_\done,\,s_f }(p'_\done, p_N') \nonumber \\
\label{kinematics_eq3}
\eea
correspond to the contributions from the diagrams (a)-(b) depicted in Fig.~\ref{unitar8}. The notation is as follows: $p'_\dtwo=(p_N'+q'_2)$ and 
$p'_\done=(p_N'+q'_1)$ are momenta of the intermediate $\DDelta$ isobar and   $s_\done (s_\dtwo)$   are its spin projections. 
Quantities  $T_{s_i,\, s_f}(\sqrt{s}, q_\sigma',p_N')$, $T_{s_i,\, s_\dtwo}(\sqrt{s}, p'_\dtwo,q_2')$ and $T_{s_i,\, s_\done}(\sqrt{s}, p'_\done,q_1')$ 
stand for  $\sigma N$ and   $\pi \DDelta $ production amplitudes obtained by solving  the scattering equation Eq.~\refe{eq6.1}. 
The kinematic of the reaction is shown in Fig.~\ref{kinematics_fig1}. The vector $\nbox{p_N'}$ lies in the $xy$-plane. All calculations are performed in the c.m. 
system of the initial $\pi N$ state. Since Eq.~\refe{eq6.1}  is  also solved in  the same reference frame no additional boost for the
 $T_{s_i,\, s_f}(\sqrt{s}, q_\sigma',p_N')$, $T_{s_i,\, s_\dtwo}(\sqrt{s}, p'_\dtwo,q_2')$ and $T_{s_i,\, s_\done}(\sqrt{s}, p'_\done,q_1')$ amplitudes
is required. The  $T_{s_i,\, s_f}(\sqrt{s}, q_\sigma',p_N')$ amplitude  is directly calculated from the $\sigma N$ partial waves as shown in Appendix~\ref{Appendix_PWA}. 
The isobar production amplitudes $T_{s_i,\, s_\dtwo}(\sqrt{s}, p'_\dtwo,q_2')$ and $T_{s_i,\, s_\done}(\sqrt{s}, p'_\done,q_1')$  are also 
calculated from the corresponding partial waves, see Appendix~\ref{Appendix_PWA}. 
Since vectors $p'_\done=(p_N'+q'_1)$ and $p'_\dtwo=(p_N'+q'_2)$ do not lie  in the $xy$-plane the obtained amplitudes
are rotated  around $z$-axis by the corresponding azimuthal angles $\phi_{p_\done}$ and  $\phi_{p_\dtwo}$ respectively \cite{Jacob:1959}.
\begin{figure}
  \begin{center}
    \includegraphics[width=8cm]{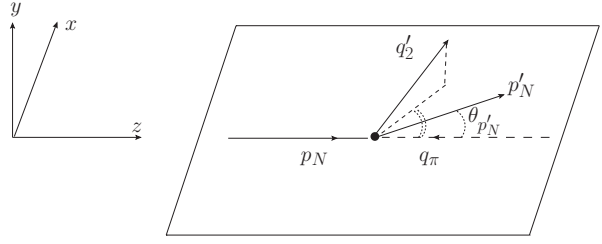}
    \caption{ Kinematics of the reaction $\pi N\to \pi\pi N$
      \label{kinematics_fig1}}
  \end{center}
\end{figure}

The $\sigma\pi\pi$ decay vertex $V^{\sigma\pi\pi}(q_\sigma, q_1',q_2')$ is obtained  from the corresponding interaction Lagrangian.
Due to the spin of the $\DDelta$-isobar  and the final nucleon the evaluation of the $\DDelta \pi N$ decay vertices  
$V^{\DDelta\pi N}_{s_\dtwo,\,s_f }(p'_\dtwo, p_N')$ and $V^{\DDelta\pi N}_{s_\done,\,s_f }(p'_\done, p_N')$ is more involved.
The vertices
\bea
V^{\DDelta\pi N}_{s_{\Delta_j},\,s_f }(p_{\DDelta_j}, p_N')= {\rm i}  \frac{f_{\pi N \DDelta}}{m_\pi} I_{\pi N\Delta}
[\bar u(s_f,p_N') u^\mu(s_{\Delta_j},p_{\Delta_j})] (p_{\Delta_j}-p_N')_\mu,
\label{kinematics_eq3.3}
\eea
are derived  from the corresponding $\pi N\DDelta$ Lagrangian and evaluated  in the c.m. of the initial $\pi N$ state. In this reference 
frame $\DDelta$-isobar is moving 
along the direction defined by a vector sum of the final nucleon $\nbox{p_N'}$ and the  momentum of the pion emitted by  $\DDelta$.  Using Eqs.~\refe{PWA_eq1} the isobar  spin-vectors are first defined in the helicity basis.
Then the decay vertex is numerically evaluated  for each helicity combination. 
    
Here $I_{\pi N\Delta}$ is an isospin factor,  
and the spin-vector $u^\mu(s_{\Delta_j},p_{\Delta_j}) $ satisfies the  Rarita-Schwinger conditions, see  Appendix~{\ref{Appendix_PWA}}. 

The four-fold  differential cross section  reads
\bea
\frac {{\rm d}  \sigma^ \ch  }{ {\rm d} E_N' {\rm d} \cos{\theta_{N'}} {\rm d}\Omega_{q_2'} } = \frac{1}{16 (2\pi)^5\sqrt{(p_N q_\pi)^2  -m_\pi^2 m_N^2}} 
| \overline{T_{\pi N \to \pi \pi N}^{\rm coher /incohr}}|^2  \frac{ {q_2'}^2 p_N' }{|A\, q_2' + C\, E_2'| },
\label{kinematics_eq4}
\eea
where $q_2'= |\nbox{q_2'}|$ and $p_N'=|\nbox{p_N'}|$, $A=2(\sqrt{s}-E_N')$, $C=2p_N' \cos{\theta_{\widehat {p_N', q_2' }}}$. $\theta_{\widehat {p_N', q_2' }}$
is the angle between the vectors $\nbox{q_2'}$ and $\nbox{p_N'}$. Defining quantities 
\bea
\alpha &=& C^2 - A^2, \nonumber\\
\beta &=& 2C\,(s+m_N^2 -2\sqrt{s}\,E_N'),\nonumber\\
\zeta &=& B^2 -A^2m_\pi^2, \nonumber\\
\mathcal D &=& \beta^2 -4\alpha\, \zeta,
\eea
one obtains the expression for $q_2'$
\bea
q_2' = \frac{\beta \pm \sqrt{\mathcal D}}{2\alpha},
\eea
provided that $\mathcal D \geq0$ and $q_2'>0$.

From the Eq.~\refe{kinematics_eq4} one can define angular and mass distributions:
\bea
\frac {{\rm d}  \sigma^ \ch  }{  {\rm d} \Omega_{N'}}
& =& \frac{1}{16 (2\pi)^5\sqrt{(p_N q_\pi)^2  -m_\pi^2 m_N^2}}
\int 
 {\rm d} E_N' {\rm d}\Omega_{q_2'} 
| \overline{T_{\pi N \to \pi \pi N}^{\rm coher /incohr}}|^2  \frac{ {q_2'}^2 p_N' }{|A\, q_2' + C\, E_2'| },\nonumber\\
\frac {{\rm d}  \sigma^ \ch  }{  {\rm d} m_{\pi\pi}^2}
 &=& \frac{1}{32 \sqrt{s}\,(2\pi)^5\sqrt{(p_N q_\pi)^2  -m_\pi^2 m_N^2}}
\int 
 {\rm d}\Omega_{q_2'} {\rm d} \Omega_{N'}
| \overline{T_{\pi N \to \pi \pi N}^{\rm coher /incohr}}|^2  \frac{ {q_2'}^2 p_N' }{|A\, q_2' + C\, E_2'| },
\label{kinematics_eq5}
\eea
where $m_{\pi\pi}^2=s-2\sqrt{s}\,E_N' +m_N^2$.

\section{Partial wave decomposition of isobar amplitudes \label{Appendix_PWA}}
The general details of the partial wave decomposition can be found in \cite{Jacob:1959}. For the $\pi N\to \pi N$, $\sigma N$ transitions we use the expressions 
which are elaborated in \cite{Penner:2002a}. Here we only consider complications related with the 
$\pi\DDelta$ channel. The spin-vectors  $u^\mu$
in the Rarita-Schwinger
formalism satisfies the set of constraints
 \bea
 (\slash p_\Delta - m_\Delta)u^{\mu}(\lambda_\Delta,p_\Delta)=0,\nonumber\\
 \gamma _\mu u^{\mu}(\lambda_\Delta,p_\Delta)=0, \nonumber\\
 \partial_\mu u^\mu(\lambda_\Delta,p_\Delta)=0.\nonumber\\
\label{PWA_eq1}
 \eea 
The spin structure of the  $(\pi/\sigma) \to \pi \DDelta$ and $\pi \DDelta \to \pi \DDelta$ transitions can be expressed through as
\bea
\bar u^{\mu}(\lambda_\Delta', p_\Delta') A^\mu(p_\Delta',q_{\pi/\sigma}';p_N) u(\lambda_N,p_N)
\eea
and
\bea
\bar u^{\mu}(\lambda_\Delta', p_\Delta') A_{\mu\, \nu }(p_\Delta',q_{\pi/\sigma}';p_\Delta)u^{\nu}(\lambda_\Delta, p_\Delta)
\eea
respectively.
In the c.m. of colliding particles the  amplitudes of the isobar production are functions of the c.m energy $\sqrt{s}$, isobar mass $\mu$, scattering
angle $\theta$ and particle  helicities : $T_{\lambda,\, \lambda'}(\sqrt{s}, p'_\done,q_2')=T_{\lambda',\lambda} (\sqrt{s};\mu, \cos{\theta})$
\bea
T_{\lambda',\lambda} (\sqrt{s};\mu, \cos{\theta}) = {\mathcal N^{-1}} \sum_{J} \frac{2J+1}{4\pi} T_{\lambda',\lambda}^{J}(\sqrt{s};\mu) d^J_{\lambda,\lambda'}(\cos{\theta}),
\label{PWA_eq2}
\eea
where $d^J_{\lambda,\lambda'}(\cos{\theta})$ is a Wigner $d$-function,  $\lambda'(\lambda)$ is a sum of particle  helicities in the final (initial) state,
and  ${\mathcal N}=-\sqrt{k, k'}/((4\pi)^2 2 \sqrt{s} )$ is an overall kinematical normalization factor with $k\,(k')$ being initial (final) c.m. momentum.
The $d$-functions is are normalized  in the conventional way: 
\bea
\int_{-1}^{+1} {\rm d}\cos{\theta}~d^J_{\lambda,\lambda'}(\cos{\theta})~d^{J'}_{\lambda,\lambda'}(\cos{\theta}) = \frac{2}{2J+1} \delta_{J\,J'}
\label{PWA_eq3}
\eea

 The same 
formulae of Eq.~\refe{PWA_eq2} is used for the decomposition of the interaction kernel $K_{\lambda',\lambda} (\sqrt{s};\mu, \cos{\theta}) $.

The inverse transformation is 
\bea
 T_{\lambda',\lambda}^{J}(\sqrt{s};\mu) = (2\pi)\, \mathcal{N} \int_{-1}^{+1} {\rm d}\cos{\theta}~
T^{}_{\lambda',\lambda} (\sqrt{s};\mu, \cos{\theta})~ d^J_{\lambda,\lambda'}(\cos{\theta}),\nonumber\\ 
 K_{\lambda',\lambda}^{J}(\sqrt{s};\mu) = (2\pi)\, \mathcal{N} \int_{-1}^{+1} {\rm d}\cos{\theta}~
K^{}_{\lambda',\lambda} (\sqrt{s};\mu, \cos{\theta})~ d^J_{\lambda,\lambda'}(\cos{\theta}).\nonumber\\
\label{PWA_eq4}
\eea
There are four (eight) independent amplitudes to describe the various helicity combinations of $\pi N \to \pi\DDelta$ ($\pi\DDelta \to \pi\DDelta$) transitions.
Due to the parity conservation in the strong interaction one can define amplitude with well defined parity $P=\pm 1$ as linear combinations:
\bea
T_{\lambda',\lambda}^{J\pm}(\sqrt{s};\mu) = T_{\lambda',\lambda}^{J}(\sqrt{s};\mu) \pm \eta\, T_{\lambda',-\lambda}^{J}(\sqrt{s};\mu) 
\eea
where $\eta=\eta_{m}\eta_{B}(-1)^{J-s_1-s_2}$ and $s_1(s_2)$ and $\eta_{\rm m}(\eta_{\rm B})$ are  the spin and the parity of the meson and baryon  in the entrance channel.

\section{Isospin decomposition of the $\pi N\to \pi\pi N $ reaction \label{Isopsin}}
Due to the isospin conservation all $\pi N \to \pi \pi N$ transitions can be expressed in term amplitudes with well defined isospin.
The 'minimal' isospin decomposition would correspond to the separation 
of the  isospin $\frac{3}{2}$ and $\frac{1}{2}$ states.
Within the isobar approximation this is already enough to separate contributions between $N^*$ and  $\Delta^*$ resonances.
For the $\pi^- p \to \pi^0 \pi^0$ the relevant isospin amplitudes are
\bea
\left< \Delta^0\pi^0 |\pi^- p \right >&=&-\frac{1}{3}\sqrt{\frac{1}{5}}\,\, T^{\fth}_{\pi\Delta}+\frac{\sqrt{2}}{3}\,\, T^\foh_{\pi\Delta},\nonumber\\
\left < \sigma n | \pi^- p    \right>&=& - \sqrt{\frac{2}{3}}\,T^{\foh}_{\sigma N}.
\label{Isospin_e1}
\eea

One can also perform an isospin  decomposition of the $\pi N \to \pi\pi N$ reaction beyond isobar approximation.
Within  the  $[1\otimes 1]\otimes \frac{1}{2}$ scheme the isospin decomposition has the form:
\bea
<\pi^0\pi^0 n|T | \pi^- p > &=& \frac{2}{3} \,\sqrt{\frac{1}{5}}\, T_2^\fth +\frac{\sqrt{2}}{3} \,T_0^\foh, \nonumber\\
<\pi^+\pi^- n|T| \pi^- p > &=& \frac{1}{3} \,\sqrt{\frac{1}{5}}\, T_2^\fth +\frac{1}{3} \,T_1^\fth
-\frac{1}{3} \,T_1^\foh -\frac{\sqrt{2}}{3} \,T_0^\foh, \nonumber\\
<\pi^0\pi^- p|T |\pi^- p > &=& -\sqrt{\frac{1}{10}} \,T_2^\fth -\frac{1}{3}\sqrt{\frac{1}{2}} \,T_1^\fth
+\frac{\sqrt{2}}{3}\sqrt{\frac{1}{2}} \,T_1^\foh,
  \nonumber\\
<\pi^+\pi^+ n| T|\pi^+ p > &=& \sqrt{\frac{4}{5}} \,T_2^\fth,   \nonumber\\
<\pi^+\pi^0 p| T| \pi^+ p > &=& -\sqrt{\frac{1}{10}} \,T_2^\fth -\sqrt{\frac{1}{2}} \,T_1^\fth, 
\label{Isospin_eq2}
\eea
where the upper subscript denotes  the total isospin and the lower one stands for the isospin of the $\pi\pi$ subsystem. 
Thus the $\rho N$-subchannel would only contribute  to the $T_1^\fth$  and $T_1^\foh$ amplitudes.   
The  independent isospin amplitudes $T_2^\fth$,  $T_1^\fth$ ,  $T_1^\foh$  ,  $T_0^\foh$ correspond to the four irreducible representations
of the isospin group and completely define isospin structure of the $\pi N \to 2\pi N$ transitions.  
The isospin amplitudes of Eq.~\refe{Isospin_eq2} can be expressed through the  isobar ones Eq.~\refe{Isospin_e1} as follows:
\bea
T_2^{\fth} &=& -\frac{1}{\sqrt{6}}\, T^{\fth}_{\pi\Delta},\nonumber\\
T_0^{\foh} &=& \sqrt{\frac{2}{3}} \,T^{\foh}_{\pi\Delta} -2\sqrt{3}\,T^{\foh}_{\sigma N}.
\eea

\section{Interaction Lagrangians \label{Appendix_Lagrangians}} 
In this Appendix we summarize the Lagrangian densities and decay widths of the baryonic resonances.
 
The $\pi NN$ Lagrangian reads
\bea
\mcl_{\pi NN} = \frac{f_{\pi NN}}{m_\pi}  \bar u_N [\gamma_\mu \gamma_5 { \boldsymbol\tau} ]u_N \partial^\mu {\boldsymbol \pi}.   
\label{Lagr_piNN}
\eea
The $\pi NN^*$ and $\sigma NN^*$ couplings  of the $J^P=\foh^+$ resonance  are
\bea
\mcl_{\varphi N N^* } = \frac{{\rm g}_{\varphi N N^* }}{m_\varphi}\bar u_{N^*}
\left( \begin{array}{c}  \gamma_5 \\ \rm i \end{array} \right)
\gamma^\mu  \tau_\varphi u_N  \partial_\mu
\varphi + h.c.;
\label{Lagr_piNNp}
\eea
 for the the $J^P=\foh^-$ resonance   they are chosen  in the form:
\bea
\mcl_{\varphi N N^* } = \frac{{\rm g}_{\varphi N N^* }}{m_\varphi}\bar u_{N^*}
\left( \begin{array}{c} 1 \\  \rm i\gamma_5 \end{array} \right)
\tau_\varphi u_N  
\varphi + h.c.,
\label{Lagr_piNNm}
\eea
where $\varphi=\pi$, $\sigma$, $\tau_\pi={\boldsymbol \tau}$,  $\tau_\sigma=1$,
and the upper(lower) factor  in the brackets correspond to the $\pi$-( $\sigma$-) meson.

The $\pi\pi\sigma$-coupling is described by 

\bea
\mcl_{\pi\pi\sigma} = \g_{\pi\pi\sigma} m_\sigma \sigma(\boldsymbol{\pi \pi} ).
\label{Lagr_pipisigma}
\eea

The $\pi N\Delta$ coupling  is defined as
\bea
\mcl_{\pi N^*\Delta  } = \frac{{\rm g}_{\pi N\Delta }}{m_\pi}\bar u_{\Delta}^\mu\,
{\boldsymbol T} \,u_N\,  
\partial_\mu {\boldsymbol \pi} + h.c.,
\label{Lagr_piDN}
\eea
and Lagrangian density for the  $N^* \to \pi \DDelta$  transitions 
is given by 
\bea
\mcl_{\pi N^*\Delta  } = \frac{{\rm g}_{\pi N^*\Delta }}{m_\pi}\bar u_{\Delta}^\mu\,
{\boldsymbol T}\left( \begin{array}{c}  1 \\ \rm i\gamma_5 \end{array} \right)\,
 \,u_{N^*}  \,
\partial_\mu {\boldsymbol \pi} + h.c.,
\label{Lagr_piDNS}
\eea
where the upper(lower) factor in the brackets stands for the positive(negative)-parity nucleon resonance.
The isospin transition factor $\boldsymbol T$ can be defined via the  Clebsch-Gordan coefficient  
$\boldsymbol T= C_{\foh,\, I_N; \,\, 1\, I_\pi}^{\fth I_\Delta}$,
where $I_N$, $I_\pi$, and $ I_\Delta$ are isospin projections of the nucleon resonance, the pion,  and  $\DDelta$ respectively.
\end{appendix} 
The decay width  of the $\sigma$-meson,  $\Gamma_\sigma(\mu^2_\sigma)$,  and $\DDelta$-isobar,  $\Gamma_{\Delta}(\mu^2_\Delta)$, are readily obtained using Lagrangian desities  Eqs.~(\ref{Lagr_pipisigma}, \ref{Lagr_piDNS})
as functions of the isobar masses. The isobar self-energy is a solution of the DSE-type equaiton  Eq.\refe{eq1.0}. In the K-matirx approximation the 
imaginary part of the isobar self-energy can be expressed in terms of the 
isobar decay width as follows:
\bea
&&\SSigma_\sigma(\mu^2_\sigma) =\sqrt{\mu^2_\sigma}\, \Gamma_\sigma(\mu^2_\sigma),\nonumber\\
&&\SSigma_\Delta(\mu^2_\Delta) =\sqrt{\mu^2_\Delta}\, \Gamma_\Delta(\mu^2_\Delta)
\eea
The partial decay width of the $N^*(1525)$ and $\NN(1440)$ states are defined through as follows
\bea
&&\Gamma_{\sigma N} = \int_{4m_\pi^2}^{(m_{N^*}-m_N)^2 }{\rm d}\mu^2_\sigma A_\sigma(\mu^2_\sigma) \Gamma_{\sigma N} (\mu^2_\sigma),\nonumber\\
&&\Gamma_{\pi\Delta} = \int_{(m_N +m_\pi)^2}^{(m_{N^*}-m_\pi)^2 }{\rm d}\mu^2_\Delta A_\sigma(\mu^2_\Delta) \Gamma_{\Delta N} (\mu^2_\Delta),
\eea
where the quantitites $\Gamma_{\sigma N} (\mu^2_\sigma)$ and $ \Gamma_{\Delta N} (\mu^2_\Delta)$ can be readily evaluated from the couplings 
Eqs.~(\ref{Lagr_piNNp},\ref{Lagr_piNNm}, \ref{Lagr_piDNS}). The coupling constants used in the calculations
are given in Table~\ref{resonance_coupling_constants}.

\begin{table}[h]
	\begin{center}
		\begin{tabular}
			{l|l|l }
			\hhline{===}
			coupling constant  &  N(1535) &  N(1440)    \\
			\hhline{===}
	$\g_{\pi NN^*}$            & 0.5627 & 7.407        \\
	$\g_{\sigma NN^*}$         & 0.00 &  -7.61       \\		
	$\g_{\pi N^*\Delta(1232)}$ & 0.00 &   7.68     \\		
			\hhline{===}
		\end{tabular}
	\end{center}
	\caption{Resonance coupling constants used in the calculatuions.
		\label{resonance_coupling_constants}}
\end{table}

\bibliographystyle{h-physrev}
\bibliography{tau1}

\end{document}